\documentclass[fleqn,usenatbib]{mnras}

\usepackage[T1]{fontenc}
\usepackage{ae,aecompl}

\DeclareRobustCommand{\VAN}[3]{#2}
\let\VANthebibliography\thebibliography
\def\thebibliography{\DeclareRobustCommand{\VAN}[3]{##3}\VANthebibliography}

\DeclareSymbolFont{matha}{OML}{txmi}{m}{it}% txfonts
\DeclareMathSymbol{\varv}{\mathord}{matha}{29}

\usepackage[normalem]{ulem} 
\usepackage{graphicx}	% Including figure files
\usepackage{amsmath}	% Advanced maths commands
\usepackage{amssymb}	% Extra maths symbols
\usepackage{color}

\newcommand{\ecr}[0]{\varepsilon_{\mathrm{cr}}}
\newcommand{\fcr}[0]{f_{\mathrm{cr}}}

\newcommand{\pcr}[0]{P_{\mathrm{cr}}}
\newcommand{\pth}[0]{P_{\mathrm{th}}}
\newcommand{\ewp}[0]{\varepsilon_{{\rm a}, +}}

\newcommand{\ewm}[0]{\varepsilon_{{\rm a},-}}

\newcommand{\ewpm}[0]{\varepsilon_{{\rm a},\pm}}
\newcommand{\pwpm}[0]{P_{{\rm a},\pm}}
\newcommand{\vcr}[0]{\varv_{\mathrm{cr}}}
\newcommand{\va}[0]{\varv_\mathrm{a}}

\newcommand{\eps}{\varepsilon}

\title[A finite volume method for two-moment CRHD]{A finite volume method for two-moment cosmic-ray hydrodynamics on a moving mesh}

\author[T. Thomas et al.]{
T. Thomas,$^{1}$\thanks{E-mail: tthomas@aip.de (TT)}
C. Pfrommer,$^{1}$
R. Pakmor$^{2}$
\\
$^{1}$ Leibniz-Institute for Astrophysics Potsdam (AIP), An der Sternwarte 16, 14482 Potsdam, Germany\\
$^{2}$ Max Planck Institute for Astrophysics, Karl-Schwarzschild-Str. 1, 85741 Garching, Germany
}

\date{Accepted XXX. Received YYY; in original form ZZZ}

\pubyear{2015}

\begin{document}
\label{firstpage}
\pagerange{\pageref{firstpage}--\pageref{lastpage}}
\maketitle

% Abstract of the paper
\begin{abstract}
We present a new numerical algorithm to solve the recently derived equations of two-moment cosmic ray hydrodynamics (CRHD). The algorithm is implemented as a module in the moving mesh \textsc{Arepo} code. Therein, the anisotropic transport of cosmic rays (CRs) along magnetic field lines is discretised using a path-conservative finite volume method on the unstructured time-dependent Voronoi mesh of \textsc{Arepo}. The interaction of CRs and gyroresonant Alfv\'en waves is described by short-timescale source terms in the CRHD equations. We employ a custom-made semi-implicit adaptive time stepping source term integrator to accurately integrate this interaction on the small light-crossing time of the anisotropic transport step. Both the transport and the source term integration step are separated from the evolution of the magneto-hydrodynamical equations using an operator split approach. The new algorithm is tested with a variety of test problems, including shock tubes, a perpendicular magnetised discontinuity, the hydrodynamic response to a CR overpressure, CR acceleration of a warm cloud, and a CR blast wave, which demonstrate that the coupling between CR and magneto-hydrodynamics is robust and accurate. We demonstrate the numerical convergence of the presented scheme using new linear and non-linear analytic solutions.
\end{abstract}

% Select between one and six entries from the list of approved keywords.
% Don't make up new ones.

%%%%%%%%%%%%%%%%%%%%%%%%%%%%%%%%%%%%%%%%%%%%%%%%%%

%%%%%%%%%%%%%%%%% BODY OF PAPER %%%%%%%%%%%%%%%%%%

\section{Introduction}

CRs are highly energetic particles that pervade most astrophysical plasmas. While CR electrons are important agents that shape the non-thermal emission seen in radio, X-ray and $\gamma$-ray observations, CR protons in the Milky Way's mid-plane contain on average 100 times more energy than CR electrons and dominate the total energy budget of CRs \citep{2013Zweibel}. Similar to our Milky Way, in many astrophysical environments the energy contained in CRs is sufficiently large that their dynamics is affecting the  thermal gas \citep{1990Boulares,2017Zweibel}. Notable examples of such a situation are winds of star-forming galaxies. Inside these winds CRs have energy densities comparable to those of magnetic fields and thermal energy densities \citep{1991Breitschwerdt,2020Buck}. In the interstellar medium (ISM) the major source of CRs are expanding supernova shocks where the diffusive shock acceleration mechanism accelerates thermal low-energy particles to relativistic energies \citep{1987Blandford,2014Caprioli}.  After they leave their acceleration side and are escaping into the ISM, CRs start to accelerate its ambient medium out of the disc and launch mass-loaded galactic size outflows \citep{2017Heckman}. The dynamics of CR driven winds are investigated using one-dimensional flux tube models  \citep{1991Breitschwerdt, 2008Everett, 2017Recchia, 2019Schmidt}, stratified box simulations targeting the scales of several parsecs \citep{2016Girichidis, 2016Simpson, 2018Farber}, idealized simulations of isolated disks \citep{2003Hanasz,2012Uhlig, 2013Hanasz, 2014Salem,2016PakmorIII, 2017Ruszkowski,2017Wiener, 2018Jacob,2018Butsky,2019Chan,2020Dashyan}, and cosmological simulations \citep{2008Jubelgas,2014SalemII,2016Salem,2020Hopkins,2020Buck}. All of these methods assume different modes of CR propagation and interactions.

CRs with energies $\gtrsim$ GeV are nearly collisionless and interact with their environment by scattering off of magnetic irregularities \citep{BookSchlickeiser}. Proposed candidates for these irregularities are turbulent magnetosonic waves \citep{1975Lee,2002Yan,2013Vukcevic} and Alfv{\'e}n waves on scales comparable to the gyroradius of the CR \citep{1969Kulsrud,1975Skilling}. Depending on the frequency of this scattering, the microphysical transport of CRs drastically changes. If the scattering is infrequent but existing then CRs start to diffusive along magnetic field lines. In addition to this diffusion, a population of CRs might drift with some non-negligible average velocity. If this drift is faster than the local speed of Alfv{\'e}n waves then CRs excite gyroresonant Alfv{\'e}n waves through the gyroresonant instability \citep{1969Kulsrud}. These waves are in turn able to scatter CRs more frequently which further amplifies the instability. This self-amplifying feedback loop stops when the average velocity coincides with the Alfv{\'e}n speed, i.e.\ when the CRs start to stream with the Alfv{\'e}n waves. Recent particle-in-cell simulations support this picture \citep{2018Lebiga,2019Holcomb,2019Bai,2019Haggerty}. On a macroscopic level, diffusing or streaming CRs spatially redistribute their energy more independently of the gas flow compared to the internal energy of the gas which seeds a plethora of new dynamics \citep{1971Skilling,2014Salem,2017Wiener}. 

CR transport is commonly described by one of three different theories. The kinetic description directly uses the Maxwell-Newton system of equations to investigate the trajectories and electromagnetic interactions of individual charged particles \citep{2020Pohl}. Fokker-Planck-type theories describe an ensemble of CRs in the full six dimensional phase space with a statistical approach \citep{BookSchlickeiser,2018Malkov}. Hydrodynamical models only describe the spatial transport while encoding the phase space information in a moment hierarchy of the distribution function \citep{1984McKenzie, 1991Breitschwerdt,1992Ko,2017Zweibel}. The hydrodynamic description of CRs is commonly employed in simulations that model large scale astrophysical applications that are not reachable with kinetic simulations. The majority of the published numerical studies uses the one-moment model where CRs are described only by the time evolution of their energy density \citep{2003Hanasz,2006Wagner,2006Pfrommer,2007Pfrommer,2007Ensslin,2008Jubelgas,2012Yang,2013Hanasz,2014Girichidis,2014Salem,2017Pfrommer,2019Dubois}.

While this approach is successful in describing the advective and diffusive transport of CRs, streaming of CRs imposes another challenge. \citet{2010Sharma} observe that a standard finite difference or finite volume discretisation of CR streaming seeds a strong numerical instability that quickly leads to unphysical results. They propose a regularization that adds numerical diffusion and tames the numerical instability. \citet{2018Jiang} recently proposed to use a second moment of the CR distribution function to simultaneously evolve the energy and flux of CRs which automatically cures the numerical instabilities of the one-moment approach.

Inspired by their idea, \citet{2019Thomas} derived equations for the CR energy and momentum flux starting from the quasi-linear Fokker-Planck theory of \citet{BookSchlickeiser}. Including equations for the energy of gyroresonant Alfv{\'e}n waves allows for a description of CR streaming and diffusion based on their gyroresonant interactions and various (collisionless or collisional) damping processes. \citet{2019Thomas} showed that an effective description including the formal order $\mathcal{O}(\bar{\nu}\va^2/c^2)$ of the scattering between CRs and Alfv{\'e}n waves is necessary in order to achieve momentum and energy conservation while complying with the first law of thermodynamics (here, $\bar{\nu}$ is the pitch-angle averaged CR scattering frequency, $\va$ is the Alfv{\'e}n velocity and $c$ is the light speed). Furthermore, the derived theory is only Galilean invariant to this order. The resulting two-moment CR-fluid theory reduces in the limit of strong scattering to previously presented one-moment theories. The emerging dynamics of CRHD was shown to be consistent with the theoretical expectation and free of the numerical instability. Applying CRHD to observations with the MeerKAT radio telescope, which discovered a unique population of faint non-thermal filaments pervading the central molecular zone close to the Galactic center, revealed compelling evidence that GeV CRs are mainly streaming with the local Alfv{\'e}n speed \citep{2020Thomas}.

\citet{2019Thomas} present one-dimensional and highly idealised simulations. In this paper we present a numerical algorithm that can be used to simulate CRHD with the moving mesh code \textsc{Arepo} in three dimensions and for a variety of applications that range from simulating supernova remnant explosions to jets from active galactic nuclei to galaxies and galaxy clusters in cosmological environments. The basis of our algorithm is a path-conservative finite volume method to accurately simulate the anisotropic transport of CRs combined with a custom-made adaptive time stepping integrator to model their gyroresonant interaction with Alfv{\'e}n waves.

Our paper is structured as follows. In Section~\ref{sec:equations} we review the CRHD equations and briefly describe the modelled physics. In Section~\ref{sec:algorithm} we detail our numerical descritization of the CRHD equations on a moving mesh and test the algorithm with various problems in Section~\ref{sec:test}. We present a formulation of CRHD in cosmological comoving coordinates in App.~\ref{app:cosmo}. As path-conservative schemes are not common in computational astrophysics, we present a derivation of those schemes in App.~\ref{app:pc}. In App.~\ref{app:ode-integrator-proofs} we present a mathematical and numerical convergence proof of our source term integrator. Throughout the paper we use Heaviside-Lorentz units and write $\mathbfit{a} \mathbfit{b}$ for the tensor product of $\mathbfit{a}$ and $\mathbfit{b}$.

\section{Equations of Cosmic Ray Hydrodynamics}
\label{sec:equations}

We use the two-fluid approximation to describe the CR-gas composite fluid which allows separate transport of the non-relativistic thermal particle population and high-energy CRs. While the thermal gas is modelled with the MHD approximation, the CRs are assumed to be ultra-relativistic with a particle speed equal to the speed of light $c$.  Additionally, CRs are described using a two-moment approach where the CR energy density and its flux density are evolved independently. We use a grey approach for the CRs that only tracks the total CR energy and flux densities and does not account for their energy-dependence (see \citealp{2020Girichidis} for an energy dependent fluid theory of diffusing CRs). In this scenario \citet{2019Thomas} derived a new set of equation for CRHD based on the Fokker-Planck theory of CRs in its quasi-linear limit. The theory accounts for the anisotropic transport of CRs along magnetic field lines, the gyro-resonant interaction of CRs and Alfv{\'e}n waves, and the coupling of Alfv{\'e}n waves and CRs to the thermal gas. The complete set of equations is:
{
\begin{alignat}{3}
    &\frac{\partial \rho}{\partial t} + \mathbf{\nabla} \mathbf{\cdot} [\rho \mathbfit{u}] = 0,  \label{eq:continuity_equation} \\
    &\frac{\partial \rho \mathbfit{u}}{\partial t} + \mathbf{\nabla} \mathbf{\cdot} [\rho \mathbfit{u} \mathbfit{u} + P_\mathrm{tot} \mathbf{1} - \mathbfit{B} \mathbfit{B}] = \mathbfit{b} \mathbf{\nabla}_\parallel P_\mathrm{cr} \nonumber \\ 
    & \hspace{140pt} + \mathbfit{g}_\mathrm{gri,+} + \mathbfit{g}_\mathrm{gri,-}, \hspace{-10pt} \label{eq:euler_equation} \\
    &\frac{\partial \mathbfit{B}}{\partial t} + \mathbf{\nabla} \mathbf{\cdot} [\mathbfit{B} \mathbfit{u} - \mathbfit{u}\mathbfit{B}] = \mathbf{0}, \\
    &\frac{\partial \varepsilon}{\partial t} + \mathbf{\nabla} \mathbf{\cdot} [\mathbfit{u} (\varepsilon + P_\mathrm{tot}) - (\mathbfit{u} \mathbf{\cdot} \mathbfit{B}) \mathbfit{B}] = (P_\mathrm{cr}  + P_\mathrm{a,+}  + P_\mathrm{a,-} )\mathbf{\nabla} \mathbf{\cdot} \mathbfit{u}  \nonumber \\
    &\hspace{35pt} +\mathbfit{u} \mathbf{\cdot} (\mathbfit{b} \mathbf{\nabla}_\parallel P_\mathrm{cr} + \mathbfit{g}_\mathrm{gri,+} + \mathbfit{g}_\mathrm{gri,-})  + Q_+ + Q_- ,\label{eq:gas_energy_equation} \\
    &\frac{\partial \varepsilon_\mathrm{cr}}{\partial t} + \mathbf{\nabla} \mathbf{\cdot} [\mathbfit{u} \varepsilon_\mathrm{cr} + \mathbfit{b} \fcr] = -P_\mathrm{cr} \mathbf{\nabla} \mathbf{\cdot} \mathbfit{u}   \nonumber \\  
    & \hspace{115pt}- \va \mathbfit{b} \mathbf{\cdot} \left(\mathbfit{g}_\mathrm{gri,+} - \mathbfit{g}_\mathrm{gri,-}\right), \label{eq:cr_energy_equation} \\ 
    &\frac{\partial f_\mathrm{cr}}{\partial t} + \mathbf{\nabla} \mathbf{\cdot} [\mathbfit{u} f_\mathrm{cr}] + c^2 \mathbf{\nabla}_\parallel \pcr = -f_\mathrm{cr} (\mathbfit{b}\mathbfit{b}) \mathbf{:} \mathbf{\nabla} \mathbfit{u}  \nonumber \\
    &\hspace{125pt}\phantom{.} - c^2 \mathbfit{b} \mathbf{\cdot} \left(\mathbfit{g}_\mathrm{gri,+} + \mathbfit{g}_\mathrm{gri,-}\right), \hspace{-10pt} \label{eq:cr_flux_equation}\\
    &\frac{\partial \varepsilon_\mathrm{a,\pm}}{\partial t} + \mathbf{\nabla} \mathbf{\cdot} [\mathbfit{u} \varepsilon_\mathrm{a,\pm}  \pm \va \mathbfit{b} \varepsilon_\mathrm{a,\pm} ] = -P_\mathrm{a,\pm} \mathbf{\nabla} \mathbf{\cdot} \mathbfit{u}\nonumber\\
    & \hspace{140pt} \pm \va \mathbfit{b} \mathbf{\cdot} \mathbfit{g}_\mathrm{gri,\pm} - Q_\pm, \label{eq:wave_equation} \hspace{-10pt}
\end{alignat}
}
where $\rho$ is the mass density of the thermal gas, $\rho \mathbfit{u}$ its momentum density, $\mathbfit{B}$ the magnetic field, $\mathbfit{b}$ the unit direction of the magnetic field, $\mathbf{\nabla}_\parallel$ the gradient along this direction, $\ecr$ is the energy density of CRs, $\fcr$ is the energy flux density of CRs along the direction of the magnetic field, $\ewpm$ is the energy density of gyroresonant Alfv{\'e}n waves, and $\va = B / \sqrt{\rho}$ is the Alfv{\'e}n velocity. The total MHD energy density is given by
\begin{equation}
   \varepsilon = \frac{1}{2} \rho u^2 + \varepsilon_\mathrm{th} + \frac{1}{2} B^2,
\end{equation}
where $\varepsilon_\mathrm{th}$ is the thermal energy density. The total pressure of the composite fluid of MHD, CRs, and Alfv{\'e}n waves is,
\begin{equation}
    P_\mathrm{tot} = P_\mathrm{th} + \frac{1}{2} B^2 + P_\mathrm{cr} + P_\mathrm{a,+} + P_\mathrm{a,-}, \\
\end{equation}
where the thermal, CR, and Alfv{\'e}n wave pressures obey the following equations of state:
\begin{align}
    P_\mathrm{th} &= (\gamma_\mathrm{th} - 1) \varepsilon_\mathrm{th},\\
    P_\mathrm{cr} &= (\gamma_\mathrm{cr} - 1) \varepsilon_\mathrm{cr},\\
    P_\mathrm{a,\pm} &= (\gamma_\mathrm{a} - 1) \varepsilon_\mathrm{a,\pm},
\end{align}
where the individual adiabatic indices are given by:
\begin{align}
    \left[\gamma_\mathrm{th}, \gamma_\mathrm{cr}, \gamma_\mathrm{a} \right] &= \left[ \frac{5}{3}, \frac{4}{3}, \frac{3}{2} \right].
\end{align}

The gyroresonant interaction between Alfv{\'e}n waves and CRs is described by
\begin{align}
    \mathbfit{g}_\mathrm{gri,\pm} &= \frac{\mathbfit{b}}{3\kappa_\pm} [\fcr \mp \va (\ecr + \pcr)] \\
    &= \frac{3\pi}{8} \frac{eB}{\gamma m c^3} \frac{\ewpm}{B^2} \mathbfit{b} [\fcr \mp \va (\ecr + \pcr)],
\end{align}
where $\gamma$ is defined as the Lorentz factor of a typical CR. We use $\gamma = 2$ to simulate a population that is dominated by GeV CR protons. The diffusion coefficients $\kappa_\pm$ are calculated based on the local scattering rate with Alfv{\'e}n waves. Thus, CRHD includes diffusion that is not assuming a fixed diffusion coefficient but follows elements of the unresolved microphysics to calculate spatially and temporally varying diffusion coefficients. The terms $\mathbfit{g}_\mathrm{gri,\pm}$ have the units of a force density and are strictly aligned with the direction of the magnetic field.  

We allow the energy of gyroresonant Alfv{\'e}n waves to be damped by including the damping terms $Q_\pm$ in Eq.~\eqref{eq:wave_equation}. Damping by ion-neutral collisions, interaction with turbulence, and various plasma-kinetic processes including non-linear Landau damping have been proposed to be important \citep{2017Zweibel}. While our code can be easily expanded to include all of effects, we here focus solely on non-linear Landau damping:
\begin{align}
    Q_\pm &= \alpha \ewpm^2 \\
    &= \frac{\sqrt{\pi}}{8B^2} \frac{2eB}{\gamma m c^2} \sqrt{\frac{(\gamma_\mathrm{th} - 1)\varepsilon_\mathrm{th}}{\rho}} \ewpm^2,
\end{align}
where $\alpha$ is the self-coupling constant of Alfv{\'e}n waves. These $Q_\pm$ are quadratic in $\ewpm$ and their numerical discretization is more difficult in comparison to damping terms for ion-neutral or turbulent damping.

Direct collisions between CR particles themselves or CR and thermal particles occur rarely. They interact via Lorentz-forces provided by large-scale (MHD) electromagnetic fields $\mathbfit{B}$ or by scatterings on small scales provided by the energy contained in gyroresonant Alfv{\'e}n waves $\ewpm$. 

Perpendicular to the magnetic field CRs interact with the thermal gas indirectly through Lorentz forces. The Lorentz force acting on the CRs in the perpendicular direction is balanced by the perpendicular CR pressure gradient. This pressure gradient would alter the momentum of the electromagnetic field if we were not using the MHD approximation. Therein the magnetic field lines are frozen in and their dynamics follows the motion of the thermal gas. For this approximation to hold true the perpendicular CR pressure gradient must exert a force on the thermal gas itself and not on the electromagnetic field. This gives rise to the $\mathbf{\nabla}_\perp \pcr = \mathbf{\nabla} \pcr - \mathbfit{b} \nabla_\parallel \pcr = (\mathbf{1} - \mathbfit{b} \mathbfit{b}) \mathbf{\cdot} \mathbf{\nabla} \pcr$ terms in the gas momentum and energy equations in Eqs.~\eqref{eq:euler_equation} and \eqref{eq:gas_energy_equation}.

In our model, CRs interact with their surroundings parallel to the magnetic field only through the gyroresonant scatterings. To lowest order the force balance parallel to the magnetic field is dominated by pitch angle scattering. In this process gyroresonant Alfv{\'e}n waves stochastically scatter CRs by changing their pitch angle. For an ensemble of gyrotropic CRs this alters only the component of the mean momentum that is aligned with the magnetic field. This momentum is transferred to the gyroresonant Alfv{\'e}n waves, which changes the momentum of the thermal particles that support the hydromagnetic Alfv{\'e}n wave on the microscopic level, and cumulatively, this leads to acceleration of the mean gas momentum on the macroscopic level \citep{1981Achterberg,2019Thomas}. This process is encoded in the $\mathbfit{g}_\mathrm{gri}$ terms in Eqs.~\eqref{eq:euler_equation} and \eqref{eq:gas_energy_equation}. In addition to a change in momentum, the gyroresonant interaction also transfers energy between CRs and Alfv{\'e}n waves. This is described by the $\pm \va \mathbfit{b} \mathbf{\cdot} \mathbfit{g}_\mathrm{gri}$ terms in Eqs.~\eqref{eq:cr_energy_equation} and \eqref{eq:wave_equation}. If this transfer is such that CRs lose energy while Alfv{\'e}n waves gain energy then this process is called gyroresonant instability. This is the case when the streaming speed of CRs exceeds the Alfv{\'e}n speed, i.e., forward propagating Alfv{\'e}n waves gain energy when
\begin{equation}
    \vcr = \frac{\fcr}{\ecr + \pcr} > \va
\end{equation}
while backward propagating Alfv{\'e}n waves gain energy when $\vcr < - \va$.

The propagation of CRs in this hydrodynamic model exhibits two extremes. The first one is ballistic transport, which is realised when CRs are not scattered frequently. In this case the $\mathbfit{g}_\mathrm{gri,\pm}$ terms can be neglected and the $\mathbf{\nabla}_\parallel \pcr$ term dominates the flux evolution in  Eq.~\eqref{eq:cr_flux_equation}. CRs propagate with a characteristic light like velocity $c / \sqrt{3}$ and do not couple to the thermal gas along the magnetic field direction. The second extreme is the streaming of CRs with Alfv{\'e}n waves when scattering by those waves dominates. In this case the $\mathbf{\nabla}_\parallel \pcr$ term in the flux equation is negligibly small and the $\mathbfit{g}_\mathrm{gri,\pm}$ terms of Eq.~\eqref{eq:cr_flux_equation} dictate the dynamics. In this regime, CR and thermal fluids are tightly coupled. If the scattering by either forward or backward propagating gyroresonant Alfv{\'e}n waves is strong enough, the steady state of CR streaming can be reached where  $\mathbfit{g}_\mathrm{gri,\pm} = 0$ or equivalently, $\vcr = \pm \va$ for one of the wave types.
If none of these two extremes is an adequate approximation, CRs diffuse along the magnetic field lines. Both the CR pressure gradient and gyroresonant interaction terms contribute to Eq.~\eqref{eq:cr_flux_equation}. If a considerable CR flux is built up by the CR gradient term then the gyroresonant instability may operate and amplify Alfv{\'e}n waves at the expense of the CRs. This increases the scattering rate as $\mathbfit{g}_\mathrm{gri,\pm} \propto \ewpm$ and subsequently decreases $\fcr$ until it finds itself in the streaming regime.  

The speed of light $c$ is larger than any other velocity in the CRHD equations (since we assume the non-relativistic limit of MHD). For applications it is beneficial to use a reduced speed of light that is smaller than the actual speed of light but larger than any other relevant velocity. In this way the hierarchy of the dynamics is preserved and physical implications of the reduced speed of light are minimised. In the following, we use $c$ to denote the reduced speed of light and only replace $c^2$ factors in Eq.~\eqref{eq:cr_flux_equation} with their reduced values. Any other occurring factors of $c$ use its actual value. Although this is an artificial approximation, it heavily reduces the computational cost because it relaxes the time step constraint for a numerically stable simulation.

\section{Numerical Algorithm}
\label{sec:algorithm}

In this section we present a finite volume method that solves a descretized version of Eqs.~\eqref{eq:continuity_equation} to \eqref{eq:wave_equation} on the moving mesh of \textsc{Arepo} \citep{2010Springel}. The majority of terms in Eqs.~\eqref{eq:continuity_equation} to \eqref{eq:wave_equation} depend on the flow velocity $\mathbfit{u}$ and describe adiabatic processes. The remaining gradient or divergence terms are derivatives along the magnetic field lines and represent the anisotropic transport of CRs. Instead of developing a unified finite volume method for both categories, we apply two different finite volume schemes: one for the adiabatic terms which use the advantageous properties of the moving mesh code and another one for the anisotropic transport. In addition to gradient and divergence terms, Eqs.~\eqref{eq:continuity_equation} to \eqref{eq:wave_equation} also contain source terms that describe the fast micro-scale dynamics of CRs and Alfv{\'e}n waves. We develop a special implicit integrator that manages the stiffness of those terms. The two finite volume methods and the source term integrator are combined in an operator-split approach to yield the full evolution for a single time step. We now describe each of those integration steps separately. 

\subsection{Adiabatic CRHD in \textsc{Arepo}}
\label{sec:adiabatics}
During this first step we solve the parts of the CRHD equations that contain all adiabatic processes:
\begin{align}
    \frac{\partial \rho}{\partial t} + \mathbf{\nabla} \mathbf{\cdot} [\rho \mathbfit{u}] &= 0, \\
    \frac{\partial \rho \mathbfit{u}}{\partial t} + \mathbf{\nabla} \mathbf{\cdot} [\rho \mathbfit{u} \mathbfit{u} + P_\mathrm{tot} \mathbf{1} - \mathbfit{B} \mathbfit{B}] &= \mathbf{0}, \\
    \frac{\partial \mathbfit{B}}{\partial t} + \mathbf{\nabla} \mathbf{\cdot} [\mathbfit{B} \mathbfit{u} - \mathbfit{u}\mathbfit{B}] &= \mathbf{0}, \\
    \frac{\partial \varepsilon}{\partial t} + \mathbf{\nabla} \mathbf{\cdot} [\mathbfit{u} (\varepsilon + P_\mathrm{tot}) - (\mathbfit{u} \mathbf{\cdot} \mathbfit{B}) \mathbfit{B}] &= (P_\mathrm{cr}  + P_\mathrm{a,+}  + P_\mathrm{a,-} )\mathbf{\nabla} \mathbf{\cdot} \mathbfit{u}, \label{eq:adiabatic_gas_energy_equation}  \\
    \frac{\partial \varepsilon_\mathrm{cr}}{\partial t} + \mathbf{\nabla} \mathbf{\cdot} [\mathbfit{u} \varepsilon_\mathrm{cr}] &= -P_\mathrm{cr} \mathbf{\nabla} \mathbf{\cdot} \mathbfit{u},  \label{eq:adiabatic_cr_energy_equation} \\
    \frac{\partial f_\mathrm{cr}}{\partial t} + \mathbf{\nabla} \mathbf{\cdot} [\mathbfit{u} f_\mathrm{cr}] &= -f_\mathrm{cr} (\mathbfit{b}\mathbfit{b}) \mathbf{:} \mathbf{\nabla} \mathbfit{u},  \\
    \frac{\partial \varepsilon_\mathrm{a,\pm}}{\partial t} + \mathbf{\nabla} \mathbf{\cdot} [\mathbfit{u} \varepsilon_\mathrm{a,\pm}] &= -P_\mathrm{a,\pm} \mathbf{\nabla} \mathbf{\cdot} \mathbfit{u} . \label{eq:adiabatic_alfven_energy_equation}
\end{align}
These equations mostly resemble those solved in \citet{2017Pfrommer} but additionally include the Lagrangian transport of $\fcr$ and $\ewpm$. 

The equations are solved on the moving mesh of \textsc{Arepo}. We use the second-order accurate time integration and gradient reconstruction scheme \citep{2016PakmorII} and extent the algorithm of \citet{2017Pfrommer}: The finite volume scheme uses an extension of the HLLD \citep{2005HLLD} approximate Riemann solver that takes the additional pressures and increased signal velocities into account. As an input state for the HLLD we use the interface-interpolated primitive variables $\rho$, $\mathbfit{u}$, $\mathbfit{B}$, $\pth$, $\pcr$ and $\pwpm$. The divergence terms on the left-hand side of Eqs.~\eqref{eq:adiabatic_cr_energy_equation} to \eqref{eq:adiabatic_alfven_energy_equation} are upwinded based on the direction of the mass flow at a given interface. The velocity gradient and divergence terms on the right-hand side of Eqs.~\eqref{eq:adiabatic_gas_energy_equation} to \eqref{eq:adiabatic_alfven_energy_equation} are discretised using Gauss' theorem. The velocities needed to evaluate the surface integral are taken to be those velocities of the HLLD Riemann solver that lay on a given interface. We use the Powell scheme for divergence control \citep{1999Powell,2013Pakmor}. The inclusion of the CR dynamics does not interfere with the Powell scheme. 

The quasi-Lagrangian nature of \textsc{Arepo} allows us to evaluate the Riemann solver in a frame that is approximately comoving with the interfaces. This drastically lowers the applied numerical dissipation \citep{2010Springel, 2011Pakmor}. Reynold's transport theorem states that we need to account for the mesh motion with an additional geometric flux in the finite volume scheme. We fully account for this flux during the current, adiabatic, step. As a consequence, we do not need to account for the mesh motion in all subsequent integration steps of this time step. While this choice has the advantage that it simplifies the algorithmic complexity of the upcoming integration steps, it comes at a cost: neglecting the mesh motion corresponds to an operator splitting between adiabatic and parallel transport, which induces an additional numerical error. 

\subsection{Path-conservative scheme for anisotropic transport}
\label{sec:pc_transport}
The gradient and divergence terms in Eqs.~\eqref{eq:continuity_equation} to \eqref{eq:wave_equation} that are aligned with the magnetic field are:
\begin{align}
    \frac{\partial \varepsilon_\mathrm{cr}}{\partial t} &+ \mathbf{\nabla} \mathbf{\cdot} (\mathbfit{b} f_\mathrm{cr}) = 0, \label{eq:parallel_ecr}\\
    \frac{\partial f_\mathrm{cr}}{\partial t} &+ c^2 \mathbfit{b} \mathbf{\cdot} \mathbf{\nabla} P_\mathrm{cr} = 0 , \label{eq:parallel_fcr}\\
    \frac{\partial \rho \mathbfit{u}}{\partial t} &- \mathbfit{b} \mathbfit{b} \mathbf{\cdot} \mathbf{\nabla} P_\mathrm{cr} = \mathbf{0},   \label{eq:parallel_mom}\\
    \frac{\partial \varepsilon_\mathrm{a,\pm}}{\partial t} &+ \mathbf{\nabla} \mathbf{\cdot} (\pm\mathbfit{b} \va \varepsilon_\mathrm{a,\pm}) = 0.  \label{eq:parallel_ewpm}
\end{align}
We omitted the gas energy equation and defer its discussion to the end of this subsection. These equations can be conveniently written into the more compact form 
\begin{align}
    \frac{\partial \mathbfit{U}}{\partial t} + \mathbf{\nabla} \mathbf{\cdot} \mathbfit{F}(\mathbfit{U}) + \mathbfss{H}(\mathbfit{U}) \mathbf{:} \nabla \mathbfit{U} = \mathbf{0}, \label{eq:non_conservative_standard_form}
\end{align}
where\footnote{The rows in the matrix $\mathbfit{F}$ and the rank-3 tensor $\mathbfss{H}$ correspond to entries in the rows of $\mathbfit{U}$. The scalar product and the double contraction in Eq.~\eqref{eq:non_conservative_standard_form} have to be applied to the columns of $\mathbfit{F}$ and $\mathbfss{H}$.}
\begin{align}
    \mathbfit{U} &= \left(\varepsilon_\mathrm{cr}, f_\mathrm{cr}, \rho \mathbfit{u}, \varepsilon_{a,+}, \varepsilon_{a,-} \right)^{T}, \\
    \mathbfit{F} &= \left(\mathbfit{b} f_\mathrm{cr}, \mathbf{0}, \mathbf{0}, +\mathbfit{b} \va \varepsilon_\mathrm{a,+} , -\mathbfit{b} \va \varepsilon_\mathrm{a,-} \right)^{T}, \\
    \mathbfss{H} &= (\gamma_\mathrm{cr} - 1) \left[ \begin{matrix} 0 & 0 & \dots \\ 
    c^2\mathbfit{b} & 0 &\dots  \\
    -\mathbfit{b} \mathbfit{b} & 0 & \dots  \\
    0 & 0 & \dots  \\
    0 & 0 & \dots 
    \end{matrix}\right],
\end{align}
where $\mathbfit{U}$ is the state vector, $\mathbfit{F}$ is the flux vector, and $\mathbfss{H}(\mathbfit{U}) \mathbf{:} \nabla \mathbfit{U}$ is the so called non-conservative product. This non-conservative product poses a conceptional challenge. Commonly employed Godunov-like finite volume methods can be readily applied to equations that only contain a flux divergence but no non-conservative product. Path-conservative finite volume methods generalize the well-established theory and methods of finite volume schemes to equations with non-conservative products \citep{2006Pares}. We provide an introduction to path-conservative schemes in App.~\ref{app:pc}. We use such a path-conservative scheme and describe its implementation on the Voronoi mesh provided by \textsc{Arepo}.
\begin{figure}
\centering{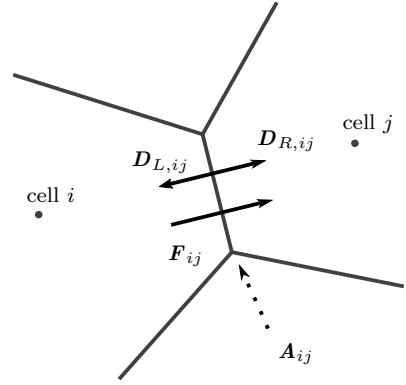}
\caption{ Assignment of the flux $\mathbfit{F}_{ij}$ and linear fluctuations $\mathbfit{D}_{LR,ij}$ for the interface $\mathbfit{A}_{ij}$ as used in the path-conservative finite volume scheme.}
\label{fig:pc_mesh}
\end{figure}
We describe this discretization for a cell with label $i$ that has a volume $V_i$ and shares an interface with an adjacent cell $j$. This interface has a label $ij$ and a vector area $\mathbfit{A}_{ij}$ that points from the inside of $i$ towards $j$. The situation is shown in Fig.~\ref{fig:pc_mesh}. The path-conservative finite volume scheme as generalized for this mesh configuration for a purely non-conservative equation reads
\begin{align}
    V_i \frac{\mathrm{d} \mathbfit{U}_i}{\mathrm{d} t} +  V_i \mathbfss{H}(\mathbfit{U}_i) \mathbf{:} \nabla_\mathrm{recon} \mathbfit{U}_i + \sum_{j} \mathbfit{A}_{ij} \mathbf{\cdot} \mathbfit{D}_{L, ij} = \mathbf{0}, \label{eq:path_conservative_i}
\end{align}
whereas the classical Godunov-like finite volume scheme reads
\begin{align}
    V_i \frac{\mathrm{d} \mathbfit{U}_i}{\mathrm{d} t} + \sum_{j} \mathbfit{A}_{ij} \mathbf{\cdot} \mathbfit{F}_{ij} = \mathbf{0}. \label{eq:finite_volume_i}
\end{align}
The corresponding equation of the path-conservative scheme for the cell $j$ reads 
\begin{align}
    V_j \frac{\mathrm{d} \mathbfit{U}_j}{\mathrm{d} t} + V_j \mathbfss{H}(\mathbfit{U}_j) \mathbf{:} \nabla_\mathrm{recon} \mathbfit{U}_j + \sum_{i} (- \mathbfit{A}_{ij}) \mathbf{\cdot} \mathbfit{D}_{R, ij} = \mathbf{0}. \hspace{-5pt}  \label{eq:path_conservative_j}
\end{align}
The individual terms have the following meaning:
\begin{itemize}
    \item The interface flux $\mathbfit{F}_{ij}$ is the flux $\mathbfit{F}$ exchanged between neighbouring cells. This flux enters only in Godunov-like finite volume schemes and is calculated with an exact or approximate Riemann solver.
    \item The factors $\mathbfit{D}_{L,R, ij}$ are called linear fluctuations and are introduced to incorporate the non-conservative product into the finite volume scheme. They ensure numerical stability and consistency of the numerical solution at discontinuities by adding numerical dissipation (see also App.~\ref{app:pc}). These fluctuations are calculated by an approximate Riemann solver. We add non-zero linear fluctuations only for those equations that contain a non-conservative product, i.e., for $\fcr$ and $\rho \mathbfit{u}$. 
    \item The term $\mathbfss{H}(\mathbfit{U}_i) \mathbf{:} \nabla_\mathrm{recon} \mathbfit{U}_i$ is the non-conservative product evaluated within the cell. This term accounts for the non-conservative product in smooth parts of the flow. The gradient $\nabla_\mathrm{recon} \mathbfit{U}_i$ is the gradient of the linear reconstruction of $\mathbfit{U}$ inside the cell $i$. We use the least-square gradient estimate of \citet{2016PakmorII}. 
\end{itemize}
For purely conservative equations the flux $\mathbfit{F}_{ij}$ can be interpreted as a physical flux, i.e. it tells us the flow rate of $\mathbfit{U}$ through the interface $A_{ij}$. This flux is naturally the same for the right and left adjacent cell of the interface. Such an interpretation is not possible for the linear fluctuations. Here, $\mathbfit{D}_{L, ij} \neq \mathbfit{D}_{R, ij}$ holds in general. 

Equations~\eqref{eq:parallel_ecr} to \eqref{eq:parallel_ewpm} contain either a flux divergence or a non-conservative product such that either Eq.~\eqref{eq:path_conservative_i} or Eq.~\eqref{eq:finite_volume_i} is sufficient to describe their evolution in the finite volume framework. It is possible to describe both a flux divergence and a non-conservative product simultaneously in the same finite volume discretization with minor modifications in the combined version of Eqs.~\eqref{eq:path_conservative_i} and \eqref{eq:finite_volume_i} \citep{2016Dumbser}.

We now describe the Riemann solver that we use to calculate the interface flux and linear fluctuations. The Riemann solver takes the states left ($L$) and right ($R$) of an interface in order to calculate the flux and linear fluctuations based on the input states. For a second-order accurate scheme these states need to be interpolated onto the interface. To this end, we assume that $\mathbfit{U}$ is a linear function inside the cell and use the least-square gradient estimate of \citet{2016PakmorII} for the interpolation. Our Riemann solver is based on the generalization of the HLLE Riemann solver for path-conservative scheme as derived in \citet{2016Dumbser}. The classical form (for a zero non-conservative product) of this Riemann solver assumes that the solution to the Riemann problem at the interface can be approximated by a constant intermediate value, denoted by $\mathbfit{U}_*$. The region of influence of the Riemann problem is assumed to be bounded by a leftwards-travelling wave with speed $S_L$ and a rightwards-travelling wave with speed $S_R$. This idea can be generalized to adopt the HLLE for hyperbolic equations with non-conservative products \citep[see][or App. \ref{app:pc}]{2016Dumbser}. We use this Riemann solver in its localized Lax-Friedrichs (LF) limit where we set $S_L = - S$ and $S_R = +S$ where $S$ denotes the absolute value of the fastest wave speed.  

The individual fluxes and linear fluctuations are calculated as follows. Equation~\eqref{eq:parallel_ecr} describes the evolution of $\ecr$ in conservative form. Consequently, the usual LF flux is
\begin{equation}
    F^{\varepsilon_\mathrm{cr}}_{ij} = \frac{(b f_{\mathrm{cr}})_L + (b f_{\mathrm{cr}})_R}{2} - \frac{S}{2} \left(\varepsilon_{\mathrm{cr},R} - \varepsilon_{\mathrm{cr},L}\right),
    \label{eq:Fecr}
\end{equation}
and the value of the intermediate state is given by
\begin{equation}
    \varepsilon^{*}_{\mathrm{cr}, ij} = \frac{\varepsilon_{\mathrm{cr},L} + \varepsilon_{\mathrm{cr},R}}{2} + \frac{(b f_{\mathrm{cr}})_L + (b f_{\mathrm{cr}})_R}{2S}.
\end{equation}
Here, $b_{L,R} = \mathbfit{A}_{ij} \mathbf{\cdot} \mathbfit{B}_{L,R} / (A_{ij} B_{L,R})$ are the projections of the magnetic field onto the interface normal. We use a notation where $q^*_{ij}$, $F^q_{ij}$, $D^q_{ij, L}$, and $D^q_{ij, R}$ denote the intermediate state, the flux, and linear fluctuations as projected on the interface normal of a quantity $q$. 

Equation~\eqref{eq:parallel_fcr} for $\fcr$ is purely non-conservative. To calculate the intermediate value of $\fcr$ we evaluate equation~(14) of \citet{2016Dumbser} and obtain
\begin{align}
    f^{*}_{\mathrm{cr}, ij} = \frac{f_{cr,L} + f_{cr,R}}{2} &+  c^2 \frac{\bar{b}_{L}}{2S} \left(P^{*}_{\mathrm{cr}, ij} - P_{\mathrm{cr},L}\right) \nonumber \\
    &- c^2  \frac{\bar{b}_{R}}{2S} \left(P^{*}_{\mathrm{cr}, ij} - P_{\mathrm{cr},R}\right), \label{eq:fstar}
\end{align}
where $P^{*}_{\mathrm{cr}, ij} = (\gamma_{cr} - 1) \varepsilon^{*}_{\mathrm{cr}, ij}$, and 
\begin{align}
    \bar{b}_{L} &= \frac{3 b_{L} + b_{R}}{4}, \label{eq:b_l} \\
    \bar{b}_{R} &= \frac{3 b_{R} + b_{L}}{4}, \label{eq:b_r}
\end{align}
are directional biased averages of the magnetic field projections. 
After calculating this value, the linear fluctuations for $\fcr$ are given by
\begin{align}
    D^{f_\mathrm{cr}}_{ij,L} &= + S (f_{\mathrm{cr},L} - f^{*}_{\mathrm{cr}, ij}), \\
    D^{f_\mathrm{cr}}_{ij,R} &= - S (f_{\mathrm{cr},R} - f^{*}_{\mathrm{cr}, ij}).
\end{align}
The description above lacks a value for the speed of the fastest wave of the Riemann problem at the interface. Physically, the fastest wave is formed by ballistically propagating CRs. The speed of this wave is light-like and given by $c \sqrt{\gamma_\mathrm{cr} - 1}$ \citep{2019Thomas}. Because CRs are transported along magnetic fields, we weight this wave speed with the local projection of the magnetic field at the interface to get:
\begin{align}
    S = c \sqrt{\gamma_\mathrm{cr} - 1} \max \left\lbrace b_{L} , b_{R} \right\rbrace. \label{eq:cr_wavespeed}
\end{align}
The equation for the gas momentum density $\rho \mathbfit{u}$ in Eq.~\eqref{eq:parallel_mom} is non-conservative. We use the algebraic similarity between the non-conservative product of $\fcr$ and $\rho \mathbfit{u}$ and define the linear fluctuations via:
\begin{align}
    D^{\rho \mathbfit{u}}_{ij,L} &= -\frac{\mathbfit{b}_L}{c^2} D^{f_\mathrm{cr}}_{ij,L}, \\
    D^{\rho \mathbfit{u}}_{ij,R} &= -\frac{\mathbfit{b}_R}{c^2} D^{f_\mathrm{cr}}_{ij,R},
\end{align}
where $\mathbfit{b}_{L,R} = \mathbfit{B}_{L,R} / B_{L,R}$ are the unit vectors along the direction of the magnetic field to the left and right of the interface. 

The remaining equation for $\ewpm$ in Eq.~\eqref{eq:parallel_ewpm} is conservative. This equation is independent of the light speed and if we used $S$ in Eq.~\eqref{eq:cr_wavespeed} as the signal velocity of $\ewpm$, this would introduce an unnecessary level of numerical dissipation. Hence, we use the classical LF flux:
\begin{equation}
    F^{\varepsilon_\mathrm{a,\pm}}_{ij} = \frac{\left(b \va \varepsilon_{\mathrm{a,\pm}})_L + (b \va \varepsilon_{\mathrm{a,\pm}}\right)_R}{2} - \frac{S_a}{2} \left(\varepsilon_{\mathrm{a,\pm},R} - \varepsilon_{\mathrm{a,\pm},L}\right)
\end{equation}
with an alfv{\'e}nic signal velocity given by:
\begin{align}
    S_a = \max \lbrace (b \va)_L , (b \va)_R \rbrace.
\end{align}

Finally, the kinetic energy changed during the momentum update. We account for this change in a conservative way by setting:
\begin{align}
    \Delta(\varepsilon_\mathrm{kin}) = \Delta(\varepsilon_\mathrm{tot}) = \varepsilon_\mathrm{kin}^{n+1} - \varepsilon_\mathrm{kin}^{n} = \frac{ (\rho \mathbfit{u}^{n+1})^2}{2 \rho} -\frac{ (\rho \mathbfit{u}^n)^2}{2 \rho}. \hspace{-10pt}
\end{align}
Writing the update of the kinetic and thus the total energy in this form does not alter the thermal energy but avoids artificial heating/cooling.

To ensure the stability of this transport step a Courant-Friedrichs-Levy (CFL) criterion on the time step must be fulfilled. We use
\begin{align}
    \Delta t_\mathrm{cr, req} =  \mathrm{CFL} \times \min_\mathrm{cells} \frac{\Delta x}{c \sqrt{\gamma_\mathrm{cr} - 1}},
\end{align}
where $\mathrm{CFL} \sim 0.3$ is the CFL-number, $\Delta x$ is a measure for the cell size, $\Delta t_\mathrm{cr, req}$ is the maximum allowed time step for the parallel transport step, and the minimum is taken over all active cells. The maximum allowed time step for the adiabatic CRHD step $\Delta t_\mathrm{mhd}$ is larger than $\Delta t_\mathrm{cr, req}$ even if a reduced speed of light is used. Conversely, a single parallel transport step is computationally less expensive than a single iteration of the adiabatic CRHD. It is beneficial to execute multiple iterations of the parallel transport step for one iteration of the adiabatic CRHD step to lower the total computational cost of the algorithm. This can be achieved by subcycling of the parallel transport step.  Subcycling also relaxes the effectively required time step criterion for the parallel transport step. We implement subcycling for the path-conservative scheme of this section with a total of $N_\mathrm{cr}$ subcycles. The subcycles are added in an operator-split fashion. To this end, we execute $N_\mathrm{cr} / 2$ subcycles before and after the adiabatic CRHD update is calculated. The overall algorithm (adiabtic CRHD + subcycled parallel transport) is executed with a time step
\begin{equation}
    \Delta t = \min( N_\mathrm{cr} \, \Delta t_\mathrm{cr, req}, \Delta t_\mathrm{mhd})
\end{equation}
while the time step for a single parallel transport subcycle is given by
\begin{equation}
    \Delta t_\mathrm{cr} = \frac{\Delta t}{N_\mathrm{cr}}.
\end{equation}
The downside of this procedure is that it induces some numerical errors. We found, however, that the errors caused by subcycling are quite small for the presented test problems provided a modest number of subcycles is used. 

\subsection{Gyroresonant interaction and wave damping}
\label{sec:source_integrator}
So far we have dealt with all terms that contain spatial gradients and need a finite volume approach for their numerical modelling. The remaining terms fall into the category of source terms. They describe the gyroresonant interaction of CRs with Alfv{\'e}n waves and the subsequent damping of Alfv{\'e}n waves. For the state vector 
\begin{align}
    \mathbfit{U} = \left(\varepsilon_\mathrm{cr}, f_\mathrm{cr}, \varepsilon_{a,+}, \varepsilon_{a,-} \right)^{T},
\end{align}
the source terms in Eqs.~\eqref{eq:continuity_equation} to \eqref{eq:wave_equation} take the form of 
\begin{align}
    \frac{\partial \mathbfit{U}}{\partial t} = \mathbfss{R}(\mathbfit{U}) \mathbfit{U},
    \label{eq:ode}
\end{align}
where the rate matrix $\mathbfss{R}$ is given by:
\begin{align}
    \mathbfss{R}\left(\mathbfit{U}\right) = \begin{pmatrix} \va^2 \chi \gamma_\mathrm{cr} T & - \va \chi D & 0 & 0 \\ c^2 \va \chi \gamma_\mathrm{cr} D & - c^2 \chi T & 0 & 0 \\ -\va^2 \chi \gamma_\mathrm{cr} \varepsilon_{a,+} & +\va \chi \varepsilon_{a,+} & - \alpha \varepsilon_{a,+} & 0 \\ -\va^2 \chi \gamma_\mathrm{cr} \varepsilon_{a,-} & -\va \chi \varepsilon_{a,-} & 0 & - \alpha \varepsilon_{a,-} \end{pmatrix}.
    \label{eq:rate_matrix}
\end{align}
We use abbreviations for the sum and the directional difference of the two Alfv{\'e}n wave energies:
\begin{align}
    T &= \varepsilon_{a,+} + \varepsilon_{a,-}, \\
    D &= \varepsilon_{a,+} - \varepsilon_{a,-},
\end{align}
and define
\begin{align}
    \chi = \frac{3\pi}{8} \frac{3}{\gamma m c^3 B}.
\end{align}

The characteristic timescales of the gyroresonant interaction and wave damping, encoded in the rate matrix, are typically short in comparison to the time scale of the MHD dynamics. To bridge the difference of these timescales, we integrate Eq.~\eqref{eq:ode} using an adaptive time step method. Although a variety of such methods exists, we opt for a custom-made method that takes the special structure of Eq.~\eqref{eq:ode} into account. There is an algorithm at the heart of any adaptive time step method that (i) calculates a suitable source integration time step $\Delta t_\mathrm{src}$ so that the final numerical error is small and (ii) performs the ODE intergration during a subcycle. With the adapative time stepping method the parallel transport time step $\Delta t_\mathrm{cr} \geq \Delta t_\mathrm{src}$ is separated into possibly smaller source integration subcycles. The source intergration is executed after every single parallel transport subcycle.

The integration extends from a given state $\mathbfit{U}^n$ to the state at the next time step $\mathbfit{U}^{n+1}$ using the following semi-implicit method:
\begin{align}
    \mathbfit{U}^p &= \mathbfit{U}^n + \Delta t \mathbfss{R}\left(\mathbfit{U}^{n}\right) \mathbfit{U}^p, \label{eq:rate_predictor}\\
    \mathbfit{U}^r &= \frac{1}{2}(\mathbfit{U}^n + \mathbfit{U}^p), \label{eq:rate_half_step}\\
    \mathbfit{U}^* &= \mathbfit{U}^n + \gamma \Delta t \mathbfss{R}\left(\mathbfit{U}^r\right) \mathbfit{U}^*, \label{eq:state_predictor} \\
    \mathbfit{U}^{n+1} &= \mathbfit{U}^n + \left(1-\gamma \right)\Delta t \mathbfss{R}\left(\mathbfit{U}^r\right) \mathbfit{U}^* \nonumber \\  & \hspace{27pt}+ \gamma \Delta t\mathbfss{R}\left(\mathbfit{U}^r\right) \mathbfit{U}^{n+1},  \label{eq:state_corrector}
\end{align}
where $\gamma =  1 - 2^{-1/2}$. The first step in Eqs.~\eqref{eq:rate_predictor} and \eqref{eq:rate_half_step} returns a first-order prediction of the Alfv{\'e}n wave energy densities at the middle of the current time step which is used in Eqs.~\eqref{eq:state_predictor} and \eqref{eq:state_corrector} to calculate a second-order accurate update of the state. We prove these statements in App.~\ref{app:ode-integrator-proofs}. If the rate matrix $R$ would be a constant, the final update in Eqs.~\eqref{eq:state_predictor} and \eqref{eq:state_corrector} would coincide with the implicit Runge-Kutta method of \citet{2005Pareschi}.

Due to the block-structured lower-triangular form of $R$, the linear equation systems in Eqs.~\eqref{eq:rate_predictor} to \eqref{eq:state_corrector} can be solved efficiently by direct backsubstitution instead of inverting the matrix $R$. For this we perform a 2x2 inversion of the $(\varepsilon_\mathrm{cr}, f_\mathrm{cr})$ equations followed by two trivial 1x1 inversions of the $(\varepsilon_\mathrm{a,+}, \varepsilon_\mathrm{a,-})$ equations.  

The source integration time step is chosen to keep the numerical error small. We use the already calculated first-order accurate solution from the predictor step Eq.~\eqref{eq:state_predictor} as a comparison solution and estimate the (non-dimensional) numerical error by:
\begin{align}
\mathtt{Err} = \max_i \left\lbrace \frac{|\mathbfit{U}_{i}^p - \mathbfit{U}_{i}^{n+1}|}{\mathtt{Atol} + \mathtt{Rtol} \max\left(|\mathbfit{U}_{i}^p|, |\mathbfit{U}_{i}^{n+1}|\right) } \right\rbrace,
\end{align}
where the maximum is taken over all components $\mathbfit{U}_i$ of the state vector $\mathbfit{U}$.
Here, $\mathtt{Rtol}$ and $\mathtt{Atol}$ are the desired relative and absolute errors. We start the adaptive time stepping with $\Delta t_{\mathrm{src}} = \Delta t_{\mathrm{cr}}$ and chose the time step for the next subcycle $\Delta t_{\mathrm{src, next}}$ as  
\begin{align}
    \Delta t_{\mathrm{src, next}} =\Delta t_{\mathrm{src}} \min\left(5, 0.9 \, \mathtt{Err}^{-1/2} \right),
\end{align}
where the numerical factors 5 and 0.9 prevent rapid changes in $\Delta t_{\mathrm{src}}$. If $\mathtt{Err} > 1$ the current subcycle is rejected and restarted. Otherwise, if $\mathtt{Err} \leq 1$ then the current subcycle is accepted, the state vector updated, and the time advanced by $\Delta t_{\mathrm{src}}$. The subcycling is stopped after the sum of $\Delta t_{\mathrm{src}}$ from all accepted subcycles equals $\Delta t_{\mathrm{cr}}$. The typical number of source term subcycles does not exceed 10 for the test problems presented in the next section.

Once the source term integration for the CR variables is completed, we update the gas state. To do so, we calculate the momentum and energy lost by CRs and Alfv{\'e}n waves, and add it to gas via:
\begin{align}
    \Delta\left(U_\mathrm{th}\right) &= \frac{\Delta\left(\varepsilon_\mathrm{th}\right)}{\rho} = - \frac{\Delta\left(\varepsilon_\mathrm{cr} + \varepsilon_\mathrm{a,+} + \varepsilon_\mathrm{a,-}\right)}{\rho}, \\
    \Delta\left(\rho \mathbfit{u}\right) &= - \Delta\left(\mathbfit{b} f_\mathrm{cr}\right), \\
    \Delta\left(\varepsilon_\mathrm{kin}\right) &= -\frac{ [\rho \mathbfit{u}^n + \Delta(\rho \mathbfit{u})]^2}{2 \rho} -\frac{ [\rho \mathbfit{u}^n]^2}{2 \rho},
\end{align}
where $U_\mathrm{th}$ is the internal energy per unit mass and $\Delta(q) = q^{n+1} - q^{n}$ is the total change of $q$ during the source term integration. Simply adding the corresponding work as $\Delta t_{\mathrm{cr}} \mathbfit{u} \mathbf{\cdot} \mathbfit{g}_\mathrm{gri} = -\mathbfit{u} \mathbf{\cdot} \Delta(\mathbfit{b} f_\mathrm{cr})$ to the total energy would cause artificial cooling and heating. Instead we do not directly update the total energy but instead update the internal energy and recalculate the kinetic energy to prevent this problem. Although the temperature increases due to wave damping, we calculate the affected coefficients in Eq.~\eqref{eq:rate_matrix} using the temperature at the start of the integration.

\section{Test Problems}
\label{sec:test}
In this section we test our algorithm with various simplified and complex problem setups. We do not only describe the performance of the algorithm in our discussions but also aim to understand the emerging physical dynamics. 

\subsection{CR-Alfv{\'e}n wave interaction}

In this first problem we test whether the source term integrator from Section~\ref{sec:source_integrator} correctly captures the interactions between CRs and Alfv{\'e}n waves. To this end we perform two simulations with homogeneous initial conditions such that all gradient terms vanish and only the source terms remain. We initialize the thermal gas with $\rho = m_\mathrm{p}\,\mathrm{cm}^{-3}$, $\mathbfit{u} = \mathbf{0}$, $P_\mathrm{th} = (10^4~\mathrm{K})  k_\mathrm{B} \, \mathrm{cm}^{-3}$ and $\va = 10\, \mathrm{km}\,\mathrm{s}^{-1}$. We employ a speed of light with $c=1000\, \mathrm{km}\,\mathrm{s}^{-1}$. We use two setups for CRs and Alfv{\'e}n waves to highlight different behaviours in their dynamics.

\begin{figure}
	\includegraphics[width=\columnwidth]{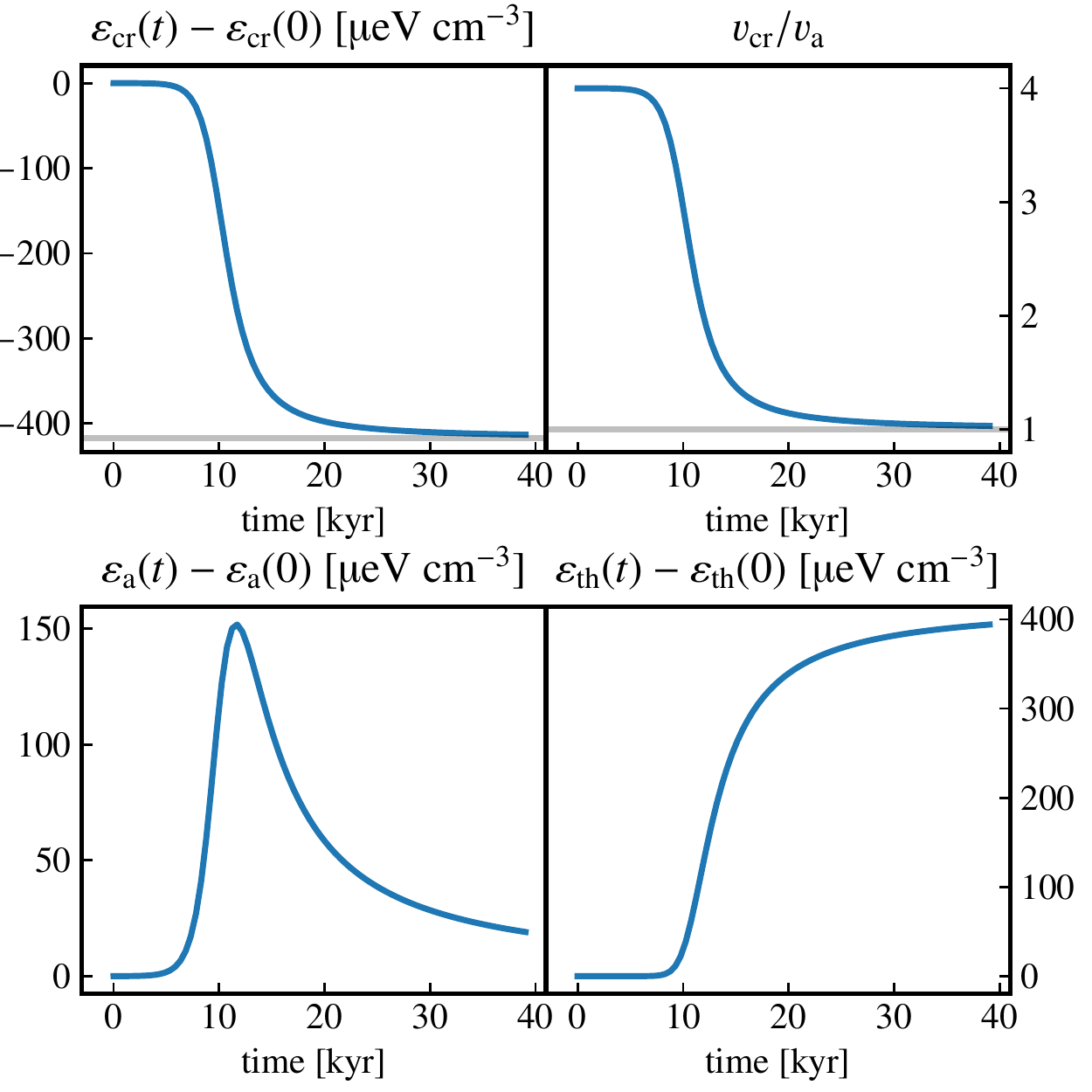}
    \caption{Time evolution of the CR, Alfv{\'e}n wave, thermal energy density, and the streaming speed for the test with initially super-alfv{\'e}nic streaming speeds. The grey line displays the analytic estimate for the asymptotic change in CR energy density and streaming speed.}
    \label{fig:one_zone_4alfv}
\end{figure}

\textbf{Super-alfv{\'e}nic streaming.} Here we investigate CRs streaming with superalfv{\'e}nic velocities. We initialize the CRs with $\ecr = 2 \eps_B$ and $\fcr = 4 \va (\ecr + \pcr)$ while the Alfv{\'e}n waves are initialized with $\ewp = 10^{-3} \ecr$ and $\ewm = 0$. We run the simulation until $t=40$kyr with $\Delta t = 10$ yr. The results are displayed in Fig.~\ref{fig:one_zone_4alfv}.

Initially the gyroresonant instability operates and transfers energy from CRs to Alfv{\'e}n waves, which decelerates the CRs. With increasing Alfv{\'e}n wave energy the growth rate of the gyroresonant instability also increases. This results in a faster deceleration and an energy transfer. Non-linear Landau damping becomes important at $t\sim10$~kyr when Alfv{\'e}n waves have accumulated sufficient energy. This damping thermalizes Alfv{\'e}n waves and in consequence, leads to an increase of thermal energy. At $t\gtrsim15$kyr, CRs are reaching the streaming speed $\vcr = \va$, the gyroresonant instability is weaker, and the deceleration of and energy transfer from CRs is slower. The instability cannot overcome the wave damping and Alfv{\'e}n wave energy is decreasing. At later times the simulation approaches the asymptotic regime where the dynamics is unaltered.

For $t\to\infty$, CRs start to stream with the Alfv{\'e}n waves $\vcr = \fcr / (\ecr + \pcr) = \va$ or $\fcr = \va (\ecr + \pcr)$. From $t=0$ to $t\to\infty$ the CRs lose a momentum density $\Delta \fcr c^{-2} = -3 \va c^{-2} (\ecr + \pcr)$. The corresponding decrease (increase) in CR (Alfv{\'e}n) energy density is $\Delta \ecr = - 3 {\va}^2 c^{-2} (\ecr + \pcr)$. As $\Delta \ecr / \ecr < 1$ the initial value of $\ecr$ can be used to evaluate the previous expressions. The values for $\Delta \ecr$ and $\Delta \fcr$ are reached in the asymptotic limit as indicated by the grey lines in Fig.~\ref{fig:one_zone_4alfv}.

\begin{figure}
	\includegraphics[width=\columnwidth]{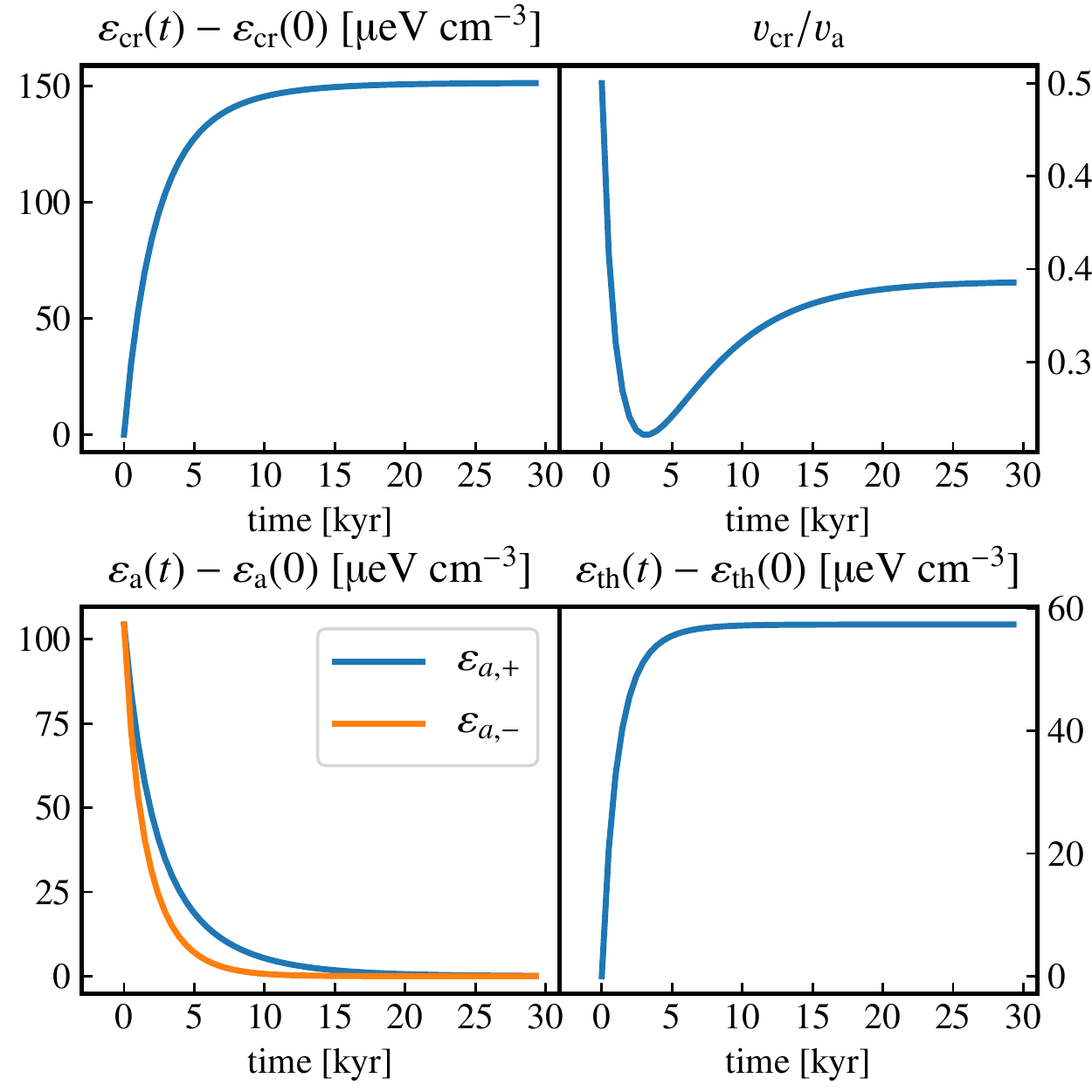}
    \caption{Time evolution of the CR, Alfv{\'e}n wave, thermal energy density, and the streaming speed for the test with an initially operating second-order Fermi process.}
    \label{fig:one_zone_2ndFermi}
\end{figure}

\textbf{Second-order Fermi process.} For this problem, we initialise a reservoir of Alfv{\'e}n waves and CRs that are streaming with sub-alfv{\'e}nic velocities. This triggers an energy transfer from Alfv{\'e}n waves to CRs which is refereed to as the second-order Fermi process \citep{1992Ko}. The CRs are initialized with $\ecr = B^2$ and $\fcr = 0.5 \va (\ecr + \pcr)$ whereas both Alfv{\'e}n wave energy densities are set to $\ewp = \ewm = 10^{-4} \ecr$. We simulate until $t = 30$kyr with a time step of $\Delta t = 10$ yr . The results are displayed in Fig.~\ref{fig:one_zone_2ndFermi}.

During the entire simulation Alfv{\'e}n waves lose energy by accelerating CRs and due to non-linear Landau damping. In turn, the CRs and thermal energy densities are monotonically increasing for all times. The gyroresonant instability cannot be active as CRs are unable to reach super-alfv{\'e}nic velocities in this setup. 
Both backward and forward propagating resonant Alfv{\'e}n waves accelerate the CRs to their respective wave frame. The timescale for these are acceleration are similar as both wave types almost identical energy densities. The situation would be symmetrical if the initial streaming velocity of CRs wouldn't be biased towards the direction of forward propagating Alfv{\'e}n waves. 

The forces exerted by gyroresonant Alfv{\'e}n waves on the CRs in Eq.~\eqref{eq:cr_flux_equation} can be rewritten in terms of difference between CR streaming velocity and the Alfv{\'e}n speed:
\begin{equation}
    \mathbfit{b} \mathbf{\cdot} \left(\mathbfit{g}_\mathrm{gri,+} + \mathbfit{g}_\mathrm{gri,-}\right) \propto \ewp \left(\vcr - \va\right) + \ewm \left(\vcr + \va\right)
\end{equation}
Initially, $\mathbfit{b} \mathbf{\cdot} \mathbfit{g}_\mathrm{gri,-} >  \mathbfit{b} \mathbf{\cdot} \mathbfit{g}_\mathrm{gri,+}$ as the velocity difference between CRs and backward propagating Alfv{\'e}n waves is larger compared to difference between CRs and forward propagating Alfv{\'e}n waves. Thus in the beginning the CRs experience a net acceleration towards $-\va$ which causes the initial decrease in the 
streaming speed. The accompanying energy transfer from Alfv{\'e}n waves to CRs is also stronger which explains the faster depletion of energy in backward propagating Alfv{\'e}n waves. At $t\sim 5$kyr, the remaining backward propagating Alfv{\'e}n waves are unable to overcome the acceleration by the forward propagating Alfv{\'e}n waves. From this point on CRs are accelerated towards $+\va$, which gives rise to an increase in streaming speed. At later times, there is no energy in Alfv{\'e}n waves left and the acceleration of CRs slows down to eventually come to a halt.

\subsection{Shock tubes with CR streaming}

\begin{figure*}
	\includegraphics[width=\textwidth]{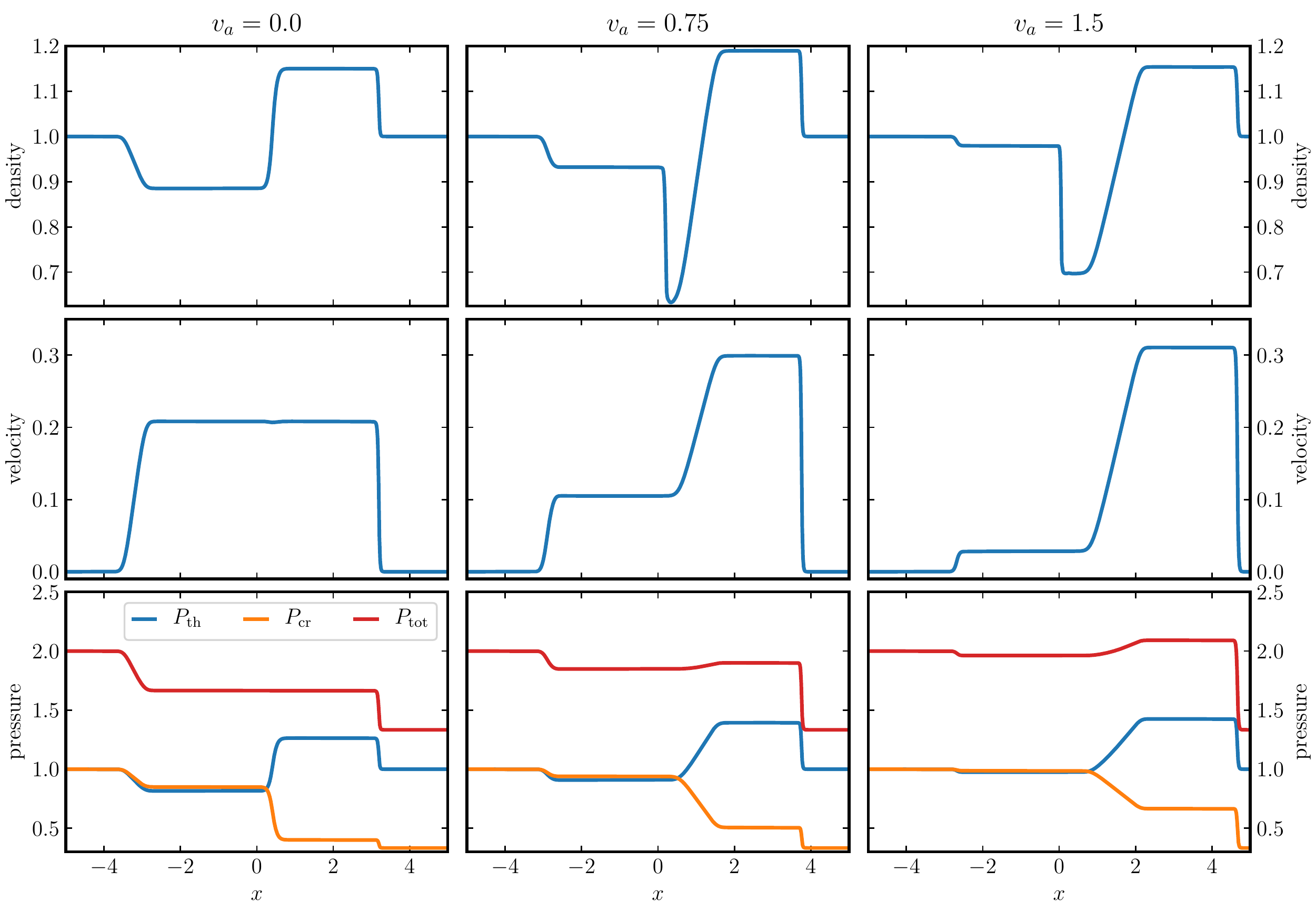}
    \caption{Density, velocity and pressure profiles for shock tubes with different values of the fixed Alfv{\'e}n speed.}
    \label{fig:shock}
\end{figure*}

Shock tubes are decisive numerical tests for the robustness and accuracy of a given numerical scheme that solves hyperbolic conservation laws. In particular, one-dimensional shock tubes unveil problems during the reconstruction stage or shortcomings of the Riemann solver. The analytic solutions to shock tube problems of hyperbolic conservation laws are self-similar and consist of sequences of rarefactions, shocks and contact discontinuities. The jumps in the conserved variables at those discontinuities are provided by the Rankine-Hugoniot jump conditions.

However, all aspects of this mature theory for hyperbolic conservation laws do not apply to our equations due the existence of and the strong dependence of the dynamics on the source terms in Eqs.~\eqref{eq:continuity_equation} to \eqref{eq:wave_equation}. Nevertheless, we can reinterpret the equations for small $\kappa_\pm$ and large $c$ as a relaxation approximation to the streaming-diffusion equations of \citet{2017Pfrommer} in the sense of \citet{2006Pares}. In this limit shock tubes solved with our equations have the same shock structure as the streaming-diffusion equations. To guarantee that this limit applies, we do not evolve $\varepsilon_\mathrm{a,\pm}$ but fix the values of $\kappa_\pm$.

We setup a shock tube with initial conditions given by: $\rho = 1$, $\mathbfit{u} = \mathbf{0}$, $\mathbfit{B} = 10^{-13} \mathbfit{e}_x$, $P_\mathrm{th} = 1$, $f_\mathrm{cr} = \va (\varepsilon_\mathrm{cr} + P_\mathrm{cr})$,
\begin{align}
    P_\mathrm{cr} = \left\lbrace \begin{matrix} 1, & x < 0, \\
    0.3333, & x > 0. \end{matrix} \right.
\end{align}
We fix the value of $\va$ independently of $\rho$ and $\mathbfit{B}$. Otherwise $\va$ would jump together with the density across any discontinuity which would make the interpretation of the results overly complicated. We use a moving one-dimensional mesh with initially 1024 equidistant cells in the domain $x \in [-5, +5]$ but we note that we get similar results for a fixed mesh. We show in Fig.~\ref{fig:shock} the result of simulations with $\va = 0, 0.75, 1.5$ at $t = 2$ with $c = 10$, $\kappa_+ = 1 / 300$, $\kappa_- = 0$, and $N_\mathrm{cr} = 4$. 

We use the case with $\va = 0$ as a reference for the discussion of the two other shocks. This case is identical to the purely adiabatic two-fluid cosmic-ray hydrodynamics with additional diffusion \citep{2017Pfrommer}. The final state of the simulation shown on the left-hand side of Fig.~\ref{fig:shock} consists of a rarefaction that is travelling to the left, a shock that travels to the right, and a central contact discontinuity over which thermal  and CR pressures experience a jump but the total pressure stays constant. The contact discontinuity is characterised by a jump in density and a constant total pressure and velocity across. The gas needs to be accelerated to reach this velocity. This is accomplished by the pressure gradients at the rarefaction and shock. Both accelerations are such that the gas velocities behind the rarefaction and shock exactly match. 

This is in contrast to the cases of $\va \neq 0$ (middle and right-hand side of Fig.~\ref{fig:shock}): here additional CR's stream across the rarefaction and decrease the total pressure gradient. This weakens the acceleration by the rarefaction and leads to a slower gas velocity behind it (i.e., to the right of the rarefaction). This would be sufficient to create a velocity mismatch at the former contact discontinuity. However, the shock on the right is also stronger with streaming CR's. Here the jump in total pressure is larger owning to a increased jump in CR pressure that have streamed across the former contact discontinuity. Consequently, the acceleration of gas at the shock is also increased which implies faster gas downstream of the shock. To compensate for the weaker acceleration to the left and stronger acceleration to the right of the former contact discontinuity, an additional rarefaction forms. This new rarefaction is moving to the right and is stronger than the already present rarefaction. It dilutes the gas efficiently, which can be seen in the $\va=0.75$ case by the variation in gas density at the position of the former contact discontinuity. 

Interestingly, the contact discontinuity still exists: it is located to the left of the new rarefaction but the density jump has switched its sign. In the $\va=0.75$ case, the central rarefaction seems to be attached to the contact discontinuity forming a compound wave as observed in the density profile. Contrarily, both rarefaction and contact discontinuity are properly separated in the $\va=1.5$ case. We argue that the observed connection in the $\va=0.75$ case is due to numerical and physical diffusion. Overall the numerical solutions are well behaved.

\subsection{Perpendicular magnetised contact discontinuity}

\begin{figure}
	\includegraphics[width=\columnwidth]{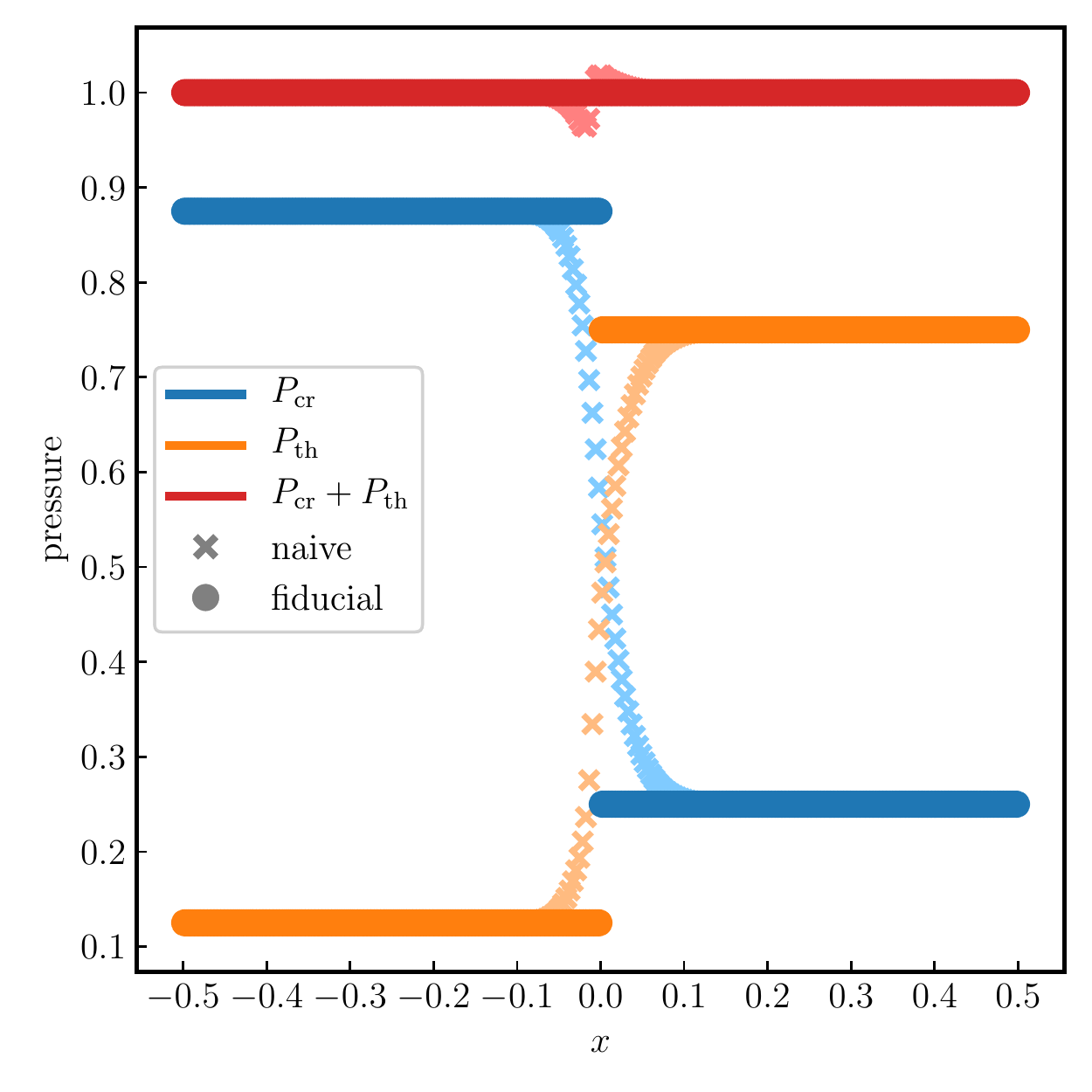}
    \caption{CR and thermal pressures of the perpendicular contact discontinuity problem. The profiles calculated with our fiducial Riemann solver are displayed with circles in opaque colours. In addition we show the profiles for the naive diffusive Riemann solver with crosses in semi-transparent colours.}
    \label{fig:perp_contact_discontinuity}
\end{figure}

In the direction perpendicular to the magnetic field interactions between Alfv{\'e}n waves and CRs are unimportant. In this direction, only the CR pressure is able to alter the momentum of the gas, meditated by the Lorentz force \citep{2019Thomas}. In a one-dimensional setting, where the magnetic field is perpendicular to the axis, the CR dynamics becomes adiabatic. The combined CR and gas system is, at its core, a two temperature fluid. A contact discontinuity in the two temperature fluid is not only characterised by a density jump, but there can also be a discontinuity in both partial pressures as long as the total pressure remains constant. We now demonstrate that the Riemann solver described in Section~\ref{sec:pc_transport} is able to accurately capture these contact discontinuities. 
We use a static numerical grid with 256 mesh points in the domain $x\in[-0.5, +0.5]$. The initial fluid velocity and CR flux are zero: $\mathbfit{u} = \mathbf{0}$ and $\fcr = 0$. The gas density is initialized with $\rho = 1$. The actual contact discontinuity is created by keeping the total pressure constant but adopting a jump in both, the CR and thermal pressure. We chose
\begin{align}
    \left[\pth, \pcr\right] = \left\lbrace \begin{matrix} \left[\frac{1}{8}, \frac{7}{8}\right] & x < 0, \\
    \left[\frac{3}{4}, \frac{1}{4}\right] & x > 0. \end{matrix} \right.
\end{align}
The magnetic field $\mathbfit{B} = \mathbfit{e}_y$ is perpendicular to the normal of the contact discontinuity. The light speed is set to $c=10$. The number of subcycles is $N_\mathrm{cr}=2$.

The pressure profiles at $t=1$ are displayed in Fig.~\ref{fig:perp_contact_discontinuity}. The results coincide with the initial conditions up to machine precision. This is by construction. The finite volume scheme of \citet{2017Pfrommer} used in Section~\ref{sec:adiabatics} to solve the adiabatic interactions uses an extension of the HLLD Riemann solver. This Riemann solver was designed to resolve the contact discontinuity without additional numerical dissipation. The finite volume scheme from Section~\ref{sec:pc_transport} describes the dynamics parallel to the magnetic field and should not have any impact in the present test. This is ensured by the implicit weighting of all numerical fluxes in Section~\ref{sec:pc_transport} with the projection of the magnetic field to the cell interface normal. This is a necessity for physical flux terms but not for terms that add artificial diffusion. Especially the weighting in the wave speed estimate in Eq.~\eqref{eq:cr_wavespeed} is important here. For cell interfaces where the magnetic field is perpendicular to the cell interface normal the wave speed is $S = 0$ due to the weighting. This is the case for the present contact discontinuity. As the wave speed is zero, the artificial diffusion term in Eq.~\eqref{eq:Fecr} is zero and no numerical diffusion is applied.

To demonstrate this, we compare our Riemann solver to an alternative that does not have this feature. Only the weighting with the projections on the magnetic field needs to be removed to accomplish this. We define a `naive' Riemann solver that uses the wave speed estimate 
\begin{equation}
    S_\mathrm{naive} = c \sqrt{\gamma_\mathrm{cr} - 1}.
\end{equation}
We rerun a simulation with the same settings but replace $S$ from Eq.~\eqref{eq:cr_wavespeed} with $S_\mathrm{naive}$. In Fig.~\ref{fig:perp_contact_discontinuity}, the thermal and CR pressure profiles calculated with the naive Riemann solver show the expected additional diffusivity around the discontinuity. It causes an increase of pressure in the low-pressure region for both, the gas and CRs. This destroys the total pressure balance at the contact discontinuity and induces prependicular fluid motion. 

To conclude, this test demonstrates that the finite volume scheme of Section~\ref{sec:pc_transport} avoids unnecessary interference with the dynamics perpendicular to the magnetic field when the proposed wave speed estimate in Eq.~\eqref{eq:cr_wavespeed} is used. 

\subsection{Anisotropic transport of a wedge}

\begin{figure*}
	\includegraphics[width=\textwidth]{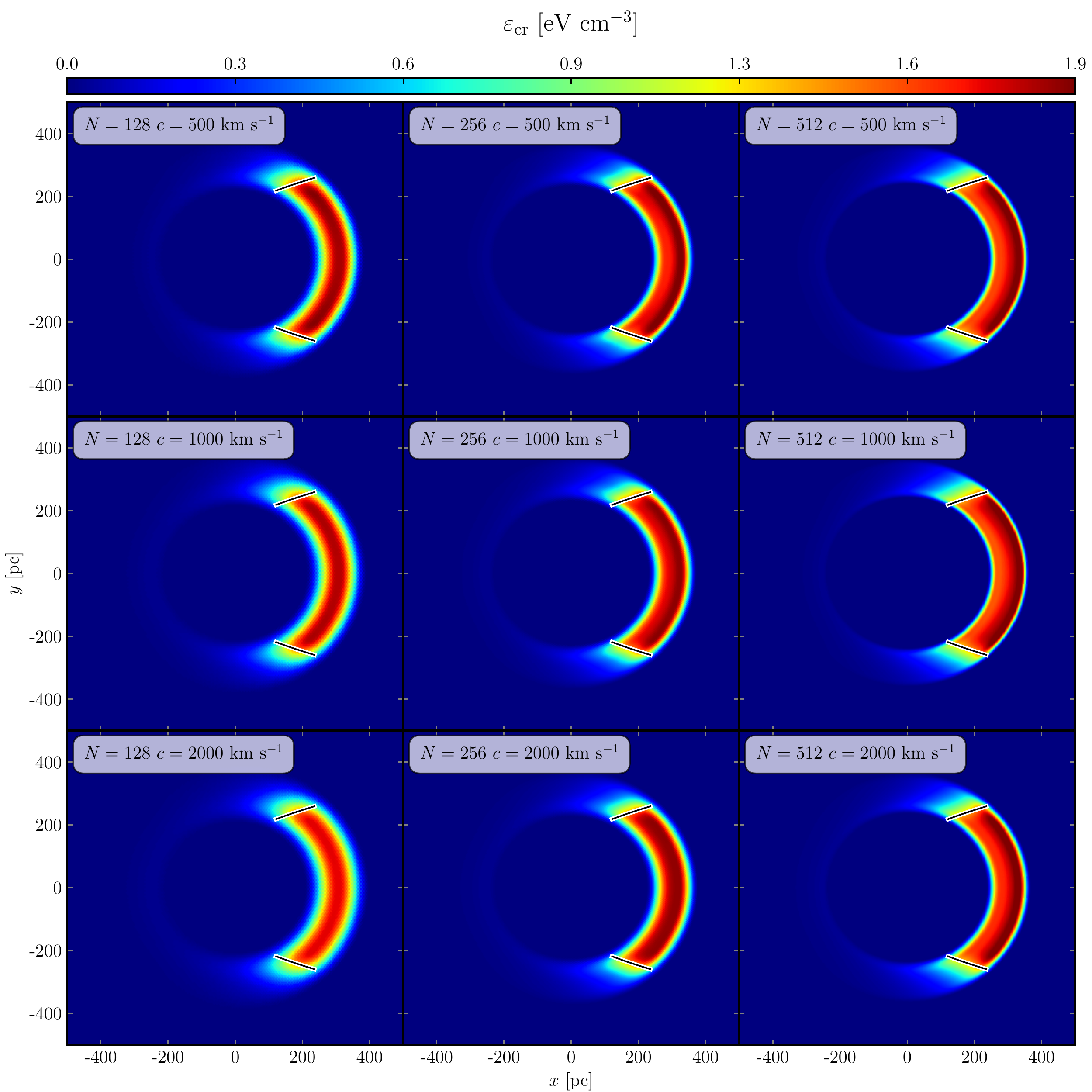}
    \caption{The anisotropic CR transport of a wedge test for a grid of varying mesh-points $N$ and light speeds $c$ at $t=5$~Myr. The black-and-white lines indicate the position of the wedge fronts if CRs were streaming exactly with the Alfv{\'e}n speed.}
    \label{fig:wedge}
\end{figure*}

The transport of a wedge along a ring magnetic field is one of the standard tests for anisotropic transport. It was used in \citet{2007Sharma} to highlight problems of standard methods that show non-physical transport perpendicular to the magnetic field. If not treated correctly, numerical discretisation effects can lead to the loss of monotonicity in diffusion problems or to a violation of the second law of thermodynamics in the context of thermal conduction.  We use this test to show that our method is able to transport CRs anisotropically without any of those problems. 

We simulate a wedge of CR energy density on a two-dimensional hexagonal, static mesh with $128^2$, $256^2$, and $512^2$  generating points in the domain $x,y \in [-500,500]$ pc. We set $\rho = m_p$ cm$^{-3}$, $\pth = c_\mathrm{sd}^2 \rho / \gamma_\mathrm{th}$, where $c_\mathrm{sd} = 30$ km s$^{-1}$ is the thermal sound speed, $\mathbfit{u}=\mathbf{0}$, and the magnetic field to be $\mathbfit{B} = \va \sqrt{\rho} \mathbfit{e}_\varphi$ where $\va = 30$ km s$^{-1}$ and $\mathbfit{e}_\varphi$ is the unit vector in polar direction. These MHD quantities are kept fixed for the simulations in this section. The CR are initialised with 
\begin{align}
    \ecr = \left\lbrace \begin{matrix} 10~ \mathrm{eV}~\mathrm{cm}^{-3} &\mathrm{for} ~|\varphi| < \frac{\pi}{12} ~\land~ \frac{r}{500 ~\mathrm{pc}} \in [0.5,0.7],  \\ 10^{-3}~ \mathrm{eV}~\mathrm{cm}^{-3} &\mathrm{else}, \end{matrix} \right.
\end{align}
where $r^2 = x^2 + y^2$ and $\varphi$ is the polar angle. In addition, we set $\fcr = 0$, $\ewpm = 10^{-4} \ecr$, and adopt $c = 500, 1000, 2000$ km s$^{-1}$.

The resulting CR energy density at $t=5$~Myr for all nine combinations of $c$ and number of mesh points is shown in Fig.~\ref{fig:wedge}. The situation along the magnetic field is reminiscent of the simulation of a rectangular CR population in \citet{2019Thomas}. The CR wedge has expanded along the magnetic field with approximately Alfv{\'e}n velocity. The two lines in each panel display the theoretical position of the edges of the wedge in the case if the CRs were streaming with exactly the Alfv{\'e}n velocity. In between those two lines $\ecr$ is mostly flat along $\varphi$ coordinate. The value of $\ecr$ increases for fixed $\phi$ along the radial coordinate from the inner edge to the outer edge due to the increasing CR energy content available at each $r$ as given by the initial conditions. At the outer edge, numerical diffusion causes a decrease of $\ecr$ in the radial direction. The radial extent of this numerical feature shrinks with increasing resolution. Additionally, CRs diffusive ahead of the wedge and start to fill the ring. We observe no artificial oscillation and monotonicity is preserved. We conclude that our scheme is able to correctly model anisotropic transport. 

Increasing the speed of light also increases the applied numerical diffusion. The numerical diffusion affects the large-$r$ edge of the wedge more strongly because in this region, the $\ecr$ gradient between wedge and background is strongest. This can be best observed by comparing the three panels of the first column in Fig.~\ref{fig:wedge}: while the maximum of $\ecr$ is located at the centre of the wedge for $c=2000$ km s$^{-3}$, it moves to larger radii for $c=500$ km s$^{-1}$. Increasing the resolution of the simulation has the expected effect: the numerical diffusion decreases if more mesh generating points are used.

\subsection{Telegrapher's equation}
\label{sec:test_telegrapher}

Linear perturbation analysis and simulations of linear waves provide another useful tool for understanding the mathematical character of equations and for code testing. In its simplest form small perturbation are introduced to a constant state that is a stable solution of the underlying equations. While no meaningful and stable state exists for our full CRHD equations, we can readily derive such a state in the telegrapher's limit of the equations.

In this limit, terms of higher order than $\mathcal{O}(1)$ in $\va / c$ are ignored. We assume that the scattering of CRs is provided by a source other than Alfv{\'e}n waves and set $\kappa_\pm$ to a fixed value, i.e., we only account for scattering centres moving in one direction.  In this scenario, Eqs.~\eqref{eq:cr_energy_equation} and \eqref{eq:cr_flux_equation} form the telegrapher's equations. This set of equations describes a non-Fickian diffusive transport of CRs \citep{2015Malkov, 2016Litvinenko, 2019Rodrigues}. With these assumptions we lost the physical interpretation of the equations (see App.~A of \citealt{2019Thomas}) but can nevertheless test the numerical performance of our code.

We linearly perturb Eqs.~\eqref{eq:continuity_equation} to \eqref{eq:wave_equation} in a one-dimensional setting by replacing every quantity $q$ by $q + \delta q$, adopt $q=\mathrm{const.}$, and neglect all second and higher-order terms in $\delta q$. The magnetic field is aligned with the axis. We neglect the transverse components of $\mathbfit{u}$ and $\mathbfit{B}$, and assume $\mathbfit{u} = \mathbf{0}$ and $\fcr = 0$ for a stable background state. The result of this perturbation procedure is:
\begin{align}
 \frac{\partial \delta \rho}{\partial t} &+ \rho \frac{\partial \delta u}{\partial x} = 0, \label{eq:linearized_tele_start} \\
\frac{\partial \delta u}{\partial t} &+ \frac{1}{ \rho }\frac{\partial \delta P_\mathrm{th}}{\partial x} = + \frac{1}{3 \rho  \kappa} \delta f_\mathrm{cr}, \\
 \frac{\partial \delta P_\mathrm{th}}{\partial t} &+ \rho c_\mathrm{sd}^2 \frac{\partial \delta u}{\partial x} = 0, \\
 \frac{\partial \delta \pcr}{\partial t} &+ \rho c_\mathrm{cr}^2 \frac{\partial \delta u}{\partial x} + \left(\gamma_\mathrm{cr} - 1\right)\frac{\partial \delta f_\mathrm{cr}}{\partial x} = 0, \\
 \frac{\partial \delta f_\mathrm{cr}}{\partial t} &+ c^2 \frac{\partial \delta P_\mathrm{cr}}{\partial x} = -\frac{c^2}{3 \kappa} \delta f_\mathrm{cr} \label{eq:linearized_tele_end},
\end{align}
where $c_\mathrm{cr}^2 = \gamma_\mathrm{cr} \pcr / \rho$. 

Solutions to those equations can be found by using the Fourier transformation of the perturbations. We use the convention $\delta q(x, t) = \delta q(k) \exp(\mathrm{i}kx - \mathrm{i} \omega t)$ for $\delta q \in [\delta \rho, \delta u, \delta \pth, \delta \pcr, \delta \fcr]$ and find the dispersion relation by solving
\begin{align}
\mathrm{det} \left[ \begin{matrix}
-\omega & \rho k & 0 & 0 & 0 \\
 0 & - \omega & k / \rho & 0 & \mathrm{i} / (3 \rho \kappa) \\
 0 & \rho c_\mathrm{sd}^2 k & -\omega & 0 & 0 \\
 0 & \rho c_\mathrm{cr}^2 k & 0 & -\omega & (\gamma_\mathrm{cr} - 1)k \\
 0 & 0 & 0 & c^2 k& -\omega -\mathrm{i} c^2 / (3 \kappa) \\
\end{matrix} \right] = 0,
\end{align}
and find
\begin{align}
 0 &= \omega \left\lbrace  \omega^2 \left(k^2 c_\mathrm{sd}^2 - \omega^2\right) \phantom{\frac{c^2}{3\kappa}} \right.  \nonumber \\ 
 &\hspace{-5pt}+\left. \frac{c^2}{3\kappa} \left[ \mathrm{i} k^2 \omega c_\mathrm{cr}^2 + \left(k^2 c_\mathrm{sd}^2 - \omega^2\right) \left( \mathrm{i} \omega  - 3k^2 \kappa (\gamma_\mathrm{cr} - 1)\right)  \right]  \right\rbrace. \hspace{-5pt} 
 \label{eq:dispersion_relation}
\end{align}
The $\omega = 0$ solution is the entropy mode known from ordinary adiabatic hydrodynamics. The remaining four solutions can be identified in the $\kappa\to\infty$ limit where gas and CR hydrodynamics are decoupled. Two solutions are sound waves waves that are modified by the presence of CRs when $\kappa$ is finite. We refer to them as `modified sound waves'. The last two waves are associated with CR dynamics. We call them `CR waves' in the following.

All simulations conducted in this section use $\rho = 1$, $\pth = 0.01 \gamma_\mathrm{th}^{-1}$, $ \pcr = 0.01 \gamma_\mathrm{cr}^{-1}$, $c=1$ and $N_\mathrm{cr} = 4$.

\subsubsection{Dispersion relation}
\begin{figure}
	\includegraphics[width=\columnwidth]{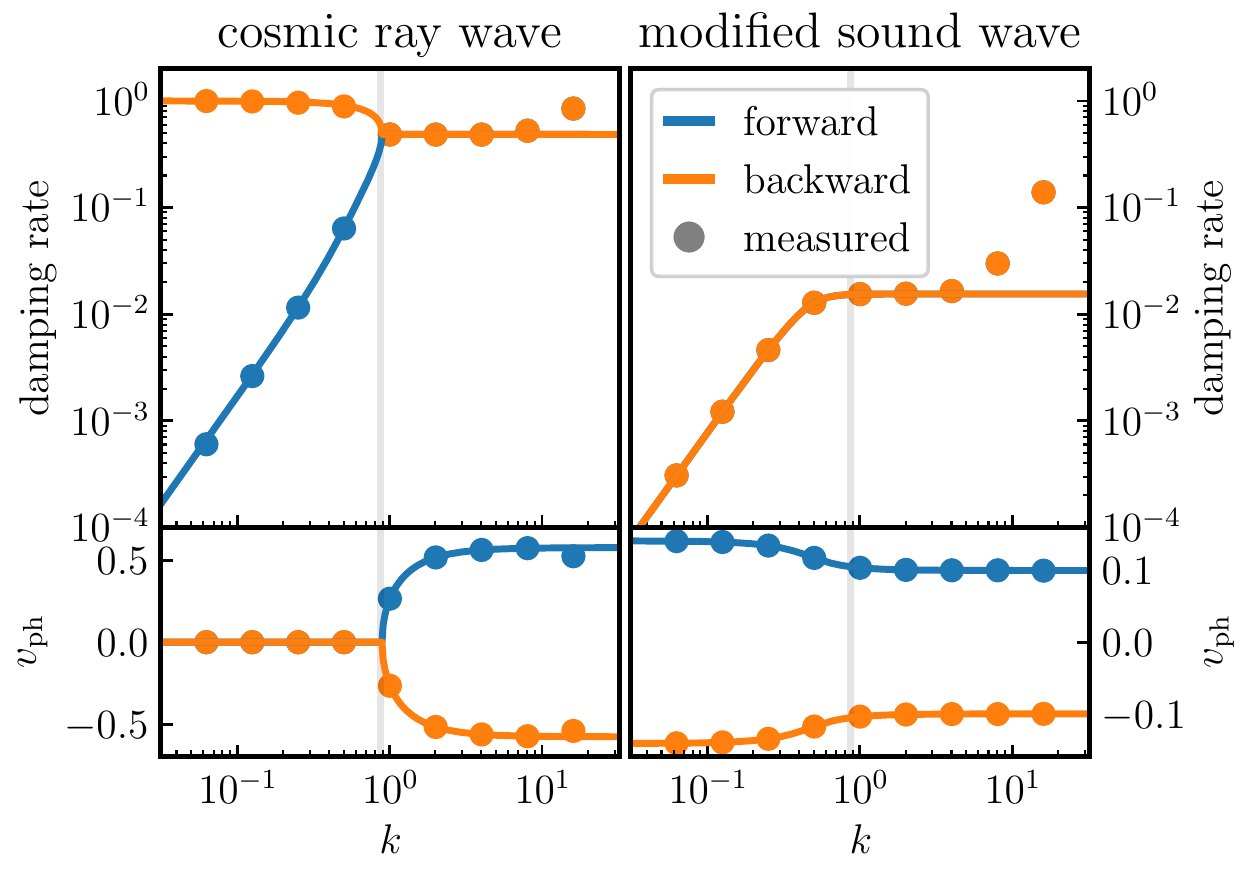}
    \caption{The dispersion relation of the hydrodynamic telegrapher's equation. We show the damping rate $-\mathrm{Im}(\omega)$ (top row) and phase velocity $\varv_\mathrm{ph} = \mathrm{Re}( \omega / k)$ (bottom row) of the CR wave (left panels) and the modified sound wave (right panels) for different wavenumbers $k$. In positive $x$-direction (forward) propagating waves are shown in blue while in negative $x$-direction (backward) propagating waves are shown in orange. The dots represent the values measured in our simulations.}
    \label{fig:dispersion_relation}
\end{figure}

In Fig.~\ref{fig:dispersion_relation} we display the damping coefficient $- \mathrm{Im}(\omega)$ and the phase velocity $\varv_\mathrm{ph} = \mathrm{Re}(\omega / k)$ for all four waves and $\kappa = 1/3$. All solutions have different behaviours below and above a certain wave number $k \sim 1$. By dropping the first term in the second line of Eq.~\eqref{eq:dispersion_relation}, we find a branching point in the CR wave solutions that is approximately located at 
\begin{equation}
    k_\mathrm{mfp} = \frac{c}{2 \sqrt{3} \kappa},
\end{equation}
which has the physical meaning of an inverse mean free path. For $k < k_\mathrm{mpf}$, the wave length is below the mean free path of CRs and the dynamics is dominated by CR scattering. The two CR wave modes are standing waves. One of them is damped at a rate $\sim 1$ while the other solution shows a diffusive-like wave frequency $\omega \sim -\mathrm{i} \kappa k^2$. The latter is a feature of the telegrapher's equations, as they are a non-Fickian description for physical diffusion. Furthermore, CRs and the thermal gas are tightly coupled as scattering is frequent. The modified sound waves are slightly damped and travel with phase speeds larger than adiabatic sound speed. For $k > k_\mathrm{mpf}$, CR scattering is inefficient and the CR and gas dynamics are decoupled. Consequently, the sound waves are travelling with the adiabatic sound velocity $c_\mathrm{sd}$. The CR waves are travelling at speeds  $\sim c/ \sqrt{3} \sim 0.57 c$. This velocity is expected in the Eddington limit adopted in the derivation of the CRHD equations \citep{2019Thomas}. 

Equipped with these analytical results, we can test whether our numerical algorithm is able to reproduce them. The simulation box has length 1 and is filled with 4096 equally spaced static mesh points. We setup 9 linear plane waves with wave numbers ranging from $k = 2\pi$ to $k = 2\pi\times512$ in powers of 2. We run simulations for both wave types and both propagation directions. The CR waves are initialised by setting $\delta \fcr(k) = 10^{-6}$ and calculating the remaining components by solving Eqs.~\eqref{eq:linearized_tele_start} to \eqref{eq:linearized_tele_end} using the respective wave frequency $\omega=\omega(k, \kappa)$ for a given $k$ and $\kappa$. The setup for the modified sound waves is similar: here we set $\delta u(k) = 10^{-6}$ and calculate the other Fourier components in terms of the velocity perturbation. After this, the real part of each component is added to the background state. This procedure ensures that the initial conditions are eigensolutions to the underlying differential equations. The simulation are run until $t=10$.

Damping rates are derived by fitting an exponential function to the amplitude of the Fourier component of $\delta u$ at the respective wave number. The real part of the wave frequency $\mathrm{Re}(\omega)$ is calculated using the average derivative of the argument of the same Fourier component. The inferred damping rates and phase velocities are also plotted in Fig.~\ref{fig:dispersion_relation}. They agree with the analytical prediction for low to intermediate wave numbers. For higher wave numbers the wave trains of the plain wave are resolved by fewer grid cells and numerical diffusion starts to affect the evolution. This can be seen in the damping rates. They are altered by a contribution originating from numerical diffusion with $\omega \to \omega + \omega_\mathrm{num} \sim \omega -\mathrm{i} (c / \sqrt{3}) \Delta x  k^2$ where $\Delta x$ is the grid spacing and $\omega_\mathrm{num}\sim-\mathrm{i} (c / \sqrt{3}) \Delta x  k^2$ is an upper limit.

\subsubsection{Parameter study}
\begin{figure}
	\includegraphics[width=\columnwidth]{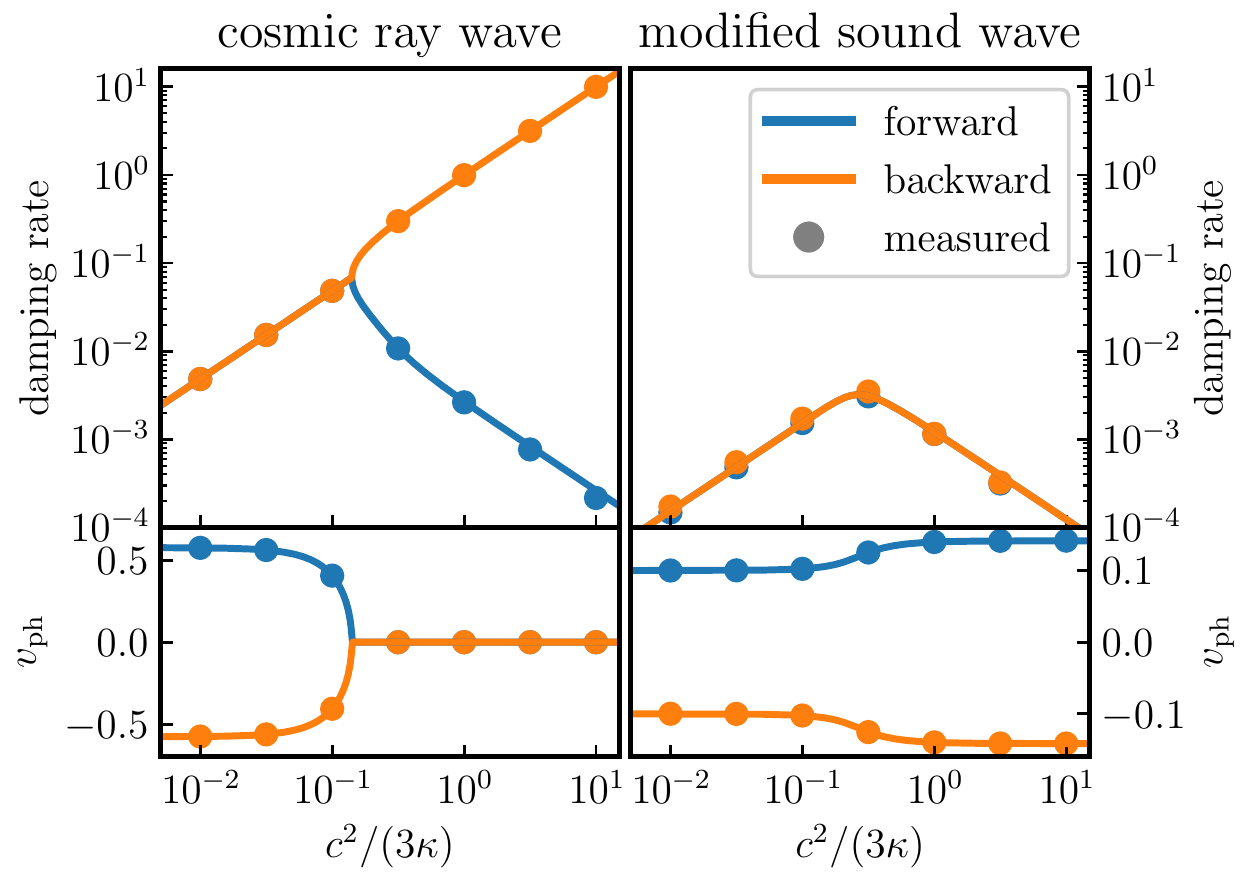}
    \caption{Same as Fig.~\ref{fig:dispersion_relation} but this time with a fixed $k$ but variable scattering rates $c^2 / (3 \kappa)$.}
    \label{fig:parameter_study}
\end{figure}

The wave number $k$ and the diffusion coefficient $\kappa$ are independent parameters in Eq.~\eqref{eq:dispersion_relation} and cannot be factored into a joint parameter. Thus the dependence of the wave frequency on $\kappa$ is different in comparison to the dependence on $k$. In Fig.~\ref{fig:parameter_study} we keep $k=2\pi \times 2$ fixed but vary $\kappa$ and show the damping rates and phase velocities of both wave types. For larger $\kappa$ scattering is inefficient and the CR and gas dynamics become decoupled. In this case, the modified sound waves are moving with the adiabatic sound speed $c_\mathrm{sd}$ while the CR waves are travelling at $c/\sqrt{3}$. The scattering rate $c^2/(3 \kappa)$ is increased for a smaller $\kappa$ and CRs are efficiently coupled to the gas. This yields standing CR waves with two distinct damping rates and an increased effective sound speed of the modified sound waves. The damping rates of the modified sound waves achieve their maximum values at $\kappa\sim k_\mathrm{mpf}$ where the transition between coupled and decoupled CRHD dynamics occurs. 
 
To see whether our code can reproduce these results, we run 7 simulations with $\kappa$ increasing from $1/30$ to $30$ in logarithmic steps for each wave type. The numerical grid consist of 4096 equally spaced points in a domain of length 1. Damping rates and phase velocities are calculated the same way as in the previous subsection and the results are shown in Fig.~\ref{fig:parameter_study}. The inferred numerical damping rates and phase velocities agree with the analytical predictions and show negligibly small deviations.

\subsubsection{Convergence study}
\begin{figure}
	\includegraphics[width=\columnwidth]{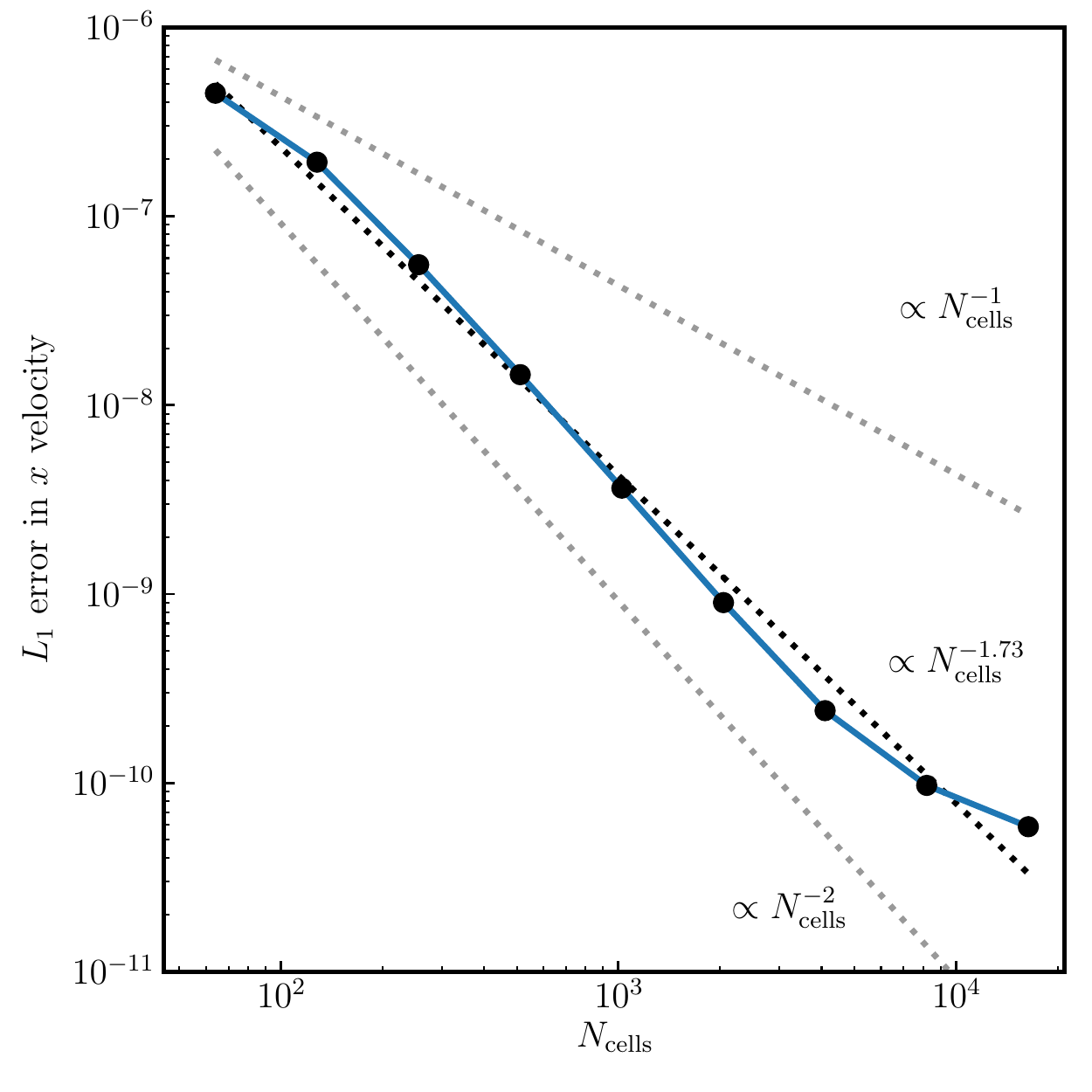}
    \caption{Convergence test for a linear forward-travelling CR wave of the telegrapher's equation.}
    \label{fig:convergence}
\end{figure}
As a last application of our linear wave analysis, we test for numerical convergence of our code. We set up an eigensolution of the forward travelling CR wave as initial condition and increase the number of cells $N_\mathrm{cells}$ from 64 to 16384 in factors of 2. The average L1-difference of the simulated $\delta u$ and the corresponding analytical eigensolution at $t=10$ is used as error measurement. This L1-error is calculated for each resolution and displayed in Fig.~\ref{fig:convergence}. The convergence order is $1.72$ and thus somewhat below second order. Although both, the CRHD and MHD modules are individually second-order accurate, we expect only a first-order convergence owing to the operator splitting of both modules. The numerical convergence rate is exceeding first order, which suggests that the measured total error is still in a regime where the error is dominated by the errors of the individual modules and not by error originating from the operator splitting. 

\begin{figure*}
	\includegraphics[width=\textwidth]{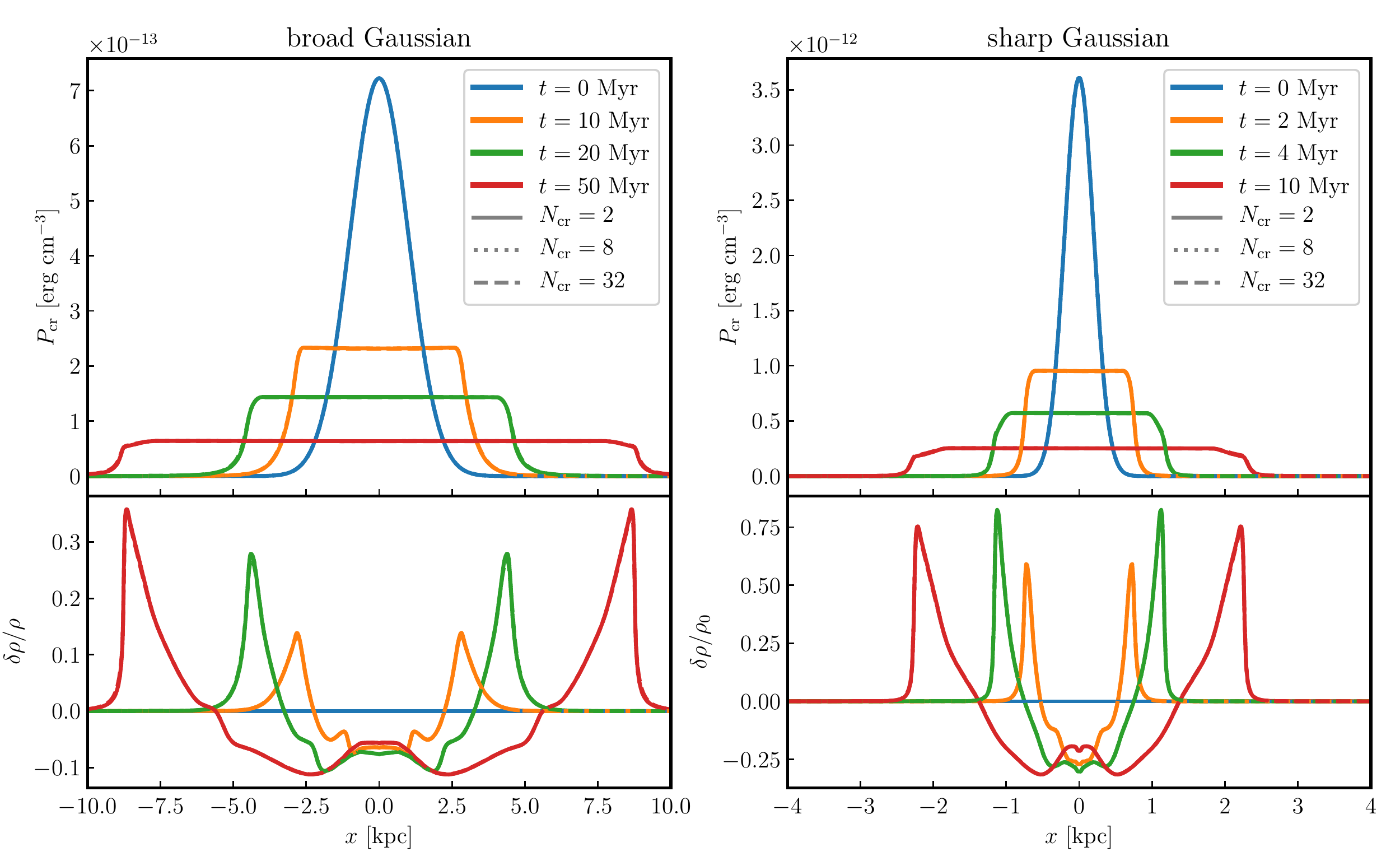}
    \caption{CR pressure (top row) and mass density fluctuation (bottom row) for a broad (left column) and a narrow Gaussian (right column) initial distribution of CRs on an otherwise homogeneous background. Note the different scales of both panels. This Figure is comparable to figure~7 in \citet{2017Wiener}.}
    \label{fig:gaussian}
\end{figure*}

\subsection{Hydrodynamic response to a Gaussian CR distribution}

We now move on to simulating the influence of a local CR overpressure on an otherwise homogeneous ambient gas distribution. The pressure-gradient of the CRs will set the gas into motion. This enables us to test the coupling between the CR and MHD modules. Our one-dimensional setup closely follows \citet{2017Wiener}. The thermal MHD fluid is initialised with a mass density $\rho_0 = 1.204 \times 10^{-24}$ g cm$^{-3}$, sound speed $c_\mathrm{sd} = 100$ km s$^{-1}$, and Alfv{\'e}n speed $\va = 100$  km s$^{-1}$. The fluid is at rest while the magnetic field points along the simulation axis. The CR flux is chosen such that the CRs stream with Alfv{\'e}n speed down their gradient. We simulate both a broad CR distribution with
\begin{equation}
    \pcr = 10^{-4} \pth + \pth \exp\left( \frac{x^2}{2 \mathrm{kpc}^2} \right),
\end{equation}
and a sharp CR distribution with 
\begin{equation}
    \pcr = 10^{-4} \pth + 5 \times \pth \exp\left( \frac{x^2}{2 (200~ \mathrm{pc})^2} \right).
\end{equation}
Initially, both Alfv{\'e}n wave energy densities are given by $\ewpm = 10^{-6} \ecr$. The simulation domain $x \in [-50, +50]~$kpc is sampled by 4096 mesh-generating points that move quasi Lagrangian. We use $c = 3000$ km s $^{-1}$ and $N_\mathrm{cr} = 2, 8, 32$.

In Fig.~\ref{fig:gaussian} we display $\pcr$ and the fractional change of the mass density for both distributions. The evolution of $\pcr$ is similar to those of the Gaussians described in \citet{2019Thomas}: both wings of the Gaussian CR distribution propagate in opposite directions while creating a flat plateau in between. At the wings the gyroresonant instability creates sufficient Alfv{\'e}n waves such that CRs and the gas are well coupled. This converts CR to gas momentum and pushes the gas away from the center. The gas reacts to this acceleration by creating a central underdensity and swept-up shells at the position of the CR gradient. The gas is rarefied between both shells.

Before we compare our results to \citet{2017Wiener}, a few technical details need to be recalled: \citet{2017Wiener} uses the implementation of \citet{2012Uhlig} to simulate the streaming of CRs in the SPH-code \textsc{Gadget}-2. In this method the streaming terms of the advection-diffusion equation for the CR energy are discretised using a parabolic SPH operator. This introduces additional numerical diffusion that damps the otherwise occurring and dominating numerical noise. By comparing figure~7 in \citet{2017Wiener} to our Fig.~\ref{fig:gaussian}, we find good overall agreement between both methods and only small differences that we will comment on in the following. We notice that the plateaus calculated with our presented method are flatter which might be caused by our smaller numerical diffusion. In consequence, our gradients in $\pcr$ at the wings of the Gaussian are steeper. Furthermore, the kinks in $\rho$ inside the central region are more pronounced in our solution albeit they are recognisable in their solutions. We attribute this to the deficiency of SPH for subsonic flows. 

We broadly vary the number of CR module subcycles $N_\mathrm{cr}$ for this test to quantify its influence on the solution. The profiles of $\pcr$ for different $N_\mathrm{cr}$ look identical while minor differences in the density are noticeable. Their origin is likely the application and subsequent subtraction of the parallel $\pcr$ gradient forces: we first add the $\nabla_\parallel \pcr$ during the adiabatic step described in Section~\ref{sec:adiabatics} and subtract it during the parallel transport step as described in Section~\ref{sec:pc_transport}. During the second step $\pcr$ changes which prevents an exact cancellation. Only if the CRs are well coupled and $\mathbfit{b} \mathbf{\cdot} ( \mathbfit{g}_\mathrm{gri,+} + \mathbfit{g}_\mathrm{gri,-}) \sim \nabla_\parallel \pcr$ then the force exerted by the gyroresonsant interaction adds the lost parallel momentum back to the gas neglecting the previous subtraction. The observed small deviations in $\delta \rho / \rho_0$ indicate that this process is mostly independent of $N_\mathrm{cr}$.

\subsection{Acceleration of a warm cloud}

\begin{figure}
	\includegraphics[width=\columnwidth]{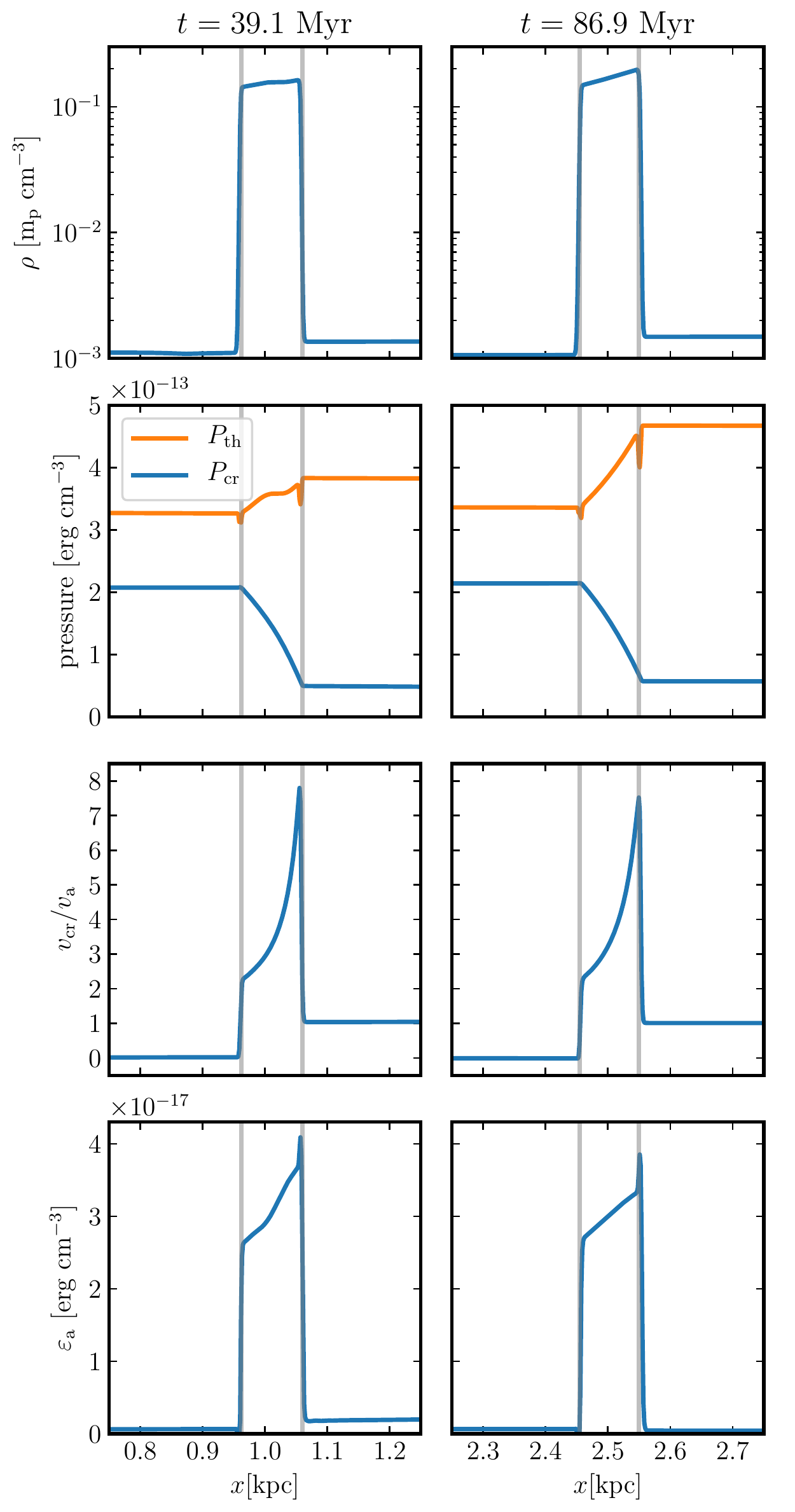}
    \caption{Profiles of density, pressure, CR streaming speed in units of the local Alfv{\'e}n speed, and energy density of forward propagating Alfv{\'e}n waves for the warm cloud test problem at two different times. Grey lines indicate the extent of the cloud.}
    \label{fig:warm_cloud}
\end{figure}

Observations employing absorption line spectroscopy of the CGM show that the CGM is multiphase medium and consists of hot $10^6$~K, low-density gas with embedded high-density, warm $10^4$~K clouds, some of which reach velocities exceeding hundreds of km s$^{-1}$ \citep{2017Tumlinson}. One of the proposed mechanisms to accelerate these clouds are CRs \citep{2017WienerII, 2019Wiener}. A flux of CRs impinging on a cloud causes the cloud to be accelerated. A reservoir of standing CRs builds up in front of the cloud and the decreasing Alfv{\'e}n speed inside the cloud causes the CRs to leave the cloud at a lower pressure but with a larger streaming speed. This effect is called the bottleneck effect \citep{1971Skilling,2017Wiener}. The `potential difference' of CR pressure between the front and the back of the cloud leads to its acceleration. We test our numerical method with a simulation of this scenario and note that other methods for two-moment CR transport have already been tested the same way \citep{2018Jiang}.

We simulate the cloud in a one-dimensional setting and mostly follow  \citet{2019Wiener}. We initialise the cloud centred at $500$~pc with mass density $\rho = 2.35\times10^{-25}~\mathrm{g}~\mathrm{cm}^{-3}$ and width $100$~pc. The mass density of the ambient medium is $\rho = 2.26\times10^{-21}~\mathrm{g}~\mathrm{cm}^{-3}$. The magnetic field is aligned with the axis of the simulation and has a strength of $1~\umu\mathrm{G}$. The thermal gas is initially at rest. The gas pressure $\pth = \rho k_B T / (\mu~m_\rmn{p})$ is uniform and set to $3.42~\mathrm{erg}~\mathrm{cm}^{-3}$. The mean molecular weight $\mu$ in units of the proton mass $m_\rmn{p}$ is assumed to be $0.6$. This pressure corresponds to a temperature $T$ of $10^{4}$~K inside the warm cloud and $1.1\times 10^{6}$~K in the hot medium. Initially, only a negligible amount of CR and Alfv{\'e}n wave energy is present.

We include optically thin cooling of the thermal gas by additionally evolving the temperature via
\begin{equation}
    \frac{\partial T}{\partial t} = \frac{\mu}{\mu_\rmn{H}} \frac{\gamma_\mathrm{th} - 1}{k_\rmn{B}} \left( \Gamma - n_\rmn{H} \Lambda \right),
\end{equation}
where $\Lambda$ is the cooling function, $\Gamma = 10^{-25}~\mathrm{erg}~\mathrm{s}^{-1}$ represents a uniform heating rate, and $\mu_\rmn{H} = 1.63$ is the mass fraction of hydrogen atoms. We use the fit to the Cloudy cooling function assuming collisional ionisation equilibrium as given in Appendix A of \citet{2018Schneider}. Cooling is implemented in an operator-split manner. The temperature is evolved in time using a subcycled Euler method. The updated temperature is allowed to vary by 1 per cent from its old value during a single cycle. We impose a temperature floor of $10^{4}$~K. Inclusion of the heating rate in addition to the gas cooling makes the hot phase, as given by the initial condition, stable.

We use constant-extrapolation boundary conditions for all quantities on the right side of the domain. The boundary conditions on the left side are set as follows: all MHD quantities are extrapolated with their constant values, the CR energy density is  $\ecr= 6.488 \times 10^{-13}~\mathrm{erg}~\mathrm{cm}^{-3}$, the forward-propagating Alfv{\'e}n wave energy density is $\ewp = 10^{-6} \ecr$, and $\fcr$ is set to be reflective by copying the value in the first cell but switching its sign.
The CR module uses $c = 500~\mathrm{km}~\mathrm{s}^{-1}$ and $N_\mathrm{cr} = 4$. The numerical grid range is $x \in [0, 3]$~kpc and consists of 2048 equally spaced mesh generating points. Because the fluid motion is mostly subsonic we keep the grid static. 

The numerical solution is displayed in Fig.~\ref{fig:warm_cloud} at $t = 39.1$ Myr and $t = 86.9$ Myr. The solution reaches a quasi-steady state where profiles have the same characteristic form over a long time while the values of individual quantities show small variations.

The bottleneck effect of CRs can be observed in the CR pressure profile: to the left of the cloud the CR pressure is higher than to its right and smoothly transitions between the two values inside the cloud. The pile-up of CRs causes them to stream with sub-alfv{\'e}nic velocities ahead of the cloud (i.e., to the left of it). Because the Alfv{\'e}n speed is reduced by a factor $\sim 10$ inside the warm cloud, this gives rise to super-alfv{\'e}nic CR streaming speeds inside the cloud. This triggers an energy transfer from CRs to Alfv{\'e}n waves via the gyroresonant instability and an effective growth of Alfv{\'e}n waves. Interestingly, the streaming speed increases as the CRs flow through the cloud. This may seem counter-intuitive as the streaming speed should decrease to the Alfv{\'e}n speed if there is sufficient energy in Alfv{\'e}n waves present. Here, the $\pcr$ gradient term and the term that describes the interaction between CRs and forward propagating Alfv{\'e}n waves are both relevant for the evolution. The CR flux decreases from $\vcr > \va$ to $\vcr \sim \va$ only if the CR-Alfv{\'e}n wave interaction dominates. But inside the cloud the $\pcr$ gradient term is dominant. As a result, the CR flux increases in the cloud because the $\pcr$ gradient is positive. The CR pressure profiles of \citet{2017WienerII} show a shallow gradient inside the cloud and jump at the right edge of the cloud. This jump is not present in our simulation. 

The acceleration by CRs is almost uniform in space so that the density only slightly increases towards the leading edge. Over the course of the simulation the cloud is not fragmented but experiences a compression by 30 per cent to $70$ pc when the initial front of CR's makes contact with it at $t\sim10$ Myr. After that event, the cloud quickly expands again to its original side length $\sim 100$ pc and approximately maintains this size for the rest of the simulation. In Fig.~\ref{fig:warm_cloud} the grey lines trace the interface between the cloud and the ambient medium. For the displayed times at $t = 39.1$ Myr and $t = 86.9$ Myr the cloud is $95$pc and $97$pc wide.

The thermal pressure to the right of the cloud is steadily increasing over time once the quasi-steady state is reached. While the CR pressure gradient dominates the overall pressure balance inside the cloud, the rising thermal pressure and its associated gradient slows down the acceleration of the cloud. The increase in $\pth$ is caused by adiabatic compression of the gas, which is the result of the acceleration of the cloud and a pile-up of gas to the right of the cloud. We confirmed this by verifying that the entropy measure $\pth \rho^{-\gamma_\mathrm{th}}$ to the right of the cloud remains nearly constant in the course of the simulation. This effect is particularly strong in our one dimensional simulation. In two or three dimensions the pile-up will likely be weaker as the flow will be able to escape in the other dimensions.

The numerical solution is well behaved and the transition between hot and warm media is sharp because we do not take into account thermal conduction in these simulations \citep{Drake2020}. The blips in the thermal pressure are caused by mixing the warm and hot phases which causes an overcooling at the interface. This seeds some small amplitude sound waves that travel inside the warm cloud.

\subsection{CRHD vortex}

\begin{figure}
	\includegraphics[width=\columnwidth]{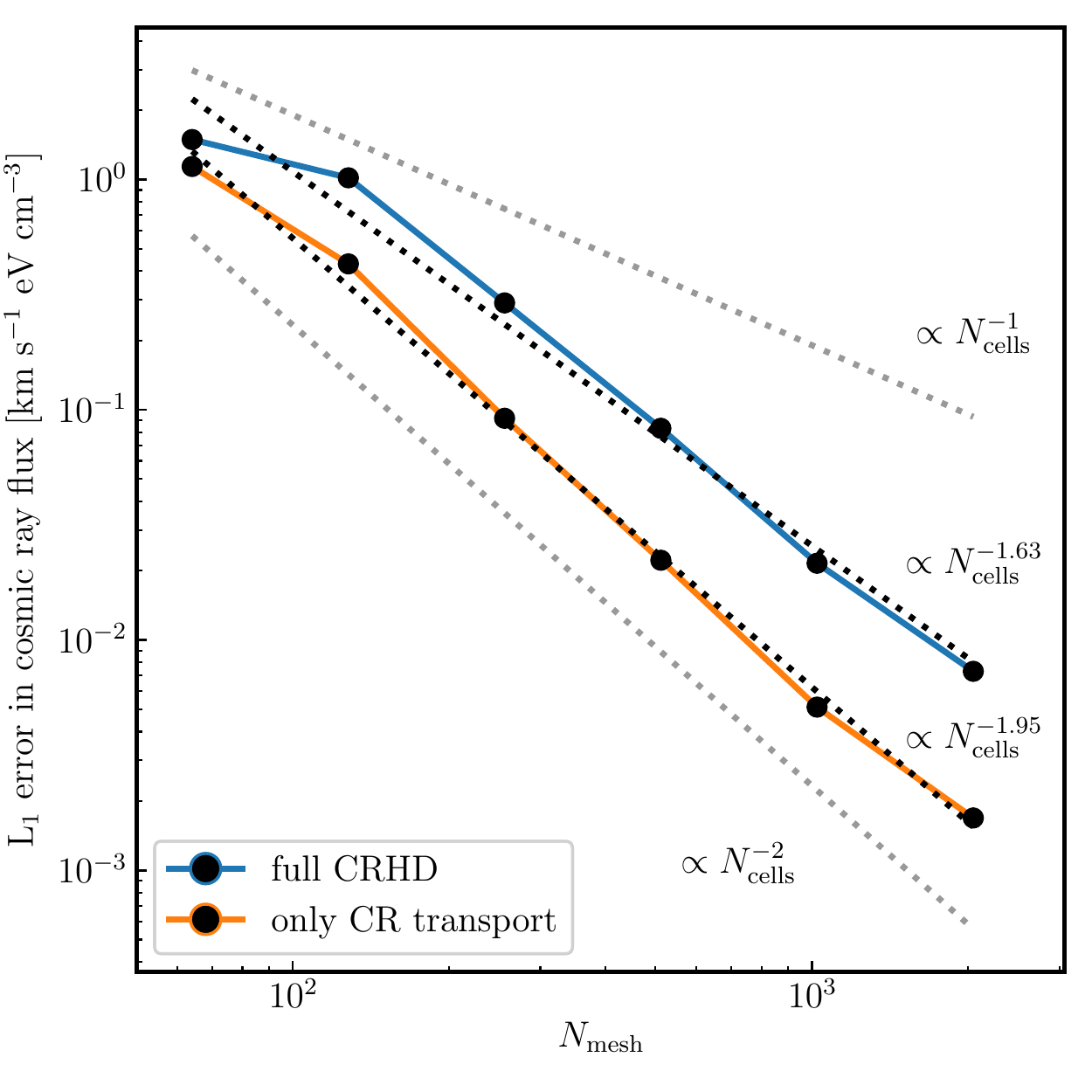}
    \caption{Convergence test for the two-dimensional isodensity vortex for the two tested modes of included CRHD physics.}
    \label{fig:convergence_vortex}
\end{figure}

To quantitatively assess the accuracy of our algorithm to anisotropically transport CR and to couple them to the gas, we simulate a steady state analytic vortex. Analytical solutions for isodensity vortices already exist for hydrodynamics and MHD \citep{1999Yee, 2004Balsara}. The principle idea in the derivation of the hydrodynamic/MHD solutions is that centrifugal force induced by the rotation of the vortex can be balanced by pressure forces or shear. This idea can be readily extended to CRHD where CRs are moving in a circular magnetic field and exert a Lorentz force with the $\nabla_\perp \pcr$ term in the momentum equation, Eq.~\eqref{eq:euler_equation}, that counteracts the centrifugal force.

Assuming an equilibrium in two-dimensional polar ($R, \varphi$) coordinates, the radial component of the Euler equation reads
\begin{align}
\frac{\partial (T_{RR} + \pcr)}{\partial R} = \frac{1}{R} \left(T_{\varphi \varphi} - T_{R R}\right), \label{eq:polar_euler}
\end{align}
where the $RR$ and $\varphi \varphi$ components of the pressure-stress tensor for a circular magnetic field are given by:
\begin{align}
    T_{RR} &= \left( \rho \mathbfit{u} \mathbfit{u} + P_\mathrm{th}\mathbf{1} + P_\mathrm{mag}\mathbf{1} - \mathbfit{B} \mathbfit{B} \right)_{RR} \\
    &= P_\mathrm{th} + \frac{B^2_\varphi}{2}, \\
    T_{\varphi \varphi} &= \left( \rho \mathbfit{u} \mathbfit{u} + P_\mathrm{th}\mathbf{1} + P_\mathrm{mag}\mathbf{1} - \mathbfit{B} \mathbfit{B} \right)_{\varphi \varphi} \\
    &= \rho u^2_\varphi + P_\mathrm{th} - \frac{B^2_\varphi}{2},
\end{align}
where $P_\rmn{mag}=B^2/2$, $u_R$, $B_R$ are the components of $\mathbfit{u}$ and $\mathbfit{B}$ in $R$ direction, and $u_\varphi$, $B_\varphi$ are the corresponding components in $\varphi$ direction. We can solve Eq.~\eqref{eq:polar_euler} by integrating the pressures after inserting a given rotation curve and a magnetic field profile. In our case there is a degeneracy in the solutions as the pressure can be provided either by the thermal gas or the CRs. We chose to keep the thermal pressure constant and integrate the CR pressure. 
Our initial conditions and steady state solutions for this setup are given by:
\begin{align}
    \rho &= m_\mathrm{p}~\mathrm{cm}^{-3}, \\
    \mathbfit{u} &= u_0\exp\left(\frac{1 - \hat{R}^2}{2} \right) \mathbfit{e}_\varphi, \\
    \mathbfit{B}  &= u_0 \sqrt{\rho} \exp\left(\frac{1 - \hat{R}^2}{2} \right) \mathbfit{e}_\varphi, \\
    \pth &= \frac{\rho u_0^2}{\gamma_\rmn{th}}, \\
    \pcr &= \rho u_0^2  \left[\frac{1}{\gamma_\mathrm{th}} - \frac{\hat{R}^2}{2}   \exp\left(1 - \hat{R}^2 \right)\right], \\
    \fcr &= \ewpm =  0,
\end{align}
where $u_0 = 30$ km s$^{-1}$ and
\begin{align}
    \mathbfit{e}_\varphi &= \frac{[-y, +x, 0]^T}{100 \mathrm{pc}}, \\
    \hat{R} &= \frac{R}{100~\mathrm{pc}}.
\end{align}
\begin{figure*}
	\includegraphics[width=\textwidth]{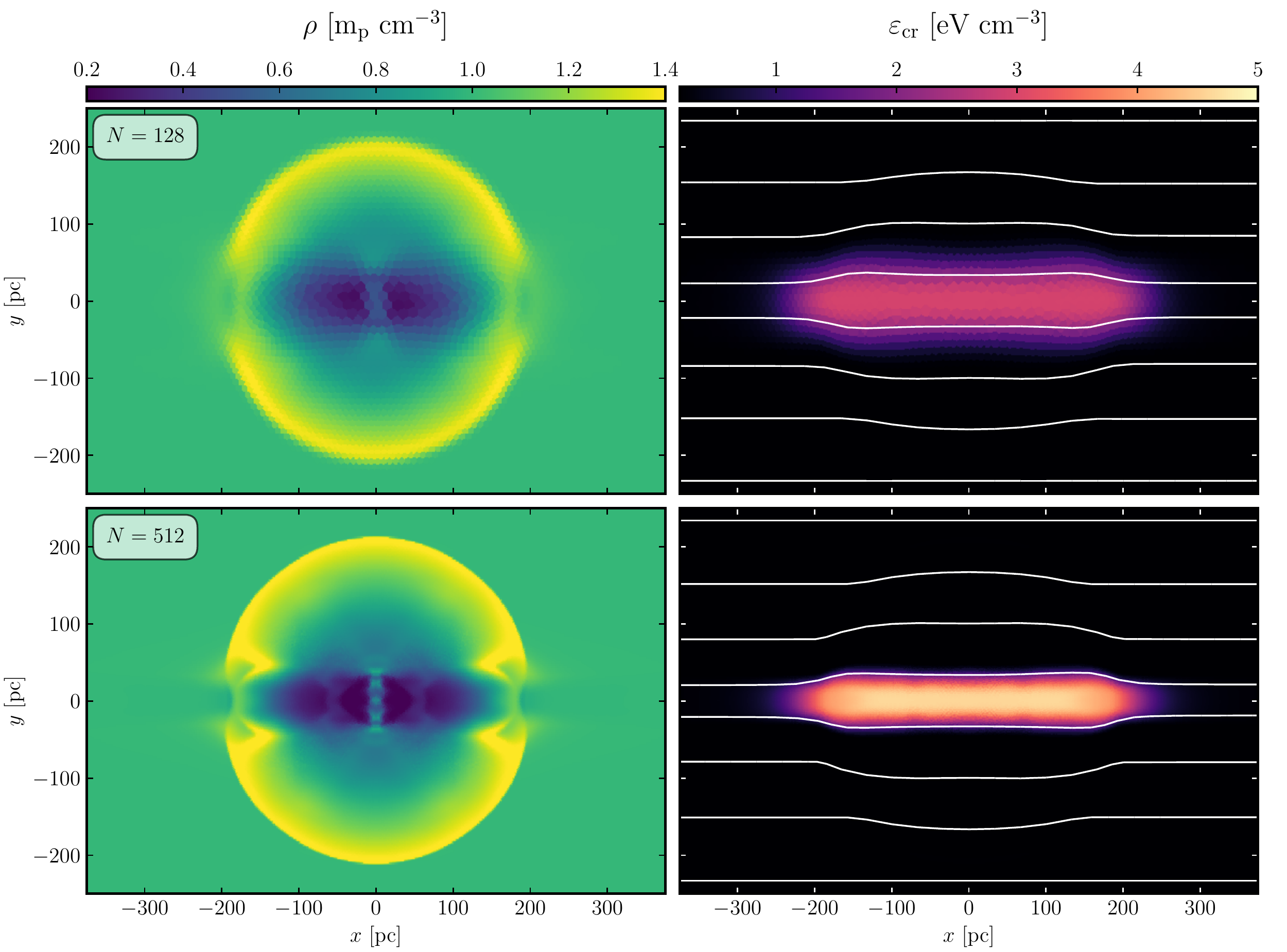}
    \caption{Mass density and CR energy density for the blast wave test problem with two different mesh resolutions. White lines trace the magnetic field.}
    \label{fig:blast_wave}
\end{figure*}
A shortcoming of this solution is that it does not contain any Alfv{\'e}n wave dynamics. This is due to two technical reasons. First, wave damping converts Alfv{\'e}n wave energy to thermal energy. Because these two components have different adiabatic indices, this would cause a pressure imbalance. Second, the gyroresonant interaction would transfer momentum from CRs to the gas along the magnetic field which would cause a additional azimuthal acceleration. In both cases the balance between the pressure gradient forces and the centrifugal force would cease to exist and the dynamical equilibrium would be lost.

We set up simulations of the vortex initial conditions on a two-dimensional hexagonal mesh in a computational domain of $x,y \in [-1, +1]$~kpc. We use $c = 1000$ km s$^{-1}$, $N_\mathrm{cr} = 10$, and constant-extrapolation boundary conditions. 

Although the initial conditions correspond to an analytical solution of this problem, numerical errors build up after the simulations has started. These errors are caused by the misalignment of mesh interfaces with the polar coordinate axes but allows us to test the convergence of our method. To perform the convergence test, we increase the number of mesh generating points from $64^2$ to $2048^2$ in powers of 2 for each dimension. With increasing mesh resolution the deviations from the analytic solution should decrease which allows us to test convergence of our scheme with a non-linear problem. We run this test with two different transport modes for the CRs.

\begin{enumerate}
    \item \textbf{Full CRHD.} Here we integrate the full set of CRHD equations on a moving mesh. This tests the code's capability to maintain the dynamical equilibrium of the analytical solution. For these simulations, we use the moving-mesh capability. 
    
    \item \textbf{Only CR transport.} Here we use the same setup, as described above, but update only $\ecr$ and $\fcr$ on a static mesh. Even with this additional restriction, the analytical solution remains valid. The resolution study tests whether the code can accurately transport CRs anisotropically and keep $\fcr=0$.
\end{enumerate}

We measure the numerical error by calculating the L1 error norm of $\fcr$ at $t=1$ Myr. The results are shown in in Fig.~\ref{fig:convergence_vortex}. Our method converges for both transport modes. In the \textit{full CRHD} case the convergence order is $\sim 1.6$ which is similar to the result found for the telegrapher's test in Section~\ref{sec:test_telegrapher}. This result is unexpectedly good because the formal convergence order is still $1$ due to the operator splitting. This suggest that the error introduced by the operator splitting is lower than the error of the individual transport modules. The convergence order for the \textit{CR transport-only} case is $\sim 2$ and corresponds to second-order convergence. This result is expected because the anisotropic transport of CRs is solely described by the parallel transport step and the order-reducing operator splitting for the coupling between CRs and gas is absent.

\subsection{Blast wave}
In this simulation we test our algorithm with a blast-wave problem triggered by a CR overpressure. We expect interesting deviations caused by the anisotropic transport of CRs in comparison to the purely spherical morphology of a Sedov-like blast wave that has been launched by a thermal overpressure. Similar tests have already been conducted for streaming and diffusing CRs \citep{2016Pakmor,2018Jiang}. Our initial conditions are as follows: the MHD quantities are set to $\rho = 1 \mathrm{m}_\mathrm{p} \mathrm{cm}^{-3}$, $\mathbfit{u} = \mathbf{0}$, $\pth = c_\mathrm{sd}^2 \rho / \gamma_\mathrm{th}$, and $\mathbfit{B} = \va \sqrt{\rho} \mathbfit{e}_x$ where sound- and Alfv{\'e}n speeds are $c_\mathrm{sd} = 10$ km s$^{-1}$ and $\va = 20$ km s$^{-1}$. The CR energy density in the background is set to $\pcr = 0.1 \pth$ and $\pcr = 100 \pth$ inside the overpressured region. This region is centred at $(x,y) = (0,0)$ and has a radius of 20~pc.
Initially, we adopt $\fcr = 0$ and $\ewpm = 10^{-4} \ecr$. We use $c=1000$ km s$^{-1}$ and $N_\mathrm{cr} = 8$ to simulate our CR dynamics. The simulations have been conducted with $128^2$ and $512^2$ moving mesh-generating points for an initially hexagonal mesh in the domain $x,y \in [-500, +500]$~pc.

The results are displayed in Fig.~\ref{fig:blast_wave}. Perpendicular and oblique to the direction of the magnetic field the evolution is adiabatic and the resulting density profile is mostly spherical. Along the magnetic field lines the evolution deviates from spherical symmetry. There the CRs start to stream away from the central overpressured region and form a mostly flat bar. This alters the shock structure and causes the density shell to break up into two discontinuities. At the inner discontinuity both the CR energy density and mass density jump while the outer discontinuity is the continuation of the spherical shell where negligible CRs are present but a density jump occurs. CRs diffuse ahead of the shock and push a small amount of gas alongside. Additionally, a central and two smaller ridges at $x\sim\pm50$~pc are observable in $\rho$. Those same ridges are also visible in figure~12 of \citet{2018Jiang}. The $y$-motion of the blast waves carries magnetic field lines along. This rarefies the CRs in the vertical direction. The perturbed magnetic field lines traces the extent of the CR distribution in Fig.~\ref{fig:blast_wave}. 

By comparing the lower to the high resolution simulation, we observe that both results look similar and show the same general features. The decreased resolution for the 128$^2$ simulations implies a higher numerical diffusivity which leads to a broader CR distribution and a lower maximum value of $\ecr$. 

\section{Summary}

In this paper we presented a new finite volume method that enables us to simulate CRHD on the moving mesh of the \textsc{Arepo} code. We extend the CRHD equations of \citet{2019Thomas} to cosmological, comoving coordinates so that the presented scheme can be applied to study CR feedback in galaxies and galaxy clusters.

Our algorithm consist of three distinct integration steps that model different parts of the included CR physics.  The first step integrates the adiabatic dynamics of MHD and CR quantities and accounts for the quasi-Lagrangian motion of the mesh. To this end, the previously available CR module \citep{2017Pfrommer} was extended to include the evolution of the CR flux and gyroresonant Alfv{\'e}n waves. In the next step a path-conservative finite volume scheme is employed to describe the anisotropic transport of CRs along magnetic field lines. We use a Lax-Friedrichs-type Riemann solver that is able to resolve contact discontinuities. This anisotropic transport step can be subcycled and uses the reduced speed of light approximation to reduce the total computational cost of the algorithm, thus enabling large-scale cosmological simulations. In the last step, the gyroresonant interaction and wave damping mechanisms are modelled in our algorithm using a source term integration step. We developed a custom-made adaptive time step semi-implicit ODE integrator to solve this short-timescale effects numerically. All three integration steps are combined in an operator-split manner.  We tested the algorithm and our implementation using multiple tests that simulate problems of varying complexity and target different aspects of the algorithm. We verified that
\begin{enumerate}
    \item multi-dimensional simulations show the expected CR streaming and diffusion modes of anisotropic CRs transport, 
    \item the algorithm is accurate by comparing the results of our simulations against solutions of the linearised and full set of CRHD equations. We showed that our implementation reaches a convergence order that ranges between first and second order, and that
    \item coupling between CR and thermal gas is correctly accounted for with simulations of shocks, blast waves, the expansion of a smooth CR distribution, and the acceleration of a warm cloud.
\end{enumerate}
The excellent performance of our method leaves us confident that the presented algorithm is versatile and allows for accurate and stable simulations of complex astrophysical environments. The combination all methods enables us to simulate the impact of CRs on the global evolution of the investigated system with little compromises on the included CR transport mechanisms.

\section*{Acknowledgements}
We thank Thomas Berlok for helpful discussions and comments on the draft of this paper. TT and CP acknowledge support by the European Research Council under ERC-CoG grant CRAGSMAN-646955. This research was supported in part by the National Science Foundation under Grant No.\ NSF PHY-1748958.

%%%%%%%%%%%%%%%%%%%%%%%%%%%%%%%%%%%%%%%%%%%%%%%%%%
%%%%%%%%%%%%%%%%%%%% REFERENCES %%%%%%%%%%%%%%%%%%
\bibliographystyle{mnras}
\bibliography{main}

\begin{thebibliography}{}
\makeatletter
\relax
\def\mn@urlcharsother{\let\do\@makeother \do\$\do\&\do\#\do\^\do\_\do\%\do\~}
\def\mn@doi{\begingroup\mn@urlcharsother \@ifnextchar [ {\mn@doi@}
  {\mn@doi@[]}}
\def\mn@doi@[#1]#2{\def\@tempa{#1}\ifx\@tempa\@empty \href
  {http://dx.doi.org/#2} {doi:#2}\else \href {http://dx.doi.org/#2} {#1}\fi
  \endgroup}
\def\mn@eprint#1#2{\mn@eprint@#1:#2::\@nil}
\def\mn@eprint@arXiv#1{\href {http://arxiv.org/abs/#1} {{\tt arXiv:#1}}}
\def\mn@eprint@dblp#1{\href {http://dblp.uni-trier.de/rec/bibtex/#1.xml}
  {dblp:#1}}
\def\mn@eprint@#1:#2:#3:#4\@nil{\def\@tempa {#1}\def\@tempb {#2}\def\@tempc
  {#3}\ifx \@tempc \@empty \let \@tempc \@tempb \let \@tempb \@tempa \fi \ifx
  \@tempb \@empty \def\@tempb {arXiv}\fi \@ifundefined
  {mn@eprint@\@tempb}{\@tempb:\@tempc}{\expandafter \expandafter \csname
  mn@eprint@\@tempb\endcsname \expandafter{\@tempc}}}

\bibitem[\protect\citeauthoryear{{Achterberg}}{{Achterberg}}{1981}]{1981Achterberg}
{Achterberg} A.,  1981, \aap, \href
  {https://ui.adsabs.harvard.edu/abs/1981A&A....98..195A} {98, 195}

\bibitem[\protect\citeauthoryear{{Bai}, {Ostriker}, {Plotnikov}  \&
  {Stone}}{{Bai} et~al.}{2019}]{2019Bai}
{Bai} X.-N.,  {Ostriker} E.~C.,  {Plotnikov} I.,   {Stone} J.~M.,  2019,
  \mn@doi [\apj] {10.3847/1538-4357/ab1648}, \href
  {https://ui.adsabs.harvard.edu/abs/2019ApJ...876...60B} {876, 60}

\bibitem[\protect\citeauthoryear{{Balsara}}{{Balsara}}{2004}]{2004Balsara}
{Balsara} D.~S.,  2004, \mn@doi [\apjs] {10.1086/381377}, \href
  {https://ui.adsabs.harvard.edu/abs/2004ApJS..151..149B} {151, 149}

\bibitem[\protect\citeauthoryear{{Blandford} \& {Eichler}}{{Blandford} \&
  {Eichler}}{1987}]{1987Blandford}
{Blandford} R.,  {Eichler} D.,  1987, \mn@doi [\physrep]
  {10.1016/0370-1573(87)90134-7}, \href
  {https://ui.adsabs.harvard.edu/abs/1987PhR...154....1B} {154, 1}

\bibitem[\protect\citeauthoryear{{Boulares} \& {Cox}}{{Boulares} \&
  {Cox}}{1990}]{1990Boulares}
{Boulares} A.,  {Cox} D.~P.,  1990, \mn@doi [\apj] {10.1086/169509}, \href
  {https://ui.adsabs.harvard.edu/abs/1990ApJ...365..544B} {365, 544}

\bibitem[\protect\citeauthoryear{{Breitschwerdt}, {McKenzie}  \&
  {Voelk}}{{Breitschwerdt} et~al.}{1991}]{1991Breitschwerdt}
{Breitschwerdt} D.,  {McKenzie} J.~F.,   {Voelk} H.~J.,  1991, \aap, \href
  {https://ui.adsabs.harvard.edu/abs/1991A&A...245...79B} {245, 79}

\bibitem[\protect\citeauthoryear{{Buck}, {Pfrommer}, {Pakmor}, {Grand}  \&
  {Springel}}{{Buck} et~al.}{2020}]{2020Buck}
{Buck} T.,  {Pfrommer} C.,  {Pakmor} R.,  {Grand} R. J.~J.,   {Springel} V.,
  2020, \mn@doi [\mnras] {10.1093/mnras/staa1960}, \href
  {https://ui.adsabs.harvard.edu/abs/2020MNRAS.497.1712B} {497, 1712}

\bibitem[\protect\citeauthoryear{{Butsky} \& {Quinn}}{{Butsky} \&
  {Quinn}}{2018}]{2018Butsky}
{Butsky} I.~S.,  {Quinn} T.~R.,  2018, \mn@doi [\apj]
  {10.3847/1538-4357/aaeac2}, \href
  {https://ui.adsabs.harvard.edu/abs/2018ApJ...868..108B} {868, 108}

\bibitem[\protect\citeauthoryear{{Caprioli} \& {Spitkovsky}}{{Caprioli} \&
  {Spitkovsky}}{2014}]{2014Caprioli}
{Caprioli} D.,  {Spitkovsky} A.,  2014, \mn@doi [\apj]
  {10.1088/0004-637X/783/2/91}, \href
  {https://ui.adsabs.harvard.edu/abs/2014ApJ...783...91C} {783, 91}

\bibitem[\protect\citeauthoryear{{Chan}, {Kere{\v{s}}}, {Hopkins}, {Quataert},
  {Su}, {Hayward}  \& {Faucher-Gigu{\`e}re}}{{Chan} et~al.}{2019}]{2019Chan}
{Chan} T.~K.,  {Kere{\v{s}}} D.,  {Hopkins} P.~F.,  {Quataert} E.,  {Su} K.~Y.,
   {Hayward} C.~C.,   {Faucher-Gigu{\`e}re} C.~A.,  2019, \mn@doi [\mnras]
  {10.1093/mnras/stz1895}, \href
  {https://ui.adsabs.harvard.edu/abs/2019MNRAS.488.3716C} {488, 3716}

\bibitem[\protect\citeauthoryear{{Dal Maso}, {LeFloch}  \& {Murat}}{{Dal Maso}
  et~al.}{1995}]{DalMaso1995}
{Dal Maso} G.,  {LeFloch} P.~G.,   {Murat} F.,  1995, Journal de mathématiques
  pures et appliquées, 74, 483

\bibitem[\protect\citeauthoryear{{Dashyan} \& {Dubois}}{{Dashyan} \&
  {Dubois}}{2020}]{2020Dashyan}
{Dashyan} G.,  {Dubois} Y.,  2020, \mn@doi [\aap]
  {10.1051/0004-6361/201936339}, \href
  {https://ui.adsabs.harvard.edu/abs/2020A&A...638A.123D} {638, A123}

\bibitem[\protect\citeauthoryear{{Drake}, {Pfrommer}, {Reynolds}, {Ruszkowski},
  {Swisdak}, {Einarsson}, {Hassam}  \& {Roberg-Clark}}{{Drake}
  et~al.}{2020}]{Drake2020}
{Drake} J.~F.,  {Pfrommer} C.,  {Reynolds} C.~S.,  {Ruszkowski} M.,  {Swisdak}
  M.,  {Einarsson} A.,  {Hassam} A.~B.,   {Roberg-Clark} G.~T.,  2020, arXiv
  e-prints, \href {https://ui.adsabs.harvard.edu/abs/2020arXiv200707931D} {p.
  arXiv:2007.07931}

\bibitem[\protect\citeauthoryear{{Dubois}, {Commer{\c{c}}on}, {Marcowith}  \&
  {Brahimi}}{{Dubois} et~al.}{2019}]{2019Dubois}
{Dubois} Y.,  {Commer{\c{c}}on} B.,  {Marcowith} A.~r.,   {Brahimi} L.,  2019,
  \mn@doi [\aap] {10.1051/0004-6361/201936275}, \href
  {https://ui.adsabs.harvard.edu/abs/2019A&A...631A.121D} {631, A121}

\bibitem[\protect\citeauthoryear{Dumbser \& Balsara}{Dumbser \&
  Balsara}{2016}]{2016Dumbser}
Dumbser M.,  Balsara D.~S.,  2016, \mn@doi [J. Comput. Phys.]
  {10.1016/j.jcp.2015.10.014}, 304, 275–319

\bibitem[\protect\citeauthoryear{{En{\ss}lin}, {Pfrommer}, {Springel}  \&
  {Jubelgas}}{{En{\ss}lin} et~al.}{2007}]{2007Ensslin}
{En{\ss}lin} T.~A.,  {Pfrommer} C.,  {Springel} V.,   {Jubelgas} M.,  2007,
  \mn@doi [\aap] {10.1051/0004-6361:20065294}, \href
  {https://ui.adsabs.harvard.edu/abs/2007A&A...473...41E} {473, 41}

\bibitem[\protect\citeauthoryear{{Everett}, {Zweibel}, {Benjamin}, {McCammon},
  {Rocks}  \& {Gallagher}}{{Everett} et~al.}{2008}]{2008Everett}
{Everett} J.~E.,  {Zweibel} E.~G.,  {Benjamin} R.~A.,  {McCammon} D.,  {Rocks}
  L.,   {Gallagher} John~S. I.,  2008, \mn@doi [\apj] {10.1086/524766}, \href
  {https://ui.adsabs.harvard.edu/abs/2008ApJ...674..258E} {674, 258}

\bibitem[\protect\citeauthoryear{{Farber}, {Ruszkowski}, {Yang}  \&
  {Zweibel}}{{Farber} et~al.}{2018}]{2018Farber}
{Farber} R.,  {Ruszkowski} M.,  {Yang} H. Y.~K.,   {Zweibel} E.~G.,  2018,
  \mn@doi [\apj] {10.3847/1538-4357/aab26d}, \href
  {https://ui.adsabs.harvard.edu/abs/2018ApJ...856..112F} {856, 112}

\bibitem[\protect\citeauthoryear{{Girichidis}, {Naab}, {Walch}  \&
  {Hanasz}}{{Girichidis} et~al.}{2014}]{2014Girichidis}
{Girichidis} P.,  {Naab} T.,  {Walch} S.,   {Hanasz} M.,  2014, arXiv e-prints,
  \href {https://ui.adsabs.harvard.edu/abs/2014arXiv1406.4861G} {p.
  arXiv:1406.4861}

\bibitem[\protect\citeauthoryear{{Girichidis} et~al.,}{{Girichidis}
  et~al.}{2016}]{2016Girichidis}
{Girichidis} P.,  et~al., 2016, \mn@doi [\apjl] {10.3847/2041-8205/816/2/L19},
  \href {https://ui.adsabs.harvard.edu/abs/2016ApJ...816L..19G} {816, L19}

\bibitem[\protect\citeauthoryear{{Girichidis}, {Pfrommer}, {Hanasz}  \&
  {Naab}}{{Girichidis} et~al.}{2020}]{2020Girichidis}
{Girichidis} P.,  {Pfrommer} C.,  {Hanasz} M.,   {Naab} T.,  2020, \mn@doi
  [\mnras] {10.1093/mnras/stz2961}, \href
  {https://ui.adsabs.harvard.edu/abs/2020MNRAS.491..993G} {491, 993}

\bibitem[\protect\citeauthoryear{{Haggerty} \& {Caprioli}}{{Haggerty} \&
  {Caprioli}}{2019}]{2019Haggerty}
{Haggerty} C.~C.,  {Caprioli} D.,  2019, \mn@doi [\apj]
  {10.3847/1538-4357/ab58c8}, \href
  {https://ui.adsabs.harvard.edu/abs/2019ApJ...887..165H} {887, 165}

\bibitem[\protect\citeauthoryear{{Hanasz} \& {Lesch}}{{Hanasz} \&
  {Lesch}}{2003}]{2003Hanasz}
{Hanasz} M.,  {Lesch} H.,  2003, \mn@doi [\aap] {10.1051/0004-6361:20031433},
  \href {https://ui.adsabs.harvard.edu/abs/2003A&A...412..331H} {412, 331}

\bibitem[\protect\citeauthoryear{{Hanasz}, {Lesch}, {Naab}, {Gawryszczak},
  {Kowalik}  \& {W{\'o}lta{\'n}ski}}{{Hanasz} et~al.}{2013}]{2013Hanasz}
{Hanasz} M.,  {Lesch} H.,  {Naab} T.,  {Gawryszczak} A.,  {Kowalik} K.,
  {W{\'o}lta{\'n}ski} D.,  2013, \mn@doi [\apjl] {10.1088/2041-8205/777/2/L38},
  \href {https://ui.adsabs.harvard.edu/abs/2013ApJ...777L..38H} {777, L38}

\bibitem[\protect\citeauthoryear{Harten, Lax  \& Leer}{Harten
  et~al.}{1983}]{1983Harten}
Harten A.,  Lax P.~D.,   Leer B.~v.,  1983, \mn@doi [SIAM Review]
  {10.1137/1025002}, 25, 35

\bibitem[\protect\citeauthoryear{{Heckman} \& {Thompson}}{{Heckman} \&
  {Thompson}}{2017}]{2017Heckman}
{Heckman} T.~M.,  {Thompson} T.~A.,  2017, {Galactic Winds and the Role Played
  by Massive Stars}.
Springer International Publishing, p.~2431,
  \mn@doi{10.1007/978-3-319-21846-5_23}

\bibitem[\protect\citeauthoryear{{Holcomb} \& {Spitkovsky}}{{Holcomb} \&
  {Spitkovsky}}{2019}]{2019Holcomb}
{Holcomb} C.,  {Spitkovsky} A.,  2019, \mn@doi [\apj]
  {10.3847/1538-4357/ab328a}, \href
  {https://ui.adsabs.harvard.edu/abs/2019ApJ...882....3H} {882, 3}

\bibitem[\protect\citeauthoryear{{Hopkins} et~al.,}{{Hopkins}
  et~al.}{2020}]{2020Hopkins}
{Hopkins} P.~F.,  et~al., 2020, \mn@doi [\mnras] {10.1093/mnras/stz3321}, \href
  {https://ui.adsabs.harvard.edu/abs/2020MNRAS.492.3465H} {492, 3465}

\bibitem[\protect\citeauthoryear{{Jacob}, {Pakmor}, {Simpson}, {Springel}  \&
  {Pfrommer}}{{Jacob} et~al.}{2018}]{2018Jacob}
{Jacob} S.,  {Pakmor} R.,  {Simpson} C.~M.,  {Springel} V.,   {Pfrommer} C.,
  2018, \mn@doi [\mnras] {10.1093/mnras/stx3221}, \href
  {https://ui.adsabs.harvard.edu/abs/2018MNRAS.475..570J} {475, 570}

\bibitem[\protect\citeauthoryear{{Jiang} \& {Oh}}{{Jiang} \&
  {Oh}}{2018}]{2018Jiang}
{Jiang} Y.-F.,  {Oh} S.~P.,  2018, \mn@doi [\apj] {10.3847/1538-4357/aaa6ce},
  \href {https://ui.adsabs.harvard.edu/abs/2018ApJ...854....5J} {854, 5}

\bibitem[\protect\citeauthoryear{{Jubelgas}, {Springel}, {En{\ss}lin}  \&
  {Pfrommer}}{{Jubelgas} et~al.}{2008}]{2008Jubelgas}
{Jubelgas} M.,  {Springel} V.,  {En{\ss}lin} T.,   {Pfrommer} C.,  2008,
  \mn@doi [\aap] {10.1051/0004-6361:20065295}, \href
  {https://ui.adsabs.harvard.edu/abs/2008A&A...481...33J} {481, 33}

\bibitem[\protect\citeauthoryear{{Ko}}{{Ko}}{1992}]{1992Ko}
{Ko} C.~M.,  1992, \aap, \href
  {https://ui.adsabs.harvard.edu/abs/1992A&A...259..377K} {259, 377}

\bibitem[\protect\citeauthoryear{{Kulsrud} \& {Pearce}}{{Kulsrud} \&
  {Pearce}}{1969}]{1969Kulsrud}
{Kulsrud} R.,  {Pearce} W.~P.,  1969, \mn@doi [\apj] {10.1086/149981}, \href
  {https://ui.adsabs.harvard.edu/abs/1969ApJ...156..445K} {156, 445}

\bibitem[\protect\citeauthoryear{{Lebiga}, {Santos-Lima}  \& {Yan}}{{Lebiga}
  et~al.}{2018}]{2018Lebiga}
{Lebiga} O.,  {Santos-Lima} R.,   {Yan} H.,  2018, \mn@doi [\mnras]
  {10.1093/mnras/sty309}, \href
  {https://ui.adsabs.harvard.edu/abs/2018MNRAS.476.2779L} {476, 2779}

\bibitem[\protect\citeauthoryear{{Lee} \& {Voelk}}{{Lee} \&
  {Voelk}}{1975}]{1975Lee}
{Lee} M.~A.,  {Voelk} H.~J.,  1975, \mn@doi [\apj] {10.1086/153625}, \href
  {https://ui.adsabs.harvard.edu/abs/1975ApJ...198..485L} {198, 485}

\bibitem[\protect\citeauthoryear{{Litvinenko} \& {Noble}}{{Litvinenko} \&
  {Noble}}{2016}]{2016Litvinenko}
{Litvinenko} Y.~E.,  {Noble} P.~L.,  2016, \mn@doi [Physics of Plasmas]
  {10.1063/1.4953564}, \href
  {https://ui.adsabs.harvard.edu/abs/2016PhPl...23f2901L} {23, 062901}

\bibitem[\protect\citeauthoryear{{Malkov}}{{Malkov}}{2018}]{2018Malkov}
{Malkov} M.~A.,  2018, \mn@doi [Nuclear and Particle Physics Proceedings]
  {10.1016/j.nuclphysbps.2018.07.023}, \href
  {https://ui.adsabs.harvard.edu/abs/2018NPPP..297..152M} {297-299, 152}

\bibitem[\protect\citeauthoryear{{Malkov} \& {Sagdeev}}{{Malkov} \&
  {Sagdeev}}{2015}]{2015Malkov}
{Malkov} M.~A.,  {Sagdeev} R.~Z.,  2015, \mn@doi [\apj]
  {10.1088/0004-637X/808/2/157}, \href
  {https://ui.adsabs.harvard.edu/abs/2015ApJ...808..157M} {808, 157}

\bibitem[\protect\citeauthoryear{{McKenzie} \& {Webb}}{{McKenzie} \&
  {Webb}}{1984}]{1984McKenzie}
{McKenzie} J.~F.,  {Webb} G.~M.,  1984, \mn@doi [Journal of Plasma Physics]
  {10.1017/S0022377800001628}, \href
  {https://ui.adsabs.harvard.edu/abs/1984JPlPh..31..275M} {31, 275}

\bibitem[\protect\citeauthoryear{{Miyoshi} \& {Kusano}}{{Miyoshi} \&
  {Kusano}}{2005}]{2005HLLD}
{Miyoshi} T.,  {Kusano} K.,  2005, \mn@doi [Journal of Computational Physics]
  {10.1016/j.jcp.2005.02.017}, \href
  {https://ui.adsabs.harvard.edu/abs/2005JCoPh.208..315M} {208, 315}

\bibitem[\protect\citeauthoryear{{Pakmor} \& {Springel}}{{Pakmor} \&
  {Springel}}{2013}]{2013Pakmor}
{Pakmor} R.,  {Springel} V.,  2013, \mn@doi [\mnras] {10.1093/mnras/stt428},
  \href {https://ui.adsabs.harvard.edu/abs/2013MNRAS.432..176P} {432, 176}

\bibitem[\protect\citeauthoryear{{Pakmor}, {Bauer}  \& {Springel}}{{Pakmor}
  et~al.}{2011}]{2011Pakmor}
{Pakmor} R.,  {Bauer} A.,   {Springel} V.,  2011, \mn@doi [\mnras]
  {10.1111/j.1365-2966.2011.19591.x}, \href
  {https://ui.adsabs.harvard.edu/abs/2011MNRAS.418.1392P} {418, 1392}

\bibitem[\protect\citeauthoryear{{Pakmor}, {Springel}, {Bauer}, {Mocz},
  {Munoz}, {Ohlmann}, {Schaal}  \& {Zhu}}{{Pakmor}
  et~al.}{2016a}]{2016PakmorII}
{Pakmor} R.,  {Springel} V.,  {Bauer} A.,  {Mocz} P.,  {Munoz} D.~J.,
  {Ohlmann} S.~T.,  {Schaal} K.,   {Zhu} C.,  2016a, \mn@doi [\mnras]
  {10.1093/mnras/stv2380}, \href
  {https://ui.adsabs.harvard.edu/abs/2016MNRAS.455.1134P} {455, 1134}

\bibitem[\protect\citeauthoryear{{Pakmor}, {Pfrommer}, {Simpson}, {Kannan}  \&
  {Springel}}{{Pakmor} et~al.}{2016b}]{2016Pakmor}
{Pakmor} R.,  {Pfrommer} C.,  {Simpson} C.~M.,  {Kannan} R.,   {Springel} V.,
  2016b, \mn@doi [\mnras] {10.1093/mnras/stw1761}, \href
  {https://ui.adsabs.harvard.edu/abs/2016MNRAS.462.2603P} {462, 2603}

\bibitem[\protect\citeauthoryear{{Pakmor}, {Pfrommer}, {Simpson}  \&
  {Springel}}{{Pakmor} et~al.}{2016c}]{2016PakmorIII}
{Pakmor} R.,  {Pfrommer} C.,  {Simpson} C.~M.,   {Springel} V.,  2016c, \mn@doi
  [\apjl] {10.3847/2041-8205/824/2/L30}, \href
  {https://ui.adsabs.harvard.edu/abs/2016ApJ...824L..30P} {824, L30}

\bibitem[\protect\citeauthoryear{{Par\'es}}{{Par\'es}}{2006}]{2006Pares}
{Par\'es} C.,  2006, \mn@doi [SIAM J. Numerical Analysis] {10.1137/050628052},
  44, 300

\bibitem[\protect\citeauthoryear{{Par\'es} \& {Muñoz-Ruiz}}{{Par\'es} \&
  {Muñoz-Ruiz}}{2009}]{2009Pares}
{Par\'es} C.,  {Muñoz-Ruiz} M.~L.,  2009, Bol. Soc. Esp. Mat. Apl. SeMA, 47,
  19–48

\bibitem[\protect\citeauthoryear{{Pareschi} \& {Russo}}{{Pareschi} \&
  {Russo}}{2005}]{2005Pareschi}
{Pareschi} L.,  {Russo} G.,  2005, \mn@doi [Journal of Scientific Computing]
  {10.1007/s10915-004-4636-4}, 25, 129

\bibitem[\protect\citeauthoryear{{Pfrommer}, {Springel}, {En{\ss}lin}  \&
  {Jubelgas}}{{Pfrommer} et~al.}{2006}]{2006Pfrommer}
{Pfrommer} C.,  {Springel} V.,  {En{\ss}lin} T.~A.,   {Jubelgas} M.,  2006,
  \mn@doi [\mnras] {10.1111/j.1365-2966.2005.09953.x}, \href
  {https://ui.adsabs.harvard.edu/abs/2006MNRAS.367..113P} {367, 113}

\bibitem[\protect\citeauthoryear{{Pfrommer}, {En{\ss}lin}, {Springel},
  {Jubelgas}  \& {Dolag}}{{Pfrommer} et~al.}{2007}]{2007Pfrommer}
{Pfrommer} C.,  {En{\ss}lin} T.~A.,  {Springel} V.,  {Jubelgas} M.,   {Dolag}
  K.,  2007, \mn@doi [\mnras] {10.1111/j.1365-2966.2007.11732.x}, \href
  {https://ui.adsabs.harvard.edu/abs/2007MNRAS.378..385P} {378, 385}

\bibitem[\protect\citeauthoryear{{Pfrommer}, {Pakmor}, {Schaal}, {Simpson}  \&
  {Springel}}{{Pfrommer} et~al.}{2017}]{2017Pfrommer}
{Pfrommer} C.,  {Pakmor} R.,  {Schaal} K.,  {Simpson} C.~M.,   {Springel} V.,
  2017, \mn@doi [\mnras] {10.1093/mnras/stw2941}, \href
  {https://ui.adsabs.harvard.edu/abs/2017MNRAS.465.4500P} {465, 4500}

\bibitem[\protect\citeauthoryear{Pohl, Hoshino  \& Niemiec}{Pohl
  et~al.}{2020}]{2020Pohl}
Pohl M.,  Hoshino M.,   Niemiec J.,  2020, \mn@doi [Progress in Particle and
  Nuclear Physics] {https://doi.org/10.1016/j.ppnp.2019.103751}, 111, 103751

\bibitem[\protect\citeauthoryear{Powell, Roe, Linde, Gombosi  \& Zeeuw]}{Powell
  et~al.}{1999}]{1999Powell}
Powell K.~G.,  Roe P.~L.,  Linde T.~J.,  Gombosi T.~I.,   Zeeuw] D. L.~D.,
  1999, \mn@doi [Journal of Computational Physics]
  {https://doi.org/10.1006/jcph.1999.6299}, 154, 284

\bibitem[\protect\citeauthoryear{{Recchia}, {Blasi}  \& {Morlino}}{{Recchia}
  et~al.}{2017}]{2017Recchia}
{Recchia} S.,  {Blasi} P.,   {Morlino} G.,  2017, \mn@doi [\mnras]
  {10.1093/mnras/stx1214}, \href
  {https://ui.adsabs.harvard.edu/abs/2017MNRAS.470..865R} {470, 865}

\bibitem[\protect\citeauthoryear{{Rodrigues}, {Snodin}, {Sarson}  \&
  {Shukurov}}{{Rodrigues} et~al.}{2019}]{2019Rodrigues}
{Rodrigues} L. F.~S.,  {Snodin} A.~P.,  {Sarson} G.~R.,   {Shukurov} A.,  2019,
  \mn@doi [\mnras] {10.1093/mnras/stz1354}, \href
  {https://ui.adsabs.harvard.edu/abs/2019MNRAS.487..975R} {487, 975}

\bibitem[\protect\citeauthoryear{{Ruszkowski}, {Yang}  \&
  {Zweibel}}{{Ruszkowski} et~al.}{2017}]{2017Ruszkowski}
{Ruszkowski} M.,  {Yang} H. Y.~K.,   {Zweibel} E.,  2017, \mn@doi [\apj]
  {10.3847/1538-4357/834/2/208}, \href
  {https://ui.adsabs.harvard.edu/abs/2017ApJ...834..208R} {834, 208}

\bibitem[\protect\citeauthoryear{{Salem} \& {Bryan}}{{Salem} \&
  {Bryan}}{2014}]{2014Salem}
{Salem} M.,  {Bryan} G.~L.,  2014, \mn@doi [\mnras] {10.1093/mnras/stt2121},
  \href {https://ui.adsabs.harvard.edu/abs/2014MNRAS.437.3312S} {437, 3312}

\bibitem[\protect\citeauthoryear{{Salem}, {Bryan}  \& {Hummels}}{{Salem}
  et~al.}{2014}]{2014SalemII}
{Salem} M.,  {Bryan} G.~L.,   {Hummels} C.,  2014, \mn@doi [\apjl]
  {10.1088/2041-8205/797/2/L18}, \href
  {https://ui.adsabs.harvard.edu/abs/2014ApJ...797L..18S} {797, L18}

\bibitem[\protect\citeauthoryear{{Salem}, {Bryan}  \& {Corlies}}{{Salem}
  et~al.}{2016}]{2016Salem}
{Salem} M.,  {Bryan} G.~L.,   {Corlies} L.,  2016, \mn@doi [\mnras]
  {10.1093/mnras/stv2641}, \href
  {https://ui.adsabs.harvard.edu/abs/2016MNRAS.456..582S} {456, 582}

\bibitem[\protect\citeauthoryear{{Schlickeiser}}{{Schlickeiser}}{2002}]{BookSchlickeiser}
{Schlickeiser} R.,  2002, {Cosmic Ray Astrophysics}.
Springer-Verlag Berlin Heidelberg

\bibitem[\protect\citeauthoryear{{Schmidt} et~al.,}{{Schmidt}
  et~al.}{2019}]{2019Schmidt}
{Schmidt} P.,  et~al., 2019, \mn@doi [\aap] {10.1051/0004-6361/201834995},
  \href {https://ui.adsabs.harvard.edu/abs/2019A&A...632A..12S} {632, A12}

\bibitem[\protect\citeauthoryear{{Schneider} \& {Robertson}}{{Schneider} \&
  {Robertson}}{2018}]{2018Schneider}
{Schneider} E.~E.,  {Robertson} B.~E.,  2018, \mn@doi [\apj]
  {10.3847/1538-4357/aac329}, \href
  {https://ui.adsabs.harvard.edu/abs/2018ApJ...860..135S} {860, 135}

\bibitem[\protect\citeauthoryear{{Sharma} \& {Hammett}}{{Sharma} \&
  {Hammett}}{2007}]{2007Sharma}
{Sharma} P.,  {Hammett} G.~W.,  2007, \mn@doi [Journal of Computational
  Physics] {10.1016/j.jcp.2007.07.026}, \href
  {https://ui.adsabs.harvard.edu/abs/2007JCoPh.227..123S} {227, 123}

\bibitem[\protect\citeauthoryear{{Sharma}, {Colella}  \& {Martin}}{{Sharma}
  et~al.}{2010}]{2010Sharma}
{Sharma} P.,  {Colella} P.,   {Martin} D.~F.,  2010, \mn@doi [SIAM Journal on
  Scientific Computing] {10.1137/100792135}, 32, 3564

\bibitem[\protect\citeauthoryear{{Simpson}, {Pakmor}, {Marinacci}, {Pfrommer},
  {Springel}, {Glover}, {Clark}  \& {Smith}}{{Simpson}
  et~al.}{2016}]{2016Simpson}
{Simpson} C.~M.,  {Pakmor} R.,  {Marinacci} F.,  {Pfrommer} C.,  {Springel} V.,
   {Glover} S. C.~O.,  {Clark} P.~C.,   {Smith} R.~J.,  2016, \mn@doi [\apjl]
  {10.3847/2041-8205/827/2/L29}, \href
  {https://ui.adsabs.harvard.edu/abs/2016ApJ...827L..29S} {827, L29}

\bibitem[\protect\citeauthoryear{{Skilling}}{{Skilling}}{1971}]{1971Skilling}
{Skilling} J.,  1971, \mn@doi [\apj] {10.1086/151210}, \href
  {https://ui.adsabs.harvard.edu/abs/1971ApJ...170..265S} {170, 265}

\bibitem[\protect\citeauthoryear{{Skilling}}{{Skilling}}{1975}]{1975Skilling}
{Skilling} J.,  1975, \mn@doi [\mnras] {10.1093/mnras/172.3.557}, \href
  {https://ui.adsabs.harvard.edu/abs/1975MNRAS.172..557S} {172, 557}

\bibitem[\protect\citeauthoryear{{Springel}}{{Springel}}{2010}]{2010Springel}
{Springel} V.,  2010, \mn@doi [\mnras] {10.1111/j.1365-2966.2009.15715.x},
  \href {https://ui.adsabs.harvard.edu/abs/2010MNRAS.401..791S} {401, 791}

\bibitem[\protect\citeauthoryear{{Thomas} \& {Pfrommer}}{{Thomas} \&
  {Pfrommer}}{2019}]{2019Thomas}
{Thomas} T.,  {Pfrommer} C.,  2019, \mn@doi [\mnras] {10.1093/mnras/stz263},
  \href {https://ui.adsabs.harvard.edu/abs/2019MNRAS.485.2977T} {485, 2977}

\bibitem[\protect\citeauthoryear{{Thomas}, {Pfrommer}  \&
  {En{\ss}lin}}{{Thomas} et~al.}{2020}]{2020Thomas}
{Thomas} T.,  {Pfrommer} C.,   {En{\ss}lin} T.,  2020, \mn@doi [\apjl]
  {10.3847/2041-8213/ab7237}, \href
  {https://ui.adsabs.harvard.edu/abs/2020ApJ...890L..18T} {890, L18}

\bibitem[\protect\citeauthoryear{{Tumlinson}, {Peeples}  \& {Werk}}{{Tumlinson}
  et~al.}{2017}]{2017Tumlinson}
{Tumlinson} J.,  {Peeples} M.~S.,   {Werk} J.~K.,  2017, \mn@doi [\araa]
  {10.1146/annurev-astro-091916-055240}, \href
  {https://ui.adsabs.harvard.edu/abs/2017ARA&A..55..389T} {55, 389}

\bibitem[\protect\citeauthoryear{{Uhlig}, {Pfrommer}, {Sharma}, {Nath},
  {En{\ss}lin}  \& {Springel}}{{Uhlig} et~al.}{2012}]{2012Uhlig}
{Uhlig} M.,  {Pfrommer} C.,  {Sharma} M.,  {Nath} B.~B.,  {En{\ss}lin} T.~A.,
  {Springel} V.,  2012, \mn@doi [\mnras] {10.1111/j.1365-2966.2012.21045.x},
  \href {https://ui.adsabs.harvard.edu/abs/2012MNRAS.423.2374U} {423, 2374}

\bibitem[\protect\citeauthoryear{{Vukcevic}}{{Vukcevic}}{2013}]{2013Vukcevic}
{Vukcevic} M.,  2013, \mn@doi [\aap] {10.1051/0004-6361/201321191}, \href
  {https://ui.adsabs.harvard.edu/abs/2013A&A...555A.111V} {555, A111}

\bibitem[\protect\citeauthoryear{{Wagner}, {Falle}, {Hartquist}  \&
  {Pittard}}{{Wagner} et~al.}{2006}]{2006Wagner}
{Wagner} A.~Y.,  {Falle} S.~A.~E.~G.,  {Hartquist} T.~W.,   {Pittard} J.~M.,
  2006, \mn@doi [\aap] {10.1051/0004-6361:20064885}, \href
  {https://ui.adsabs.harvard.edu/abs/2006A&A...452..763W} {452, 763}

\bibitem[\protect\citeauthoryear{{Wiener}, {Oh}  \& {Zweibel}}{{Wiener}
  et~al.}{2017a}]{2017WienerII}
{Wiener} J.,  {Oh} S.~P.,   {Zweibel} E.~G.,  2017a, \mn@doi [\mnras]
  {10.1093/mnras/stx109}, \href
  {https://ui.adsabs.harvard.edu/abs/2017MNRAS.467..646W} {467, 646}

\bibitem[\protect\citeauthoryear{{Wiener}, {Pfrommer}  \& {Oh}}{{Wiener}
  et~al.}{2017b}]{2017Wiener}
{Wiener} J.,  {Pfrommer} C.,   {Oh} S.~P.,  2017b, \mn@doi [\mnras]
  {10.1093/mnras/stx127}, \href
  {https://ui.adsabs.harvard.edu/abs/2017MNRAS.467..906W} {467, 906}

\bibitem[\protect\citeauthoryear{{Wiener}, {Zweibel}  \& {Ruszkowski}}{{Wiener}
  et~al.}{2019}]{2019Wiener}
{Wiener} J.,  {Zweibel} E.~G.,   {Ruszkowski} M.,  2019, \mn@doi [\mnras]
  {10.1093/mnras/stz2007}, \href
  {https://ui.adsabs.harvard.edu/abs/2019MNRAS.489..205W} {489, 205}

\bibitem[\protect\citeauthoryear{{Yan} \& {Lazarian}}{{Yan} \&
  {Lazarian}}{2002}]{2002Yan}
{Yan} H.,  {Lazarian} A.,  2002, \mn@doi [\prl]
  {10.1103/PhysRevLett.89.281102}, \href
  {https://ui.adsabs.harvard.edu/abs/2002PhRvL..89B1102Y} {89, 281102}

\bibitem[\protect\citeauthoryear{{Yang}, {Ruszkowski}, {Ricker}, {Zweibel}  \&
  {Lee}}{{Yang} et~al.}{2012}]{2012Yang}
{Yang} H. Y.~K.,  {Ruszkowski} M.,  {Ricker} P.~M.,  {Zweibel} E.,   {Lee} D.,
  2012, \mn@doi [\apj] {10.1088/0004-637X/761/2/185}, \href
  {https://ui.adsabs.harvard.edu/abs/2012ApJ...761..185Y} {761, 185}

\bibitem[\protect\citeauthoryear{{Yee}, {Sandham}  \& {Djomehri}}{{Yee}
  et~al.}{1999}]{1999Yee}
{Yee} H.~C.,  {Sandham} N.~D.,   {Djomehri} M.~J.,  1999, \mn@doi [Journal of
  Computational Physics] {10.1006/jcph.1998.6177}, \href
  {https://ui.adsabs.harvard.edu/abs/1999JCoPh.150..199Y} {150, 199}

\bibitem[\protect\citeauthoryear{{Zweibel}}{{Zweibel}}{2013}]{2013Zweibel}
{Zweibel} E.~G.,  2013, \mn@doi [Physics of Plasmas] {10.1063/1.4807033}, \href
  {https://ui.adsabs.harvard.edu/abs/2013PhPl...20e5501Z} {20, 055501}

\bibitem[\protect\citeauthoryear{{Zweibel}}{{Zweibel}}{2017}]{2017Zweibel}
{Zweibel} E.~G.,  2017, \mn@doi [Physics of Plasmas] {10.1063/1.4984017}, \href
  {http://adsabs.harvard.edu/abs/2017PhPl...24e5402Z} {24, 055402}

\makeatother
\end{thebibliography}

\appendix
\section{Cosmological Equations}
\label{app:cosmo}

The CRHD equations laid down in Eqs.~\eqref{eq:continuity_equation} to \eqref{eq:wave_equation} are valid for a static space time. \textsc{Arepo} has the capability to run simulations in an expanding universe. To facilitate simulations of CRHD in an expanding universe the CRHD equations have to be adapted to eliminate the homogeneous Hubble expansion via comoving coordinates. To this end, we assume that the expansion can be described by a time-dependent scale factor $a(t)$ that obeys Friedmann's equations. We parameterise space using the comoving coordinate $\mathbfit{x}$ and eliminate  the physical coordinate $\mathbfit{r}$ from the CRHD equations. We also define comoving analogues to previous used quantities (denoted by a subscript c) to simplify the algebraic complexity and to transform the comoving equations into a form similar to their static counterparts. We use
\begin{alignat}{4}
    &\mathbfit{r} &&= a \mathbfit{x},  &&\mathbfit{u}_\mathrm{c} &&= \mathbfit{u} - \dot{a} \mathbfit{x}, \\
    &\rho &&= \rho_\mathrm{c} a^{-3},   &&\mathbfit{B} &&= \mathbfit{B}_\mathrm{c} a^{-2}, \\ 
    &\varepsilon_\mathrm{th} &&= \varepsilon_\mathrm{th,c} a^{-3},  &&\ecr &&= \varepsilon_\mathrm{cr,c} a^{-4}, \\
    &\ewpm &&= \varepsilon_\mathrm{a,\pm,c} a^{-9/2} \quad  &&\fcr &&= f_\mathrm{cr, c} a^{-4},
\end{alignat}
where $\mathbfit{u}_\mathrm{c}$ is the peculiar velocity. The ideal gas laws for the thermal gas, CRs, and Alfv{\'e}n waves remain the same and define their corresponding comoving pressures via
\begin{alignat}{3}
    &P_\mathrm{th,c} &&= (\gamma_\mathrm{th} - 1) &&\varepsilon_\mathrm{th,c},\\
    &P_\mathrm{cr,c} &&= (\gamma_\mathrm{cr} - 1) &&\varepsilon_\mathrm{cr,c},\\
    &P_\mathrm{a,\pm,c} &&= (\gamma_\mathrm{a} - 1) &&\varepsilon_\mathrm{a,\pm,c}.
\end{alignat}
The total comoving energy density contained in the MHD fluid is defined as:
\begin{equation}
    \varepsilon_\mathrm{c} = \frac{\rho_\mathrm{c} \mathbfit{u}^2_\mathrm{c}}{2} + \varepsilon_\mathrm{th,c} + \frac{\mathbfit{B}_\mathrm{c}^2}{2a}
\end{equation}
while the total comoving pressure is:
\begin{equation}
    P_\mathrm{tot,c} = P_\mathrm{th,c} + \frac{\mathbfit{B}_\mathrm{c}^2}{2a} + \frac{1}{a} P_\mathrm{cr,c} + \frac{1}{a^{3/2}} \left(P_\mathrm{a,+} + P_\mathrm{a,-}\right). \\
\end{equation}
With those definitions the continuity equation can be written as
\begin{align}
    \frac{\partial \rho_\mathrm{c}}{\partial t} + \frac{1}{a} \mathbf{\nabla}_\mathbfit{x} \mathbf{\cdot} \left(\rho_\mathrm{c} \mathbfit{u}_\mathrm{c} \right) = 0,
\end{align}
where $\nabla_\mathbfit{x}$ is the gradient with respect to the comoving coordinate $\mathbfit{x}$. Euler's equation takes the form:
\begin{align}
    \frac{\partial (a \rho_\mathrm{c} \mathbfit{u}_\mathrm{c})}{\partial t} &+ \frac{1}{a} \mathbf{\nabla}_\mathbfit{x} \mathbf{\cdot} \left[ a \left(\rho_\mathrm{c} \mathbfit{u}_\mathrm{c} \mathbfit{u}_\mathrm{c} + P_\mathrm{tot,c} \mathbf{1} - \frac{1}{a} \mathbfit{B}_\mathrm{c} \mathbfit{B}_\mathrm{c}\right)\right] = \nonumber \\ &+\frac{1}{a}\mathbfit{b}_\mathrm{c}\mathbf{\nabla}_{\mathbfit{x},\parallel} P_\mathrm{cr,c} + a^4\left(\mathbfit{g}_\mathrm{gri,+} + \mathbfit{g}_\mathrm{gri,-}\right),
\end{align}
where $\mathbf{\nabla}_{\mathbfit{x},\parallel} = \mathbfit{b}_\mathrm{c} \mathbf{\cdot}\nabla_\mathbfit{x}$ is the gradient in comoving coordinates projected onto the direction of the magnetic field $\mathbfit{b}_\mathrm{c} = \mathbfit{B}_\mathrm{c} / B_\mathrm{c}$. The comoving magnetic field is evolved using Faraday's law:
\begin{align}
     \frac{\partial \mathbfit{B}_\mathrm{c}}{\partial t} + \frac{1}{a} \mathbf{\nabla}_\mathbfit{x} \mathbf{\cdot} [\mathbfit{B}_\mathrm{c} \mathbfit{u}_\mathrm{c} - \mathbfit{u}_\mathrm{c}\mathbfit{B}_\mathrm{c}] &= \mathbf{0}.
\end{align}
Using the Hubble function $H = \dot{a} / a$, the MHD energy equation can be written as:
\begin{align}\frac{\partial (a^2\varepsilon_\mathrm{c})}{\partial t} &+ \frac{1}{a}\mathbf{\nabla}_\mathbfit{x} \mathbf{\cdot} \left\lbrace a^2\left[\mathbfit{u}_\mathrm{c} (\varepsilon_\mathrm{c} + P_\mathrm{tot,c}) - \frac{1}{a} (\mathbfit{u}_\mathrm{c} \mathbf{\cdot} \mathbfit{B}_\mathrm{c}) \mathbfit{B}_\mathrm{c}\right] \right\rbrace = \nonumber \\
    &+\left(P_\mathrm{cr,c}  + \frac{1}{a^{1/2}} P_\mathrm{a,+,c}  + \frac{1}{a^{1/2}} P_\mathrm{a,-,c} \right) \mathbf{\nabla}_\mathbfit{x} \mathbf{\cdot} \mathbfit{u}_\mathrm{c} \nonumber \\
    &+\mathbfit{u}_\mathrm{c} \mathbf{\cdot} \left[ a^2 \mathbfit{b}_\mathrm{c} \mathbf{\nabla}_{\mathbfit{x},\parallel} P_\mathrm{cr,c} +  a^5\left(\mathbfit{g}_\mathrm{gri,+} + \mathbfit{g}_\mathrm{gri,-}\right)\right] \nonumber \\ &- a^5 Q_\pm + \frac{aH}{2}\mathbfit{B}_\mathrm{c}^2,
\end{align}
while the CR energy equation becomes (assuming $\gamma_\mathrm{cr}=4/3$):
\begin{align}
    \frac{\partial \varepsilon_\mathrm{cr,c}}{\partial t} &+ \frac{1}{a} \mathbf{\nabla}_\mathbfit{x} \mathbf{\cdot} [\mathbfit{u}_\mathrm{c} \varepsilon_\mathrm{cr,c} + \mathbfit{b}_\mathrm{c} f_\mathrm{cr,c}] = -P_\mathrm{cr,c} \frac{1}{a} \mathbf{\nabla}_\mathbfit{x} \mathbf{\cdot} \mathbfit{u}_\mathrm{c}  \nonumber \\
    &\phantom{.} - a^4 \varv_\mathrm{a} \mathbfit{b}_\mathrm{c} \mathbf{\cdot} \left(\mathbfit{g}_\mathrm{gri,+} - \mathbfit{g}_\mathrm{gri,-}\right) .
\end{align}
The CR energy flux equations in comoving coordinates is:
\begin{align}
\frac{\partial f_\mathrm{cr,c}}{\partial t} &+ \frac{1}{a}\mathbf{\nabla}_\mathbfit{x} \mathbf{\cdot} [\mathbfit{u}_\mathrm{c} f_\mathrm{cr,c}] +\frac{c^2}{a} \mathbf{\nabla}_{\mathbfit{x},\parallel} P_\mathrm{cr,c} =   \nonumber \\
    &\hspace{-15pt}-\frac{1}{a}f_\mathrm{cr,c} (\mathbfit{b}_\mathrm{c}\mathbfit{b}_\mathrm{c}) \mathbf{:} \mathbf{\nabla}_\mathbfit{x} \mathbfit{u}_\mathrm{c} - a^4 c^2 \mathbfit{b}_\mathrm{c} \mathbf{\cdot} \left(\mathbfit{g}_\mathrm{gri,+} + \mathbfit{g}_\mathrm{gri,-}\right).
\end{align}
Finally, the equation for the energy contained in gyroresonant Alfv{\'e}n waves is given by:
\begin{align}
\frac{\partial \varepsilon_\mathrm{a,\pm,c}}{\partial t} &+ \frac{1}{a}\mathbf{\nabla}_\mathbfit{x} \mathbf{\cdot} [\mathbfit{u}_\mathrm{c} \varepsilon_\mathrm{a,\pm,c}  \pm \varv_\mathrm{a} \mathbfit{b} \varepsilon_\mathrm{a,\pm,c} ] =  \nonumber\\
    &-P_\mathrm{a,\pm,c} \frac{1}{a}\mathbf{\nabla}_\mathbfit{x} \mathbf{\cdot} \mathbfit{u}_\mathrm{c} + a^{9/2} \left(\pm  \va \mathbfit{b} \mathbf{\cdot} \mathbfit{g}_\mathrm{gri,\pm} - Q_\pm \right).
\end{align}
We use the Alfv{\'e}n speed with the variables $\mathbfit{B}$ and $\rho$ instead of their comoving counterparts because it simplifies the equation above. The Alfv{\'e}n speed can be conveniently calculated using the comoving variables via
\begin{equation}
    \va = B_\mathrm{c} \rho_\mathrm{c}^{-1/2}  a^{-1/2}.
\end{equation}
These equations reduce to Eqs.~\eqref{eq:continuity_equation} to \eqref{eq:wave_equation} in the case of a static universe, i.e.\ when $a = 1$ and $H=0$.

\section{Path-Conservative Schemes}
\label{app:pc}
In this appendix we provide a short introduction to path-conservative schemes. Because path-conservative schemes share many properties and generalise Godunov-type finite-volume schemes for conservation laws, we first recall the derivation of the latter to explain the former. 

\begin{figure}
\centering
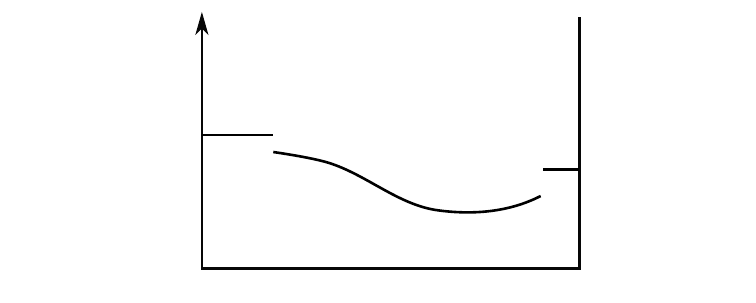
\caption{Structure of the flux $\mathbfit{F}(x)$ inside the cell $i$ that consists of a smooth inner part $\left.\mathbfit{F}\right\vert_\mathrm{smooth}$ and two constant interface fluxes $\mathbfit{F}_{*, i \pm 1/2}$. Gray dashed lines represent characteristics that travel with $S_L$ and $S_R$ away from the interfaces.}
\label{fig:fluxes}
\end{figure}

The solution to the one-dimensional conservation law
\begin{align}
    \frac{\partial \mathbfit{U}}{ \partial t} + \frac{\partial \mathbfit{F}(\mathbfit{U})}{\partial x} = \mathbf{0}
\end{align}
is described in the finite volume framework using cell averages $\dots, \mathbfit{U}_{i-1}, \mathbfit{U}_{i}, \mathbfit{U}_{i+1}, \dots$ of the state vector $\mathbfit{U}$. The time evolution of those is described by
\begin{align}
    \mathbf{0} &= \frac{\mathrm{d} \mathbfit{U}_i}{\mathrm{d} t} + \frac{1}{\Delta x} \int_{i} \mathrm{d}x \frac{\partial \mathbfit{F}}{\partial x}   \label{eq:finite_volume}, \\
    &= \frac{\mathrm{d} \mathbfit{U}_i}{\mathrm{d} t} + \frac{1}{\Delta x} \left(  \mathbfit{F}_{*,i + 1/2} -  \mathbfit{F}_{*,i - 1/2} \right)  \label{eq:godunov},
\end{align}
where we used the divergence theorem to evaluate the integral and $\mathbfit{F}_{*,i \pm 1/2}$ are the fluxes evaluated at the interfaces $i \pm 1/2$. Let us concentrate the discussion on one of those interfaces. At the interface itself the state vector may be discontinuous and may have different values to its left and right side denoted by $\mathbfit{U}_{L,R}$. Formally, this situation is similar to the initial conditions of a Riemann problem that is centred on the interface. Riemann solvers calculate a full or approximate solution to the Riemann problem to obtain the value for the flux $\mathbfit{F}_*$ through this interface using the given information. One of the most popular and simplest Riemann solvers is the HLLE Riemann solver \citep{1983Harten}. This Riemann solver approximates the solution to the Riemann problem with a single constant intermediate state $\mathbfit{U}_*$. The region where $\mathbfit{U}_*$ is realized is separated from the $\mathbfit{U}_{L,R}$ regions by one left- and one rightwards travelling discontinuity. We denote the speed of those discontinuities by $S_L$ and $S_R$, respectively. The intermediate state $\mathbfit{U}_*$ can be calculated using the Rankine-Hugoniot jump conditions for these discontinuities: 
\begin{align}
    S_L (\mathbfit{U}_* - \mathbfit{U}_L) &= \mathbfit{F}_* - \mathbfit{F}_L\\
    S_R (\mathbfit{U}_* - \mathbfit{U}_R) &= \mathbfit{F}_* - \mathbfit{F}_R.
\end{align}
Both equations can be readily solved for $\mathbfit{U}_*$ and $\mathbfit{F}_*$. A trivial modification of those equations is:
\begin{align}
     \mathbfit{F}_* &= \mathbfit{F}_L + S_L (\mathbfit{U}_* - \mathbfit{U}_L) \label{eq:fstar_l}\\
                  &= \mathbfit{F}_R + S_R (\mathbfit{U}_* - \mathbfit{U}_R) \label{eq:fstar_r},
\end{align}
which states that the HLL flux can be expressed in terms of left- or right-handed fluxes and states once $\mathbfit{U}_*$ is known. Inserting Eqs.~\eqref{eq:fstar_l} and \eqref{eq:fstar_r} into Eq.~\eqref{eq:godunov} gives:
\begin{align}
    \frac{\mathrm{d} \mathbfit{U}^i}{\mathrm{d} t} &+ \frac{1}{\Delta x} \left(  \mathbfit{F}_{i,R} -  \mathbfit{F}_{i,L} \right) \nonumber \\ &+ \frac{S_R}{\Delta x} \left(\mathbfit{U}_{*,R} - \mathbfit{U}_{i,R}\right) - \frac{S_L}{\Delta x} \left(\mathbfit{U}_{*,L} - \mathbfit{U}_{i,L}\right) = \mathbf{0} \hspace{-5pt} \label{eq:godonuv_with_jump},
\end{align}
where we switched from the face-centred back to the cell-centred meaning of left ($L$) and right ($R$). This equation can be interpreted in a mathematical, distributional sense with the help of the integral form of the finite volume method in Eq.~\eqref{eq:finite_volume} and Fig.~\ref{fig:fluxes}. In Fig.~\ref{fig:fluxes} we display $\mathbfit{F}(x)$ inside the cell. At the cell interface the otherwise smooth flux jumps from its value `inside' the cell to the HLL flux. The derivative of this profile inside the cell is given by:
\begin{align}
    \frac{\partial \mathbfit{F}} {\partial x} &= \left. \frac{\partial \mathbfit{F}}{\partial x} \right\vert_{\mathrm{smooth}} + \delta(x - x_{i+1/2} - S_R t) \, S_R \left(\mathbfit{U}_{*,R} - \mathbfit{U}_{i,R}\right) \nonumber \\&\hspace{30pt} - \delta(x - x_{i-1/2} - S_L t) \, S_L \left(\mathbfit{U}_{*,L} - \mathbfit{U}_{i,L}\right), \label{eq:flux_derivative}
\end{align}
where the Dirac $\delta$-distributions account for interface jumps, which result from the waves that travel from the interface into the cell. Inserting this expression into Eq.~\eqref{eq:finite_volume} and evaluating the integral at positive but infinitely small $t$ yields Eq.~\eqref{eq:godonuv_with_jump}. The first term can be interpreted as the contribution of the smooth component to the total flux while the last two terms are the singular contributions of the jumps of $\mathbfit{F}$ at the cell interface. 

%CP

Now, the finite volume method for non-conservative equations of the form
\begin{align}
    \frac{\partial \mathbfit{U}}{\partial t} + \mathbfss{H} \frac{\partial \mathbfit{U}}{\partial x} = \mathbf{0} \label{eq:nonconsertive_eq}
\end{align}
reads in the finite volume framework as
\begin{align}
    \frac{\mathrm{d} \mathbfit{U}_i}{\mathrm{d} t} + \frac{1}{\Delta x} \int_{i} \mathrm{d}x \, \mathbfss{H} \frac{\partial \mathbfit{U}}{\partial x} = \mathbf{0}.
\end{align}
The divergence theorem cannot be applied directly and above's methods for conservation laws cannot be used without modification. 
The central idea of how to solve this integral with path-conservative schemes is to impose a strict analogy in each discretisation and calculation step to conservative equations despite their formal differences. A justification for this procedure is that the path-conservative scheme should reduce to an ordinary Godunov-scheme for $\mathbfss{H} = \partial_\mathbfit{U} \mathbfit{F}$, i.e., when the non-conservative equation coincides with a conservation law. 

We assume that $\partial_x \mathbfit{U}$ has the same functional form as $\partial_x \mathbfit{F}$ in Eq.~\eqref{eq:flux_derivative}, namely, that it is composed of an inner smooth gradient and singular contributions travelling away from the interfaces. If we adopt a linear approximation for $\mathbfit{U}$ inside the cell then the integral can be easily solved to yield the path-conservative scheme for Eq.~\eqref{eq:nonconsertive_eq}:
\begin{align}
     \frac{\mathrm{d} \mathbfit{U}_i}{\mathrm{d} t} &+ \mathbfss{H}_i \left.\frac{\partial \mathbfit{U}}{\partial x}\right\vert_\mathrm{smooth} \nonumber \\ &\hspace{-5pt}+ \frac{S_R}{\Delta x} \left(\mathbfit{U}_{*,R} - \mathbfit{U}_{i,R}\right) - \frac{S_L}{\Delta x} \left(\mathbfit{U}_{*,L} - \mathbfit{U}_{i,L}\right)  = \mathbf{0},
\end{align}
where $\mathbfss{H}_i$ is the cell-average of $\mathbfss{H}$. It is sufficient to use the midpoint value $\mathbfss{H}_i$ for $\mathbfss{H}(x)$ and the gradient of $\mathbfit{U}$ as calculated via the standard piecewise-linear approximation to get an $\mathcal{O}(\Delta x^2)$ accurate scheme. The singular contributions are the principle components of path conservative schemes that allow for stable and shock-capturing numerical simulations. We define 
\begin{align}
    \mathbfit{D}_{L} &= S_L (\mathbfit{U}_{L} - \mathbfit{U}_{*}), \\
    \mathbfit{D}_{R} &= S_R (\mathbfit{U}_{R} - \mathbfit{U}_{*}),
\end{align} 
and call the $\mathbfit{D}_{L,R}$ linear fluctuations. With this definition the path-conservative scheme reduces to
\begin{align}
     \frac{\mathrm{d} \mathbfit{U}_i}{\mathrm{d} t} &+ \mathbfss{H}_i \left.\frac{\partial \mathbfit{U}}{\partial x}\right\vert_\mathrm{smooth} \nonumber \\ &+ \frac{1}{\Delta x} \left(\mathbfit{D}_{i+1/2,L} - \mathbfit{D}_{i - 1/2,R}\right) = \mathbf{0},
\end{align}
which is the one-dimensional form of the more general Eq.\eqref{eq:path_conservative_i}.

The fluctuation terms are still unspecified as we do not have an expression for $\mathbfit{U}_*$. In the case of conservation laws we use Rankine-Hugoniot jump conditions at the cell interface to calculate $\mathbfit{U}_*$. These jump conditions are not applicable for non-conservative equations. Yet, a permitted generalization for the jump conditions is presented in \citet{DalMaso1995}. These conditions read for our case: 
\begin{align}
    S_L (\mathbfit{U}_* - \mathbfit{U}_L) &= \int_0^1 \mathrm{d}s\, \mathbfss{H}(\mathbfit{U}_L(s)) \frac{\partial \mathbfit{U}_L(s)}{\partial s}, \\
    S_R (\mathbfit{U}_* - \mathbfit{U}_R) &= \int_0^1 \mathrm{d}s\, \mathbfss{H}(\mathbfit{U}_R(s)) \frac{\partial \mathbfit{U}_R(s)}{\partial s},
\end{align}
where $\mathbfit{U}_{L,R}(s)$ with $s \in [0,1]$ are \textit{paths} connecting all states at the interface. We impose boundary conditions for these paths with
\begin{align}
    \mathbfit{U}_L(0) &= \mathbfit{U}_L \mathrm{~ and ~} \mathbfit{U}_L(1) = \mathbfit{U}_*, \\
    \mathbfit{U}_R(0) &= \mathbfit{U}_R \mathrm{~ and ~} \mathbfit{U}_R(1) = \mathbfit{U}_*.
\end{align}
These jump conditions are true generalisations as they reduce to the usual Rankine-Hugoniot conditions for $\mathbfss{H} = \partial_\mathbfit{U} \mathbfit{F}$. In this case both integrals can be solved independently of the chosen path. The results are the usual jump conditions for conservation laws in Eqs.~\eqref{eq:fstar_l} and \eqref{eq:fstar_r}. The required consistency between conservative and non-conservative schemes is achieved. 

For a path-conservative scheme to converge, almost arbitrary paths can be chosen. However, the realised numerical solutions for different paths will differ at shocks. The choice of a path dictates the jump conditions and thus the solution at the shock \citep{2009Pares}. A well-motivated path can be chosen if the underlying physical model can be expanded to include physical viscosity. The paths describing the solutions in the inviscid case can then be calculated as the limit of vanishing viscosity of a steady state solution of the viscous equation. The construction of those paths is cumbersome and rarely carried out. A more simplistic yet tractable choice is to assume linear paths:
\begin{align}
    \mathbfit{U}_L(s) &= \mathbfit{U}_L + (\mathbfit{U}_* - \mathbfit{U}_L)\, s, \\
    \mathbfit{U}_R(s) &= \mathbfit{U}_R + (\mathbfit{U}_* - \mathbfit{U}_R)\, s.
\end{align}
For those paths the generalised Rankine-Hugoniot jump conditions read:
\begin{align}
    S_L (\mathbfit{U}_* - \mathbfit{U}_L) &= \mathbfss{H}_L (\mathbfit{U}_* - \mathbfit{U}_L), \\
    S_R (\mathbfit{U}_* - \mathbfit{U}_R) &= \mathbfss{H}_R (\mathbfit{U}_* - \mathbfit{U}_R),
\end{align}
where
\begin{align}
    \mathbfss{H}_L = \int_0^1 \mathrm{d}s\, \mathbfss{H}(\mathbfit{U}_R(s)), \\
    \mathbfss{H}_R = \int_0^1 \mathrm{d}s\, \mathbfss{H}(\mathbfit{U}_L(s)).\,
\end{align}
Note that also here, the intermediate state $\mathbfit{U}_*$ implicitly enters these equations through $\mathbfit{U}_{L,R}(s)$. An iterative procedure to calculate $\mathbfit{U}_*$ based on the Newton-Raphson method is proposed in \citet{2016Dumbser}.

We apply the path-conservative scheme to model the transport of CRs along the magnetic field. In this case the fastest wave at any given interface is the light-like wave that travels with velocity $\sim c / \sqrt{3} (\mathbfit{b} \mathbf{\cdot} \mathbfit{n})$, where $\mathbfit{n}$ is the interface normal. The magnitude of this velocity hardly differs between the left and right states and both values can be assumed to be equal. We continue by assuming that $S_L$ and $S_R$ have the same magnitude but different signs, i.e.
\begin{align}
    S_L &= -S, \\
    S_R &= +S,
\end{align}
where $S$ is given by Eq.~\eqref{eq:cr_wavespeed}. In this case, the solution for $\mathbfit{U}_*$ is given by:
\begin{align}
    \mathbfit{U}_* = \frac{\mathbfit{U}_L + \mathbfit{U}_R}{2} &+ \frac{ \mathbfss{H}_R}{2 S} (\mathbfit{U}_* - \mathbfit{U}_R) \nonumber\\ &-\frac{ \mathbfss{H}_L}{2 S} (\mathbfit{U}_* - \mathbfit{U}_L).
\end{align}
The expression for $f_{\mathrm{cr}}^*$ in Eq.~\eqref{eq:fstar} is derived using this equation. For our application, $\mathbfss{H}$ only depends on the direction of the magnetic field. For our assumed operator-splitting, this is a constant during the parallel transport step. Thus we can readily calculate the $\mathbfss{H}_L$ and $\mathbfss{H}_R$ without any iterative solution. The expression for $b_L$ and $b_R$ in Eqs.~\eqref{eq:b_l} and \eqref{eq:b_r} are $\mathbfss{H}_L$ and $\mathbfss{H}_R$ terms evaluated for $\fcr$. They can be derived by solving the corresponding integrals assuming that
\begin{align}
    (\mathbfit{b} \mathbf{\cdot} \mathbfit{n})_* = \frac{1}{2} \left[(\mathbfit{b} \mathbf{\cdot} \mathbfit{n})_L + (\mathbfit{b} \mathbf{\cdot} \mathbfit{n})_R\right].
\end{align}

\section{ODE-integrator Convergence Proofs}
\label{app:ode-integrator-proofs}
Numerical solutions of an ordinary differential equation (ODE) converge to an analytical solution provided the integrator satisfies the \textit{consistency conditions} for the ODE. The consistency conditions are derived by Taylor expanding $\mathbfit{U}^{n+1}=\mathbfit{U}(t^n + \Delta t)$ for small $\Delta t$ and by subsequently substituting derivatives by the ODE itself. The consistency conditions up to third-order in $\Delta t$ of the ODE in Eq.~\eqref{eq:ode} read:
\begin{align}
    \mathbfit{U}^{n+1} &= \left[ \mathbf{1} + \Delta t \mathbfss{R} \phantom{\frac{\Delta t^2}{2}}  \right. \nonumber \\  &\hspace{22.25pt}\left. + \frac{\Delta t^2}{2} \left(\mathbfss{R}^2 + \mathbfss{R}_\mathbfit{U} \mathbfss{R}\mathbfit{U}^{n} \right)+ \mathcal{O}\left(\Delta t^3\right) \right] \mathbfit{U}^{n},
    \label{eq:consistency_eq}
\end{align}
where $\mathbfss{R} = \mathbfss{R}(\mathbfit{U}^{n})$ and $\mathbfss{R}_\mathbfit{U} = \mathrm{grad}_\mathbfit{U} \mathbfss{R}(\mathbfit{U}^{n})$. We proof that our integrator in Eqs.~\eqref{eq:rate_predictor} to \eqref{eq:state_corrector} fulfils the consistency conditions by Taylor expanding every integrator stage for small $\Delta t$. We obtain for the stared stage:
\begin{align}
    \mathbfit{U}^* &= [\mathbf{1} - \gamma \Delta t \mathbfss{R}(\mathbfit{U}^r) ]^{-1} \mathbfit{U}^n  \nonumber\\
                 &= \left[\mathbf{1} + \gamma \Delta t \mathbfss{R}(\mathbfit{U}^r) + \gamma^2 \frac{\Delta t^2}{2} \mathbfss{R}^2_r + \mathcal{O}\left(\Delta t^3\right) \right] \mathbfit{U}^n,
    \label{eq:star_stage_expansion}
\end{align}
while the result for the final stage is:
\begin{align}
    \mathbfit{U}^{n+1} &= [\mathbf{1} - \gamma \Delta t \mathbfss{R}(\mathbfit{U}^r) ]^{-1} \left[ \mathbfit{U}^n + (1 - \gamma) \mathbfit{U}^*\right] \nonumber\\
                     &= [\mathbf{1} - \gamma \Delta t \mathbfss{R}(\mathbfit{U}^r) ]^{-2}  \left[\mathbf{1}  +  (1 - 2 \gamma) \Delta t \mathbfss{R}(\mathbfit{U}^r)\right]  \mathbfit{U}^n \nonumber \\
                     &= \left[\mathbf{1} + \Delta t \mathbfss{R}(\mathbfit{U}^r) \right. \nonumber\\ &\hspace{20pt} \left.+ \left(2 \gamma - \gamma^2\right) \Delta t^2 \mathbfss{R}(\mathbfit{U}^r)^2 + \mathcal{O}\left(\Delta t^3\right)\right]  \mathbfit{U}^n.
    \label{eq:final_stage_expansion}
\end{align}
It is sufficient to expand the predicted rate matrix $\mathbfss{R}(\mathbfit{U}^r)$ up to $\mathcal{O}\left(\Delta t^2\right)$ to reach the desired overall third-order accuracy because it always enters Eqs.~\eqref{eq:final_stage_expansion} and \eqref{eq:star_stage_expansion} together with an additional factor of $\Delta t$. We get:
\begin{align}
    \mathbfss{R} (\mathbfit{U}^r)  &= \mathbfss{R} \left(\mathbfit{U}^n + \frac{\Delta t}{2} \mathbfss{R} \mathbfit{U}^p \right)  \nonumber \\
    &= \mathbfss{R}\left(\mathbfit{U}^n + \frac{\Delta t}{2}  \mathbfss{R} \mathbfit{U}^n + \mathcal{O}(\Delta t^2) \right)  \nonumber \\
    &= \mathbfss{R} + \frac{\Delta t}{2} \mathbfss{R}_\mathbfit{U} \mathbfss{R} \mathbfit{U}^n + \mathcal{O}\left(\Delta t^2\right).
    \label{eq:rate_matrix_expansion}
\end{align}
We conclude that the integrator is second-order consistent if and only if
\begin{equation}
    \gamma_\pm = 1 \pm \frac{1}{\sqrt{2}}
\end{equation}
after substituting Eq.~\eqref{eq:rate_matrix_expansion} into Eq.~\eqref{eq:final_stage_expansion} and comparing the result to Eq.~\eqref{eq:consistency_eq}. We discarded the '+'-solution because in this case the star stage would predict a solution at $t^* = t^n + \gamma_+ \Delta t  > t^n + \Delta t$ which would limit the overall stability of the integrator.

\begin{figure}
	\includegraphics[width=\columnwidth]{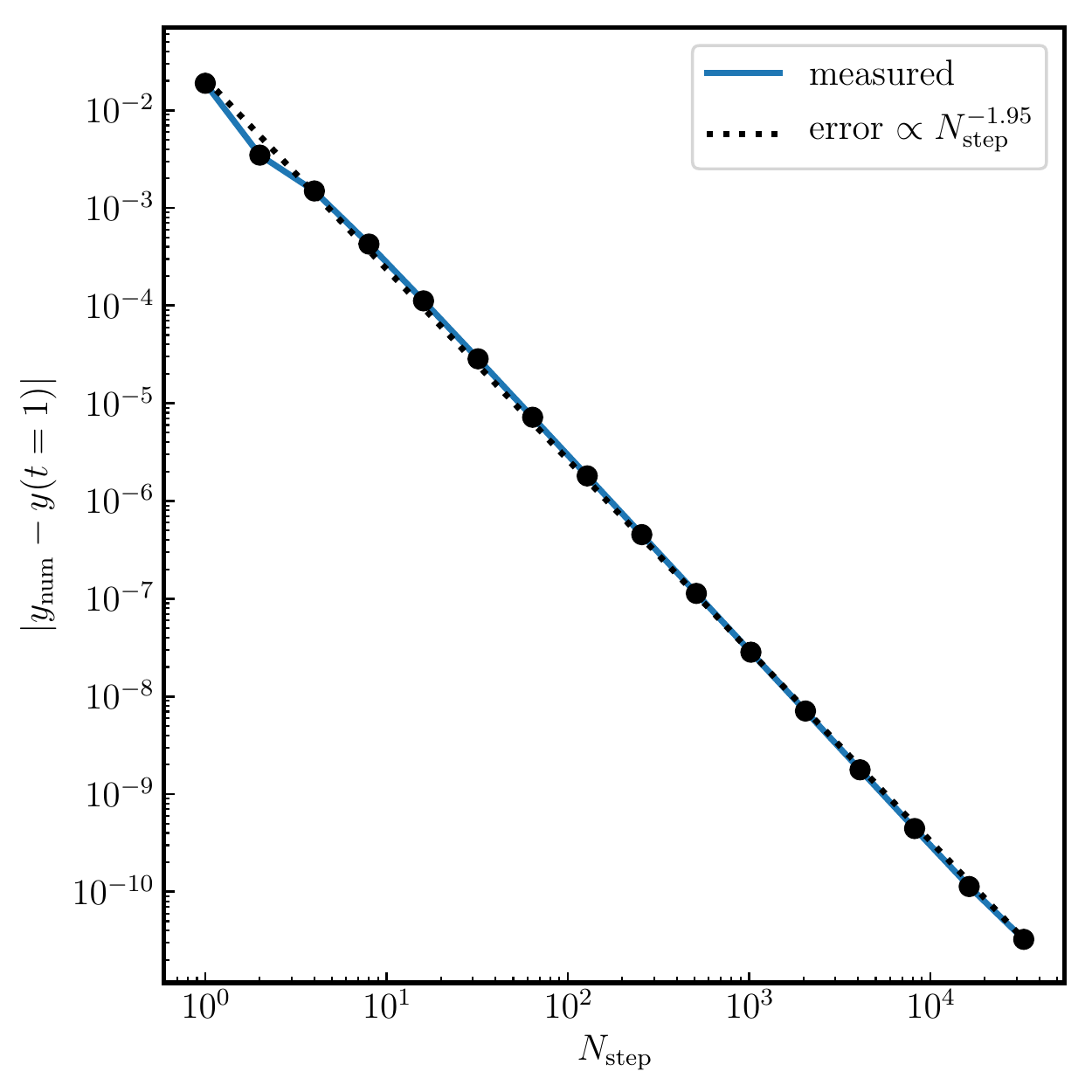}
    \caption{Convergence test for our ODE integrator applied to the initial value problem provided in Eq.~\eqref{eq:integator_test_ivp}. We display the absolute error of the numerical solution using a blue line and black dots for different $\Delta t = 1 / N_\mathrm{step}$. The black dotted line shows a power law fit to these errors.}
    \label{fig:integrator_convergence_test}
\end{figure}

We numerically test our integrator by applying it to the initial value problem:
\begin{align}
    \frac{\mathrm{d} y}{\mathrm{d} t} &= y - y^2,
    \label{eq:integator_test_ivp} \\
    y(0) &= 2,
\end{align}
which has the analytical solution:
\begin{equation}
    y(t) = \frac{\exp{(t)}}{\left(\frac{1}{2} - 1\right) + \exp{(t)}}.
    \label{eq:integator_test_analytical_solution}
\end{equation}
This differential equation is a scaled version of the equation for Alfv{\'e}n wave energy density for fixed values of $\ecr$ and $\fcr$. The performance of the integrator for this reduced problem is thus indicative of its accuracy for the entire system of equations.

We numerically integrate $y$ to $t=1$ using $\Delta t = 1 / N_\mathrm{step}$ and vary $N_\mathrm{step}$ from 1 to 32768. In Fig.~\ref{fig:integrator_convergence_test} we show the absolute difference of the numerical solution and analytical solution of  Eq.~\eqref{eq:integator_test_analytical_solution}. The numerical errors behave slightly worse than expected and scale with $\Delta t^{1.95}$.

% Don't change these lines
\bsp	% typesetting comment
\label{lastpage}
\end{document}